Griffith University, Gold Coast, Australia

# Explaining the Mechanism of Growth in the Past Two Million Years – Vol. I

The generally accepted fundamental postulates are contradicted by data

Prof. Ron W. Nielsen
(aka Jan Nurzynski)

2017

# Explaining the Mechanism of Growth in the Past Two Million Years
# Vol. I

The generally accepted fundamental postulates are contradicted by data

Prof. Ron W. Nielsen
(aka Jan Nurzynski)
Griffith University, Gold Coast, Qld, 4222. Australia
ronwnielsen@gmail.com

2017

**About the Author:** http://home.iprimus.com.au/nielsens/ronnielsen.html

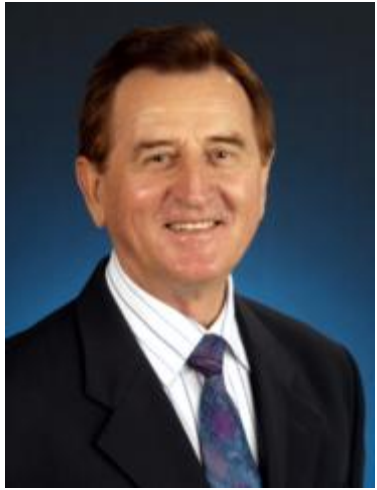

**Abstract:** Economic growth and the growth of human population in the past 2,000,000 years are extensively examined. Data are found to be in a clear contradiction of the currently accepted explanations of the mechanism of growth, which revolve around two fundamental but incorrect doctrines: (1) the doctrine of stagnation (inappropriately labelled also as Malthusian stagnation, because Malthus never claimed that his positive checks would cause a long-lasting and wide-spread stagnation) and (2) the doctrine of explosion described also as a takeoff, sprint, spike or by other similar attributes. These doctrines and other related postulates are contradicted even by precisely the same data, which are used in the economic research and by the research results published in a prestigious scientific journal as early as in 1960. The generally accepted explanations are not based on a rigorous analysis of data but on impressions created by the easily misleading features of hyperbolic distributions. Two leading theories: the Demographic Transitions Theory (or Model) and the Unified Growth Theory are fundamentally incorrect. Descriptions of the past socio-economic conditions are not questioned. They might have been harsh, difficult and primitive but they are not reflected in the growth trajectories. They did not create stagnation in the economic growth and in the growth of population. Likewise, impacts of the Industrial Revolution on many aspects of life are not questioned. It is only demonstrated that this event had absolutely no impact on shaping growth trajectories. A general law of growth is formulated and used to explain the mechanism of growth of human population and of economic growth. The growth was predominantly hyperbolic. Such a growth is described by exceptionally simple mathematical function and the explanation of the mechanism of growth turns out to be also simple.

**Contents**

**Preface**

The spontaneous and unconstrained economic growth and the growth of human population is hyperbolic. Unfortunately, this type of growth, if not rigorously analysed, can easily lead to incorrect conclusions and even experienced researchers can be easily mislead by its features.

Hyperbolic growth is slow over a long time and fast over a short time but it is still the same, *monotonically increasing*, growth. Its interpretations should never be based on dividing it into two or three distinctly different components governed by distinctly different mechanisms. Unfortunately, this mistake is repeatedly demonstrated in scientific literature. The slow growth is interpreted as stagnation and the fast growth as explosion. Different mechanisms of growth are suggested for the two conceived but non-existing parts. They belong to the same, monotonically increasing, trajectory.

Hyperbolic distributions have to be interpreted as a whole. The same mechanism should be applied to the slow growth and to the fast growth because they belong to the same distribution. If we apply the mechanism of stagnation to the slow growth, we should apply precisely the same mechanism to the fast growth, which would be obviously incorrect. Likewise, if we apply the mechanism of explosion to the fast growth, we should also apply the same mechanism to the slow growth.

Even though hyperbolic distributions appear to be made of two distinctly different components, it is impossible to locate on them a transition from the slow growth to the fast growth. There is no transition, or more precisely, the transition occurs slowly over the *whole range* of hyperbolic distribution.

Unfortunately, current interpretations of the mechanism of economic growth and of the growth of population are *not* based on the rigorous analysis of data by on the incorrect and misleading impressions created by hyperbolic distributions. These distributions have to be carefully and methodically analysed.

Rigorous analysis of data point to a fundamentally serious problem with the currently accepted interpretations of the economic growth and of the growth of population. It is not just some minor corrections or adjustments that have to be made but rather it is the need of turning the research around and basing it on radically new concepts. The fundamental doctrines have to be abandoned and replaced by the evidence-based interpretations. They have to be based on accepting hyperbolic growth.

I am presenting here a compilation of my articles aimed at correcting the current misconceptions about the economic growth and about the growth of population. The presented chapters are arranged approximately in the order of the publication of the relevant articles but there are some exceptions. Notably, Chapter 2 represents the last publication in these series. It is it presented at the beginning of this compilation because it contains a convenient overview of conclusions based on this extensive research. All these chapters explain why economic and demographic research has been on the wrong track for a long time and why it is unavoidable to turn it around and to follow new directions as indicated by data. This is how science works. Even the most cherished and the best-established doctrines have to be abandoned if they are contradicted by data, and the sooner it is done the better.

*A guide to chapters*

Chapter 1.

- Reasons for this series of investigations.
- Hyperbolic illusions.
- Examples of scientifically unacceptable conjectures.

Chapter 2.

- Method of reciprocal values.
- First examples showing how the Unified Growth Theory (Galor, 2005, 2011) is contradicted by data.
- An overview of the results presented in the remaining chapters.

Chapters 3 and 6

- Analysis of the global and regional growth of the Gross Domestic Product (GDP) during the AD era.

- Economic growth was hyperbolic.
- There was no stagnation and no Malthusian trap in the economic growth.
- There were no takeoffs at any time to a distinctly faster growth.
- Industrial Revolution did not shape growth trajectories.
- Unified Growth Theory (Galor, 2005, 2011) contradicted by the same data, which were used during its development.

Chapter 4

- Growth of the world population in the past 12,000 years was hyperbolic. The early observation of von Foerster, Mora and Amiot (1960) for the AD era have been now confirmed and extended to the BC era.
- There was no stagnation in the growth of the world population and no transition from stagnation to growth.
- What is commonly interpreted as the population explosion is just the natural continuation of the hyperbolic growth.
- There was only one major transition during that time but it was not the usually claimed transition from stagnation to growth but from hyperbolic growth to hyperbolic growth.
- There was only one example when demographic catastrophes might have had impact on shaping the growth trajectory. It was a unique event when there was an unusual convergence of five major demographic catastrophes. However, their combined impact was small and short-lasting.

Chapter 5

- Demographic Transitions Theory is contradicted repeatedly by data, even by the same data, which are used in its defence.
- Fluctuations in birth and death rates, as well as fluctuations in the growth rate, cannot be used as the evidence of the so-called Malthusian stagnation because they do not shape the population growth trajectories.

Chapter 6

- The concept of differential takeoffs in the economic growth proposed by Galor (2005, 2008, 2011, 2012) is examined. Differential takeoffs never happened.
- Industrial Revolution had absolutely no impact on shaping economic growth trajectories, global or regional

Chapter 7

- Three regimes of growth postulated by Galor (2005, 2011) did not exist. His postulate is based on his habitually distorted presentations of data (Maddison, 2001). Precisely the same data, when analysed, contradict his fundamental postulates.
- Industrial Revolution had no impact on shaping economic growth trajectories.

Chapter 8

- Growth of population, global and regional, during the AD era was hyperbolic.
- There was no stagnation in the growth of population and no transition from stagnation to growth.
- Unified Growth Theory is contradicted by the same data, which were used but not analysed during its formulation.

Chapter 9

- Puzzling properties of the *growth rate* of income per capita, described by Galor as a mystery of the growth process (Galor, 2005), are explained. This mystery was created by his habitually distorted presentation of data.
- The growth rate of income per capita (GDP/cap) was increasing monotonically. There was no sudden spike claimed by Galor.

- Industrial Revolution had no impact on shaping the GDP/cap trajectory.

Chapter 10

- Gross Domestic Product per capita (GDP/cap), global and regional, was increasing monotonically by following the linearly-modulated hyperbolic distributions.
- There was no stagnation in the growth of income per capita and no takeoffs from stagnation to growth.
- Industrial Revolution had absolutely no impact on changing economic growth trajectories. There was no boosting in the economic growth. On the contrary, shortly after the Industrial Revolution, economic growth was diverted to slower trajectories.

Chapter 11

- The following claim is examined: "The mind-boggling phenomenon of the Great Divergence in income per capita across regions of the world in the past two centuries, that accompanied the take-off from an epoch of stagnation to a state of sustained economic growth, presents additional unresolved mysteries about the growth process" (Galor, 2005, p. 220).
- This single sentence is overflowing with misinformation.
- The great divergence never happened. This feature was created by the habitually distorted presentations of data.
- The epoch of stagnation did not exist.
- There was no takeoff from stagnation to growth but a monotonically increasing hyperbolic growth.
- Various regions are on different levels of economic growth but they never diverged to distinctly different trajectories.
- It is shown that we can take any set of hyperbolic distributions and construct a great divergence by selecting a few strategically located points on these distributions. This constructed great divergence has nothing to do with the mechanism of the economic growth but reflects only the distortions introduced to the purely mathematical distributions.

Chapter 12

- Effects of Malthusian positive checks (Malthus, 1798) are investigated using the UN data (UNDP, 2011).
- His observations are confirmed.
- Malthusian positive checks are responsible for increasing the death rate but they also trigger a process of regeneration.
- Malthusian positive checks do not produce stagnation but stimulate growth.

Chapter 13

- Using the data of Maddison (2010) it is shown that Industrial Revolution had absolutely no impact on shaping the trajectories describing economic growth and the growth of population in the United Kingdom, the centre of this revolution.

Chapter 14

- The fine structure of growth trajectories is explained.
- It is incorrect to claim that fluctuations in the growth rate can be taken as the manifestation of the so-called Malthusian stagnation (see for instance Artzrouni & Komlos, 1985; Lagerlöf, 2003) because these fluctuations have no impact on shaping growth trajectories.

Chapter 15

- Various attempts of explaining the mechanism of hyperbolic growth are discussed.

Chapter 16

- The general law of growth is formulated. This law links directly the force of growth with the growth trajectories and thus allows for an easy way of studying the mechanism of growth.

Chapter 17

- The mechanism of hyperbolic growth of population and of economic growth is explained.

Chapter 18

- The growth of the ancient population in Australia is examined because a sudden transition from a slow to a fast growth, described as the intensification of growth, was claimed by Johnson and Brook (2011).
- It is demonstrated that the claimed intensification never happened. The growth of population in Australia was increasing monotonically.
- The claimed intensification represents a typical mistake made with the interpretation of hyperbolic distributions but this time it was made in a discipline, which is outside the mainstream of the demographic research.

Chapter 19

- Puzzling properties of income per capita (GDP/cap) are explained. They are purely mathematical properties of dividing two hyperbolic distributions.
- The GDP/cap data are described by the linearly modulated hyperbolic distributions.

Chapter 20

- Growth of population was hyperbolic not just in the past 12,000 years as demonstrated in Chapter 4, but also in the past 2,000,000 years.
- This analysis confirms the earlier observation of Deevey (1960) that the growth of population over such a long time was in three stages. However, Deevey expected that each stage was leading to an equilibrium. Data show diametrically different shapes: each stage was represented by a hyperbolic growth, which if continued would increase to infinity at a fixed time.
- The commonly claimed epoch of stagnation never existed. There was no stagnation in the growth of population in the past 2,000,000 years.
- The repeatedly claimed takeoffs from stagnation to growth never happened because there was no stagnation. There were only transitions from hyperbolic growth to hyperbolic growth.
- Hyperbolic growth was remarkably stable. There were only two major demographic transitions in the past 2,000,000 but they were not transitions from stagnation to growth but from hyperbolic growth to hyperbolic growth.
- Using the characteristic properties of income per capita, mathematical description of the economic growth is also extended over the past 2,000,000 years. The pattern is the same as for the growth of population.

Chapter 21

- A survey of demographic catastrophes shows that individually they were too weak to have a significant impact on the growth of human population.
- Contrary to the established knowledge, demographic catastrophes did not shape the growth of population and the economic growth.

*New directions*

1. There is no need to explain the mechanism of the so-called Malthusian stagnation because there was no stagnation in the economic growth and in the growth of human population. The concept of stagnation is inapplicable to the description and to the explanation of the mechanism of the economic growth and of the growth of human population.

2. The concept of the so-called Malthusian stagnation was not proposed by Malthus and neither was the concept of the so-called Malthusian trap. Malthus has never claimed any form of a prolonged

    and widespread stagnation in the growth of population or in the economic growth. These concepts were erroneously introduced much later and are incorrectly linked with Malthus.

3. Malthus noticed the dual effects of his positive checks: they increase the rate of mortality but they also stimulate growth (Nielsen, 2016a, Malthus, 1798). He also suggested that this stimulating effect should be further investigated. The combined impact of Malthusian positive checks is that they *increase* the growth rate. Malthusian positive checks do not cause stagnation but stimulate growth.

4. Discussions of the past socio-economic conditions might be interesting but there is no obvious connection between them and the growth trajectories describing economic growth and the growth of population. Growth trajectories are immune to the changing socio-economic conditions. They cannot explain the mechanism growth.

5. There is no need to explain the escape from the so-called Malthusian trap because there was no trap in the economic growth and in the growth of population.

6. There is no need to explain the transition from stagnation to growth because there was no such transition. The growth was hyperbolic in the past and continued to be hyperbolic at the time of the alleged transitions. The alleged transitions represent the natural continuation of the monotonically increasing hyperbolic distributions.

7. Hyperbolic growth was remarkably stable. There were only two major transitions in the past 2,000,000 but they were not transitions from stagnation to growth but from hyperbolic growth to a new hyperbolic growth.

8. There is no need to explain the boosting effects of the Industrial Revolution because Industrial Revolution had no impact on shaping the economic growth trajectories and the trajectories describing the growth of population.

9. There is no need to explain the mechanism of takeoffs because they never happened.

10. There is no need to explain the differential timing of takeoffs (Galor, 2005, 2011) because there were no takeoffs.

11. There is no need to explain the numerous mysteries of growth listed by Galor in his Unified Growth Theory because there are no such mysteries. They were created by a repeatedly distorted presentations of data. Analysis of the same data shows that there were no mysteries.

12. There is no need to explain the "mind boggling phenomenon of the Great Divergence" (Galor, 2005, p. 220) because the great divergence never happened. It was again created by a distorted presentation of data. Any set of monotonically increasing hyperbolic distributions can be used to create great divergence when they are suitably distorted as repeatedly done in the Unified Growth Theory and in other similar publications. Economic growth in various regions is now on different levels of development but they all follow closely similar trajectories governed by the same mechanism of growth.

13. There is no need to explain the sudden spike (Galor, 2005, 2011) in the growth rate of income per capita because there was no spike.

14. There is no need to explain a transition from a nearly horizontal to a nearly vertical growth of income per capita (GDP/cap) because there was no transition. These perplexing and puzzling features have now been explained: they represent nothing more than just the mathematical property of dividing two hyperbolic distributions but they also represent the monotonically increasing distributions without any sudden transition between the two apparently different patterns of growth.

15. Demographic Transitions Theory and the Unified Growth Theory are contradicted by data and should not be used in the demographic and economic research.

16. Demographic and economic research has to be based on accepting that the historical growth of population and the economic growth were following hyperbolic distributions.

17. Growth of population and economic growth were predominantly hyperbolic in the past 2,000,000 years and, as first noticed by Deevey (1060), it was in three major staged. Economic growth followed a similar pattern.

18. The mechanism of the hyperbolic economic growth and of the growth of population has been now explained (Nielsen, 2016b). Hyperbolic growth is the simplest unconstrained growth prompted by the fundamental forces: the indispensable force of procreation in the case of the growth of population and the simplest market force in the case of the economic growth. These were the dominating forces in the past.

19. Demographic catastrophes did not shape the growth of population or the economic growth. The only example when they might have caused a minor and insignificant disturbance in the growth trajectory was when there was an unusual and unique combination of five major demographic catastrophes.

20. The currently accepted interpretations of the mechanism of economic growth and of the growth of population are not only in conflict with data used in the economic research but they are also dangerously misleading because they create the sense of security by claiming that after a long stage of stagnation, extending over thousands of years, we have finally escaped the Malthusian trap and entered into a sustained growth regime. Scientific evidence shows that the opposite is true. The past growth was stable, secure and sustainable, as demonstrated by the largely stable hyperbolic distributions. In contrast, it is the current growth, which is potentially unsustainable. For the first time in human existence, our ecological footprint is larger than ecological capacity and it continues to increase. For the first time in human history we support our existence on the continually increasing ecological deficit.

21. Conjectures and inspiration are acceptable in science but they have to be controlled and moderated by data. When conjectures are supported by conjectures, the created system of doctrines and explanations becomes quickly scientifically untenable.

Ron W. Nielsen  
Gold Coast, Australia  
September, 2017

# Scientifically unacceptable established knowledge in demography and in economic research

*By* Ron W. NIELSEN †

**Abstract.** The established knowledge in demography and in the economic research is based on the concept of the so-called Malthusian stagnation and on the associated concept of the escape from the Malthusian trap. These two fundamental concepts were gradually enforced by numerous other related postulates all aimed at explaining the mechanism of the historical growth of population and of the historical economic growth. Examples of publications based on the established knowledge are closely examined. They are used to show why the established knowledge is scientifically unacceptable. It is also pointed out that the established knowledge is contradicted by data and by their analyses. Interpretations of the historical economic growth and of the historical growth of population has to be based on accepting hyperbolic growth. However, the discussed examples point to a more serious problem in these two fields of research. It is a fundamental systemic problem, the problem associated with the way research is conducted. Doctrines, interpretations and declarations used by the established knowledge have to be often accepted by faith. Data are either ignored or manipulated to support preconceived ideas. Contradicting evidence is methodically ignored. To be recognised as science, demographic and economic research has to adhere to the scientific rules of investigation.

**Keywords.** Economic growth, Population growth, Gross Domestic Product, Hyperbolic growth, Malthusian stagnation, Malthusian trap, Malthusian positive checks, Malthusian oscillations, Fertility rate, Mortality rate, Famines, Pestilence, Wars
**JEL.** A10, A12, A23, B22, B41, C12, Y80.

## 1. Introduction

T wo fields of research, economic growth and the growth of population, which might appear to be distinctly different, are in fact closely related for at least three reasons. *First*, there is obviously no economic growth without humans. *Second*, there is a close correlation between economic growth and the growth of human population (Nielsen, 2016a, 2016b). *Third*, in order to understand the growth of income per capita, measured by the Gross Domestic Product per capita (GDP/cap), it is obviously necessary to study not only the economic growth but also the growth of human population. It is *inter alia* for these reasons, that the best source of information about the historical economic growth, compiled by the world-renown economist, includes not only the data describing the growth of the GDP but also the growth of population (Maddison, 2001, 2010).

## 2. The established knowledge

The established knowledge in demography and in the economic research revolves around two fundamental concepts: the so-called Malthusian stagnation and the explosion, which is supposed to have marked a dramatic escape from the so-called Malthusian trap. Gradually and by accretion, in the process extending over many years, these two fundamental concepts were adorned by various additional explanations, speculations and conjectures all adding to the now established knowledge based on the scientifically unacceptable doctrines and beliefs. These two fundamental regimes of growth, stagnation and explosion, are described as Stage 1 and Stage 2, respectively, in the Demographic Transition Theory

† AKA Jan Nurzynski, Griffith University, Environmental Futures Research Institute, Gold Coast Campus, Qld, 4222, Australia.
☎. +61407201175
✉. ronwnielsen@gmail.com

(see Nielsen, 2016c and references therein). The epoch of stagnation was supposed to have lasted for many thousands of years and was allegedly strongly controlled by the Malthusian positive checks (Malthus, 1798) generating an unstable stage of growth characterised by irregular Malthusian oscillations. The mechanism of growth is claimed to have changed dramatically at the time of the alleged population explosion when the growth was supposed to have changed from slow to fast. The transition from stagnation to explosion is described as the great escape from the Malthusian trap.

We have already demonstrated that the established knowledge is convincingly contradicted by the relevant data and by their analyses (Biraben, 1980; Clark,1968; Cook,1960; Durand, 1974; Gallant, 1990; Haub, 1995; Kapitza, 2006; Kremer, 1993; Lehmeyer, 2004; Livi-Bacci, 1997; Maddison, 2001, 2010; Mauritius, 2015; McEvedy & Jones, 1978; Nielsen, 2014, 2015a, 2016a, 2016b, 2016c, 2016d, 2016e, 2016f, 2016g, 2016h, 2016i; Podlazov, 2002; Shklovskii, 1962, 2002; Statistics Mauritius, 2014; Statistics Sweden, 1999; Taeuber & Taeuber, 1949; Thomlinson, 1975; Trager, 1994, United Nations, 1973, 1999, 2013; von Hoerner, 1975, von Foerster, Mora & Amiot, 1960; Wrigley & Schofield, 1981). The aim of this publication is (1) to outline briefly the origin of the established knowledge, (2) to explain why the established knowledge is so strongly established, (3) to explain the deceptive evidence in data, which can be used in support of the established knowledge, (4) to give a few examples of how strongly the established knowledge is established and (5) to explain why the established knowledge as illustrated by these examples is scientifically unacceptable.

## 3. Evidence in data

Data describing the historical growth of population and the historical economic growth are hardly ever analysed. Recently, attempts were made to use some of these data (Maddison, 2001) but they were presented in grossly distorted and misleading diagrams, which appear to be supporting the established knowledge (Ashraf, 2009; Galor, 2005a, 2005b, 2007, 2008a, 2008b, 2008c, 2010, 2011, 2012a, 2012b, 2012c; Galor & Moav, 2002; Snowdon & Galor, 2008). Data were not analysed to learn from them but manipulated to support preconceived ideas. Such approach to research is scientifically unacceptable. Data have to be carefully and methodically analysed to avoid drawing incorrect conclusions. Their superficial examination creates strong impression of stagnation followed by explosion but when closely analysed they show that the apparent explosion was just the natural continuation of the past hyperbolic growth.

Global population in 10,000 BC is estimated at only between 1 and 10 million (McEvedy & Jones, 1978, Thomlinson, 1975). Now the population of this size can be located in just a single city. By AD 1, global population increased to only a few hundred million. The estimated values vary between 170 and 400 million (Biraben, 1980; Durand, 1974; Haub, 1995; McEvedy & Jones, 1978; Thomlinson, 1975; United Nations, 1973, 1999). Now, the population of this size or even larger can be found in just a single country.

The first billion of global population was reached around AD 1800 (Biraben, 1980; Durand, 1974; McEvedy & Jones, 1978; Thomlinson, 1975; United Nations, 1973, 1999) and from that time on the growth was progressing exceedingly fast. The origin of *Homo Sapiens* is usually claimed at around 200,000 years ago, but it might have been even earlier (Weaver, Roseman & Stringer, 2008). Thus, it took many thousands of years for the world population to increase to one billion but after reaching the first billion, the second billion was added in just only about 130 years (United Nations, 1999). The process of many hundreds of thousands of years was suddenly compressed to just over 100 years. The consumption of natural resources and the stress on the environment started to increase rapidly.

If adding one billion in just 130 years sounds too fast, the next billion was added in just 29 years, the next in 15 years, the next in 13 years, and the next in 12 years, increasing the size of global population to 6 billion (US Census Bureau, 2016). The last billion, which increased global population to 7 billion, was added in 13 years (US Census Bureau, 2016). We call it the slowing-down growth but obviously the slowing down process is still too slow.

Assuming a medium-intensity growth, the size of the world population is projected to increase to 8.39 billion in 2030 and 9.63 billion in 2050 reaching a maximum of 10.48 billion around 2080 (Nielsen, 2006). These projections are in good agreement with the US Census Bureau (2016) projections of 8.34 billion in 2030 and 9.41 billion in 2050. It is what we hope for, but the high intensity growth could lead to 12.26 billion by the end of the current century (Nielsen, 2006), assuming that such a growth can be supported by the availability of natural resources.

© RON W. NIELSEN, 2017, Explaining the Mechanism of Growth in the Past Two Million Years
______________________________________________________________________________________Similar surprising pattern of a slow growth in the past and a fast growth in recent years is reported for the growth of the Gross Domestic Product (Maddison, 2001, 2010). The first trillion dollars ($10^{12}$) of the GDP (expressed in the 1990 international Geary-Khamis dollars) was reached in 1870. The next trillion was added in just 51 years, the next in 19 years and the next in only 10 years, increasing global GDP to $4 trillion in 1950. By 1998, global GDP increased to $34 trillion. The latest estimate for 2014 is $91 trillion (World Bank, 2016) and the projected value for 2050 is $118 trillion (Nielsen, 2015b).

Using such numbers, it would be easy to conclude that there was a long epoch of stagnation in the past economic growth and in the growth of human population and that this stagnation was followed by a sudden explosion. However, such a conclusion, which is the corner stone of the established knowledge in demography and in the economic research, would be unscientific because impressions can be misleading. Scientific research has to be conducted scientifically. If economic and demographic research is supposed to be recognised as science they have to adhere to the scientific rules of investigation.

In science, data have to be methodically analysed. This fundamental requirement in scientific research appears to have been ignored in economic and demographic research. Hasty conclusion about stagnation followed by explosion is also clearly incorrect and scientifically unacceptable because over 50 years ago, von Foerster, Mora and Amiot, (1960) demonstrated that the growth of population during the AD era was hyperbolic. This crucial contribution to science should not have been ignored. It should have been further investigated because hyperbolic growth rules out the interpretations based on the assumption of stagnation followed by explosion.

Postulates of the established knowledge are also unacceptable because hyperbolic growth have been recognised and confirmed by other independent investigations (Kapitza, 2006; Kremer, 1993; Podlazov, 2002; Shklovskii, 1962, 2002; von Hoerner, 1975). Accepting the fundamental postulates of established knowledge is scientifically unjustified because for a long time now there was a large body of data describing the growth of population not only during the AD era but also during the BC era (Biraben, 1980; Clark,1968; Cook,1960; Durand, 1974; Gallant, 1990; Haub, 1995; Livi-Bacci, 1997; McEvedy & Jones, 1978; Taeuber & Taeuber, 1949; Thomlinson, 1975; Trager, 1994, United Nations, 1973, 1999, 2013). These data should have been analysed to check the earlier claims about the hyperbolic growth.

Fundamental postulates of the established knowledge are now contradicted by the excellent new data describing economic growth and the growth of population (Maddison, 2001, 2010). These postulates are scientifically unacceptable because they are consistently contradicted by the analysis of relevant data (Nielsen, 2013a, 2013b, 2013c, 2014, 2015a, 2016a, 2016b, 2016c, 2016d, 2016e, 2016f, 2016g, 2016h, 2016i).

Data describing birth *and* death rates and the associated growth of population are limited (Lehmeyer, 2004; Mauritius, 2015; Statistics Mauritius, 2014; Statistics Sweden, 1999; Wrigley & Schofield, 1981) but they also show consistently that the established knowledge, as expressed in the Demographic Transition Theory, is contradicted by their analysis (Nielsen, 2016c). We do not even have to analyse these data mathematically to see that they are in contradiction of the established knowledge because even though the birth and death rates and the associated growth rates were fluctuating, their time-dependence does not fit into the patterns claimed by the Demographic Transition Theory. Furthermore, the corresponding distributions describing the growth of population do not display any form of stagnation during the alleged Stage 1 or a transition to the alleged Stage 2, which is supposed to represent the explosion. Data show no such patterns.

Demographic Transition Theory is based on a persistent and blatant disregard for relevant data. This theory is supported by largely meaningless presentations of data for birth *or* death rates. These rates have to be studied *together* and they should show the expected behaviour, as claimed by the Demographic Transition Theory, that the gap between them is approximately zero during the alleged Stage 1 and that it increases during the alleged Stage 2. Such patterns are not confirmed by the best available data (Lehmeyer, 2004; Mauritius, 2015; Statistics Mauritius, 2014; Statistics Sweden, 1999; Wrigley & Schofield, 1981), which show that the Demographic Transition Theory is contradicted by the data describing birth and death rates and by the associated data describing the growth of population. Paradoxically, when methodically analysed, data used in support of the Demographic Transition Theory are in fact in its clear contradiction.

A theory contradicted by just a single set of data is scientifically unacceptable and the Demographic Transition Theory was first contradicted by the results of von Foerster, Mora and Amiot (1960) who demonstrated that the growth of human population during the AD era was hyperbolic and thus that Stages 1 and 2 claimed by this theory did not exist. The Demographic Transition Theory should have been rejected or at least fundamentally modified about 50 years ago. Its continuing use over such a long time has been scientifically unjustified.

The postulate of the so-called Malthusian stagnation followed by explosion, and all other associated postulates and explanations of the historical economic growth and of the historical growth of population followed by a mythical escape from the Malthusian trap have no place in science. They may, however, have a place in the history of science.

## 4. Hyperbolic growth

Hyperbolic distributions are strongly deceptive and it is easy to make a mistake with their interpretation. Fortunately, however, analysis of hyperbolic distributions is also trivially simple (Nielsen, 2014) and it is easy to avoid making an easy mistake.

Examples of two hyperbolic distributions, a hyperbolic distribution describing the growth of the world population during the AD era and the distribution describing the world economic growth, are shown in Figures 1 and 2. Their analysis is based on using the method of reciprocal values (Nielsen, 2014). For a sufficiently wide range of data, hyperbolic distributions can be *uniquely* identified using this method because if the reciprocal values are decreasing linearly, then the growth is hyperbolic. There is no other option. It is something similar to the unique identification of the exponential growth. For a sufficiently large range of good quality data, exponential growth can be uniquely identified by the linear distribution of the logarithm of the size of a growing entity.

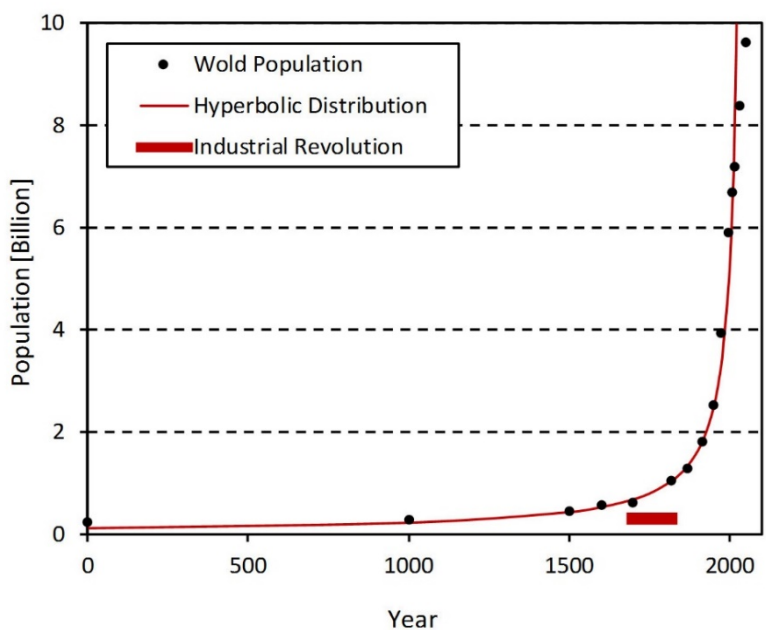

**Figure 1.** *Data describing the growth of the world population (Maddison, 2010) are compared with hyperbolic distribution.*

Figures 1 and 2 show that the growth of human population and economic growth were indeed slow over a long time, but it was hyperbolic growth, which *is* slow over a long time and fast over a short time. It is still the same, *monotonically*-increasing, growth. It is *impossible* to divide such a growth into distinctly-different components and the best way to see it, is to examine the reciprocal values of the size of the growing entity, in our case the reciprocal values of the GDP or of the size of the population (Nielsen, 2014).

Hyperbolic distributions have to be analysed and interpreted as a whole. The same mechanism has to be applied to the slow and fast growth. If we apply the mechanism of the so-called Malthusian stagnation to the slow growth, we have to apply precisely the same mechanism to the fast growth. If we apply the mechanism of explosion to the fast growth, then precisely the same mechanism should be applied to the slow growth, which obviously is incorrect because explosion has to be triggered by something and there was clearly no explosion along the slow growth.

The usually assumed event that was supposed to have triggered population explosion or a sudden takeoff in economic growth or in the growth of population is the Industrial Revolution but as we can see in Figures 1 and 2, there was no sudden explosion during the Industrial Revolution or at any other time. The growth was increasing monotonically. Transition from slow to fast growth takes place all the time. We could demonstrate this monotonic growth even more clearly by using reciprocal values of data or by the semilogarithmic display (Nielsen, 2014, 2016a, 2016b, 2016d, 2016e, 2016f, 2016g, 2016h, 2016i) but the primary aim of presenting these two diagrams is to illustrate the deceptive character of hyperbolic distributions. They can easily lead to incorrect interpretations particularly when they are not analysed but only used to quote certain, well-selected numbers or when they are deliberately manipulated and distorted (Ashraf, 2009; Galor, 2005a, 2005b, 2007, 2008a, 2008b, 2008c, 2010, 2011, 2012a, 2012b, 2012c; Galor & Moav, 2002; Snowdon & Galor, 2008) to support preconceived ideas. Hyperbolic distributions have to be analysed.

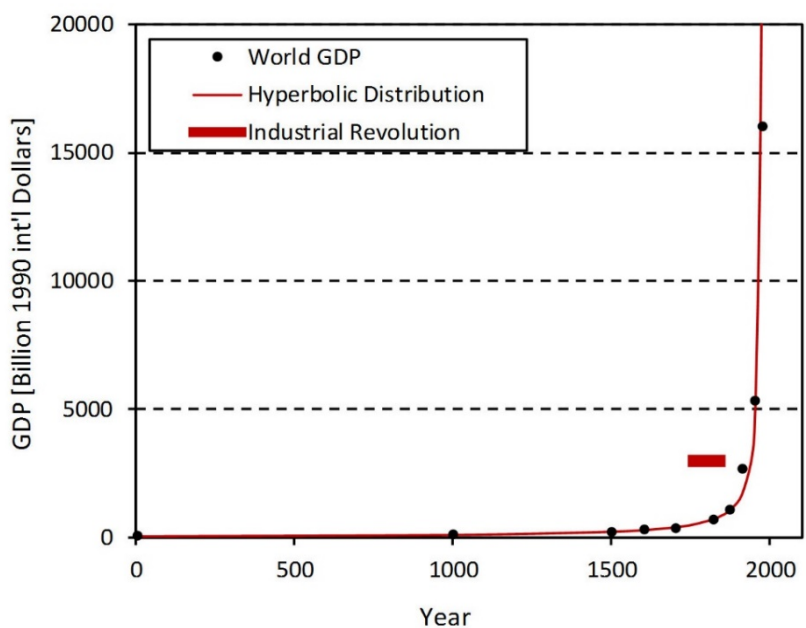

**Figure 2.** *Data describing the growth of the world Gross Domestic Product (Maddison, 2010) are compared with hyperbolic distribution.*

Figure 1 shows also that the growth of population is not yet levelling off. It is still following closely the fast-increasing historical hyperbolic distribution. Maddison's data end in 2008. The point representing the size of the population in 2014 is from the US Bureau of Census (2016) while the last two points are the predicted values (Nielsen, 2006). Not until 2030 or maybe even until 2050 could we expect a clear departure from the historical hyperbolic trend. The future of the population growth is uncertain, in much the same way as the future of the world economic growth (Nielsen, 2015b).

## 5. The origin of the concept of stagnation

Two features make the concept of the so-called epoch of Malthusian stagnation deceptively attractive: (1) it is strongly believable and (2) it is supposed to have originated over 200 years ago. It is believable because the growth of human population and the economic growth over thousands of years were indeed slow, so slow that they appear to have been stagnant. It is also an old concept because its origin is traced, to Malthus (1798), perhaps inaccurately because Malthus never used the word *stagnation* in his book.

> The Malthusian theory, as was outlined initially by Malthus (1978), captures the main attributes of the epoch of Malthusian stagnation that had characterized most of human existence… (Galor, 2005, p. 221).
>
> The idea of multiple equilibria, or poverty traps, can be retraced back to Malthus (Wang, 2005, p. 36).

The work of Malthus was the first well-documented attempt to understand and explain the mechanism of growth of human population but it appears that this is also precisely where it ended. Considering the time when Malthus was writing his book, it was a remarkable achievement, but his work should have been not only checked but also extended using a large body of data, which were not available to Malthus but which are readily available to us.

> The history of population theory can be summarized in three words: pre-Malthusian, Malthusian, and post-Malthusian. Hardly ever in intellectual history does one man so dominate a field as does the Reverend Thomas Robert Malthus in demographic theory. To paraphrase a quotation attributed to Newton, Malthus' shoulders *must* be climbed (Thomlinson, 1965, p. 47. Italics in the original text.).
>
> …the demographic transition experiences three regimes: the 'Malthusian Regime,' the 'Post-Malthusian Regime,' and the 'Modern Growth Regime.' *Any theory* attempts (sic) to describe the process of demographic transition *must* include these three periods (Wang, 2005, p. 3. Italics added.).

Claiming, suggesting or assuming that something *must* be accepted just because it comes from a certain source is not acceptable in science. Any theory can be questioned and even should be questioned, and if necessarily corrected or rejected. The sooner it is done, the better it is for science. If Malthus's shoulders must be climbed it is only for the same reason as climbing the shoulders of any giant of human intellect: to see better and further ahead. It is not just to have a comfortable ride.

However, we are not even climbing Malthus's shoulders. Attaching his name to the concept of stagnation and calling it Malthusian stagnation sounds like defamation. It is questionable whether Malthus would be pleased with such a dubious distinction. We are putting our interpretations into his work and we are claiming that he did it.

If we read his publication carefully, we can find that he was writing not only about the destructive or impeding effects of positive checks but also about their stimulating effects (Nielsen, 2013b; Malthus, 1798). He even suggested that these stimulating effects should be further investigated. Given enough time he would have probably studied this issue further. Discussions of the impeding effects of positive checks, which we label inaccurately as Malthusian stagnation should be balanced by the discussion of their stimulating effects, which Malthus mentions in his book. Malthusian positive checks cannot cause stagnation because their combined effect is to increase the growth rate (Nielsen, 2013b).M

The phrase "Malthusian stagnation" is a misnomer because Malthus never claimed that positive checks would produce prolonged and wide-spread stagnations in the growth of population and because we know now that Malthusian positive checks, even if present, were not producing such effects (Kapitza, 2006; Kremer, 1993; Nielsen, 2013a, 2014, 2016b, 2016c, 2016d; Podlazov, 2002; Shklovskii, 1962, 2002; von Hoerner, 1975, von Foerster, Mora & Amiot, 1960). They appear to have been generally either too weak or their destructive impacts were effectively compensated by the well-known, natural process of regeneration (Nielsen, 2013a, 2013b, 2013c).

It would be interesting to search for impacts of Malthusian positive checks on the growth of population by investigating the growth of local populations. Generally, there appears to have been no impact. The only known example (Nielsen, 2016d) is a minor distortion in the growth of the world population between AD 1200 and 1400, which appears to be correlated with the convergence of *five* major demographic catastrophes: Mongolian Conquest (1260-1295) with the total estimated death toll of 40 million; Great European Famine (1315-1318), 7.5 million; the 15-year Famine in China (1333-1348), 9 million; Black Death (1343-1352), 75 million; and the Fall of Yuan Dynasty (1351-1369), 7.5 million. In general, demographic catastrophes were too weak to disturb the growth of global population (Nielsen, 2013c).

Looking for convincing evidence of impacts of Malthusian positive checks on the growth of population would not be easy because we would have to demonstrate not only clear discontinuities in the growth of population but also that these discontinuities are correlated with the records of demographic catastrophes. We would have to know the intensity of these demographic catastrophes not just in the number of deaths but in their relative impact. However, even then we would have to be aware of the possibility of spurious correlations.

Malthus never claimed that his concepts must be accepted. On the contrary, he was open to new ideas. Referring to himself in the third person he wrote:
> If he should succeed in drawing the attention of more able men to what he conceives to be the principal difficulty in the way to the improvement of society and should, in consequence, see this difficulty removed, even in theory, *he will gladly retract his present opinions and rejoice in a conviction of his error* (Malthus, 1798, p. viii. Italics added.)

It is interesting that Malthus used arithmetic and geometric progressions to support his arguments but it is not certain whether he was familiar with the hyperbolic growth, let alone that he appreciated the difference between hyperbolic and exponential (geometric) types of growth. Even now, hyperbolic distributions are repeatedly misinterpreted and exponential growth is used to explain the growth of just about anything.

Malthus claimed that "Population, when unchecked, increases in a geometrical ratio" (Malthus, 1978, p. 4). Now we know that this is not true. Population, when unchecked does not increase in a geometrical ratio (exponentially) but hyperbolically (Kapitza, 2006; Kremer, 1993; Nielsen, 2016b, 2016d; Podlazov, 2002; Shklovskii, 1962, 2002; von Hoerner, 1975, von Foerster, Mora & Amiot, 1960).

Malthus did not base his claims on a rigorous analysis of data. If he lived long enough to have better data, he would have probably discovered that the growth of population is not characterised by a constant doubling time and consequently that it could not have been increasing exponentially. If he were familiar with hyperbolic growth, he would have probably discovered that population increases hyperbolically. However, Malthus did not live long enough, he did not have access to good data and he was probably unfamiliar with hyperbolic growth. Those who lived after him and those who live now are more privileged.

## 6. Examples of questionable claims

### 6.1. The alleged Law of Population

During the alleged but non-existent epoch of the so-called Malthusian stagnation, birth rates are claimed to have been high because new generations were needed to support many tiresome and mundane activities such as hunting, gathering, cultivating crops, caring for children and generally for coping with harsh living conditions.
> According to Classical economists, and early Neo-Classical economists as well, population size was determined by the demand for labor. This was the Law of Population which constantly operated behind the seemingly random variations in fertility and mortality induced by epidemic, famine, and war (Lee, 1997, p. 1063).

Claims:
1. Population size was determined by the demand for labour.
2. This is the Law of Growth.
3. This law has been accepted by Classical and early Neo-Classical economists.
4. There were seemingly random variations in fertility and mortality.
5. Random variations were caused by epidemics, famine and war.
6. This law operated constantly behind these seemingly random variations.

It is interesting how much is claimed in this single paragraph and it does not matter whether Lee agrees with all these claims or just describes them. This quotation represents a typical set of questionable claims often encountered in publications related to the concept of the epoch of the so-called Malthusian stagnation. Can we prove them or do we have to accept them by faith?

To prove this "Law of Population" we would have to have data about the demand for labour and about the growth of population extending over thousands of years, and we would have to prove that there is a correlation between the demand for labour and the size of human population. We would have to prove that population size was determined by the demand for labour. We cannot prove it because we do not have such data, but we can show that the population data (Nielsen, 2016b, 2016d) do not display any features that could be linked with this alleged "Law of Population." This law has to be accepted by faith but this law is also in contradiction with data and with their analysis.

It is easy to imagine and claim, without a proof, that there were random variations in the fertility and mortality. It would be probably more difficult to expect that there were no variations but we have no information about these variations. We can only imagine them but we cannot analyse them.

We have reliable data about the *size* of human population (Biraben, 1980; Clark,1968; Cook,1960; Durand, 1974; Gallant, 1990; Haub, 1995; Livi-Bacci, 1997; Maddison, 2001, 2010; McEvedy & Jones, 1978; Taeuber & Taeuber, 1949; Thomlinson, 1975; Trager, 1994, United Nations, 1973, 1999, 2013) over thousands of years but we have *no matching data for fertility and mortality (birth rates and death rates)*. We also have no matching data about epidemics, famines and wars to study how they were correlated with "random variations in fertility and mortality." We have absolutely no way of proving that "the Law of Population" "constantly operated behind the seemingly random variations in fertility and mortality induced by epidemic, famine, and war." This claim is *unscientific* because we can never expect to verify it by data but also it is scientifically unacceptable because data and their analysis give no support for such declarations.

It should be also noted that growth of population is not determined directly by birth and death rates but by the *difference* between these two quantities. This difference determines the *growth rate*. More precisely, it determines the rate of natural increase but generally migrations rates are relatively small and consequently the difference between birth and death rates can be taken as determining the growth rate.

A constant (non-zero) difference (constant growth rate) produces *exponential* growth. A zero difference produces *constant* population. However, variable difference between birth and death rates (i.e. the variable growth rate) does not necessarily produce a variable *size* of the population. In fact, even large fluctuations in the growth rate are not readily reflected in the growth of population. They might be reflected only as small and negligible variations (Nielsen, 2016c).

*Fluctuations in birth and death rates have no impact on the mechanism of growth because they do not change population growth trajectories.* We can see it even without analysing data. We can easily check that even for data characterised by large fluctuations in birth and death rates, and consequently by large fluctuations in the growth rate, the corresponding data, which describe the growth of population are not affected by such fluctuations. Fluctuations in birth and death rates do not change the general character of the distributions describing the growth of population (Lehmeyer, 2004; Mauritius, 2015; Statistics Mauritius, 2014; Statistics Sweden, 1999; Wrigley & Schofield, 1981). These data are well known. Some of them are even repeatedly used to defend the erroneous Demographic Transition Theory but no-one cared to check the population data published in the same sources, which list the fluctuating birth and death rates. While the fluctuating birth and death rates are taken as the confirmation of the established knowledge, the data describing the growth of population, data coming from precisely the same sources as the data for birth and death rates, are methodically ignored. Data describing the growth of population are in contradiction of the Demographic Transition Theory and in contradiction of the established knowledge.

*6.2. The alleged losing battle*

According to the concept of the so-called epoch of Malthusian stagnation, as soon as the population started to increase, it was significantly reduced by numerous factors associated with severe living conditions.

> During the first [stage of the demographic transition], fertility is assumed to have been sufficiently high to allow a population to grow slowly even in the face of a rather high level of mortality. However, periodic epidemics of plague, cholera, typhoid and other infectious diseases would *in one or two years wipe out the gains made over decades. Over long periods of time there would, consequently, be almost no population growth at all* (van de Kaa, 2010, p. 87. Italics added.).

Claims:
1. During the first stage of the demographic transition, fertility and mortality are assumed to have been high.
2. Population was growing slowly.
3. Population growth was strongly controlled by periodic epidemics of plague, cholera, typhoid and other infectious diseases.
4. Periodic epidemics of plague, cholera, typhoid and other infectious diseases would in *one or two years* wipe out the gains made over *decades.*
5. Over long periods of time there was no population growth at all.

Van de Kaa describes the first of the four stages of growth claimed by the classical Demographic Transition Theory, the stage corresponding to the mythical but non-existent epoch of the so-called Malthusian stagnation (Nielsen 2016b, 2016c, 2016d).

Here we have a vivid description of what was happening so long ago and over a long time; not only a vivid description but also an explanation. In science, one would have to do a lot of solid work in order to be able to make such a sweeping declaration. We would have to prove that our conclusions are supported by data. We would have to give frequent examples that *the growth of population* was indeed *controlled* by "periodic epidemics of plague, cholera, typhoid and other infectious diseases." We would have to demonstrate convincingly that there were *frequent correlations* between "periodic epidemics of plague, cholera, typhoid and other infectious diseases" and the growth of population. Ideally, we would also have to prove that these frequent irregularities were *caused* by "periodic epidemics of plague, cholera, typhoid and other infectious diseases" because even observed correlations could be spurious.

Van de Kaa produces no such proof. He does not even give reference to such research. As far as we can tell, no-one has ever carried out such systematic and well-documented research.

His claims have to be accepted by faith and even more importantly, by a fixated faith because they are contradicted by data (Biraben, 1980; Clark,1968; Cook,1960; Durand, 1974; Gallant, 1990; Haub, 1995; Livi-Bacci, 1997; Maddison, 2001, 2010; McEvedy & Jones, 1978; Taeuber & Taeuber, 1949; Thomlinson, 1975; Trager, 1994, United Nations, 1973, 1999, 2013). With only one exception in the past 12,000 years, between AD 1200 and 1400 (Nielsen, 2016d), there is no convincing evidence of generally occurring "long periods of time" when there was "almost no population growth at all" and that the growth was controlled by "periodic epidemics of plague, cholera, typhoid and other infectious diseases." The only way we could hope to give support to his claims would be to find exceptions to the generally observed regularities in the growth of population but even then, his claims would not have a general application. The established knowledge may sound plausible and convincing but it has to be accepted by faith.

It is scientifically incorrect to take an easy way out by assuming that something happened, which we *think* could have happened and claim with such absolute certainty that it *did* happen. We might feel or think that our descriptions are true; we might wish for them to be true, but we should test them by following the generally accepted process of scientific investigation.

*6.3. The alleged food-controlled homeostatic equilibrium*

Harsh living conditions, and in particular the availability of food, are supposed to have a suppressive influence on the growth of human population but these intuitive expectations are again contradicted by data (UNDP, 2011) showing that growth rate is not directly proportional to the level of affluence but to the level of deprivation (Nielsen, 2013b). There is also convincing evidence that harsh living conditions in the distant past did not shape the growth of population (Nielsen, 2016b, 2016d, von Foerster, Mora & Amiot, 1960). Again, it is scientifically inexcusable to take an easy way out, ignore data and try to mould science in the image of our wished-for interpretations.

> …the food-controlled homeostatic equilibrium had prevailed since time immemorial (Komlos, 2000, p. 320).
> …the population tends to oscillate in a homeostatic mechanism resulting from the conflict between the population's natural tendency to increase and the limitations imposed by the availability of food (Artzrouni & Komlos, 1985, p. 24).

Claims:
1. There was a food-controlled homeostatic equilibrium.
2. This equilibrium prevailed since time immemorial.
3. Population tends to oscillate in a homeostatic mechanism.
4. Oscillations are caused by the natural tendency of the population to increase and by the limitations imposed by the availability of food.

It is easy to *assume* that "the food-controlled homeostatic equilibrium had prevailed since time immemorial" but it is more difficult to *prove* it. It is easy to *claim* that "the population tends to oscillate in a homeostatic mechanism resulting from the conflict between the population's natural tendency to increase and the limitations imposed by the availability of food" but it is more difficult to *prove* it.

Authors of these confident declarations do not prove anything nor do they give reference to such a proof because such a proof does not exist. These declarations are in harmony with the established knowledge but the established knowledge is in conflict with science (Kapitza, 2006; Kremer, 1993; Nielsen, 2016b, 2016d; Podlazov, 2002; Shklovskii, 1962, 2002; von Hoerner, 1975, von Foerster, Mora & Amiot, 1960).

In order to have these declarations supported by science we would have to work a little harder. We would have to design a model with the homeostatic equilibrium. We would have to have data for the availability of food "since time immemorial." We would have to have corresponding data describing the growth of population. These data would have to be at small time intervals in order to detect the postulated oscillations. We would have to demonstrate convincingly that there were oscillations in the growth of population and that there was a correlation between the recorded oscillations in the growth of population and the oscillations in the availability of food. We would have to prove that the oscillations in the growth of population were *caused* by the oscillations in the availability of food. Acceptable evidence would have to be in demonstrating that our mathematical model reproduces all these oscillations. This would have been science but what we are offered is just a story, which has to be accepted by faith.

It is easy to claim many things but it is more difficult to prove them. Our postulates and explanations might sound plausible but they would have to be verified by the rigorous process of scientific investigation. Data (Biraben, 1980; Clark,1968; Cook,1960; Durand, 1974; Gallant, 1990; Haub, 1995; Livi-Bacci, 1997; Maddison, 2001, 2010; McEvedy & Jones, 1978; Taeuber & Taeuber, 1949; Thomlinson, 1975; Trager, 1994, United Nations, 1973, 1999, 2013) give no support for the existence of the claimed fluctuations or oscillations.

There is no scientific basis for claiming that "food-controlled homeostatic equilibrium had prevailed since time immemorial." This claim has to be accepted by faith. We have to accept by faith that "population tends to oscillate in a homeostatic mechanism resulting from the conflict between the population's natural tendency to increase and the limitations imposed by the availability of food." It all might sound plausible but we cannot prove it. However, even if it sounds plausible it is contradicted by the rigorous analysis of data (Nielsen, 2016b, 2015d).

Artzrouni and Komlos (1985) carried out model calculations, which incorporated the assumed mechanism of the so-called Malthusian stagnation. Their contribution is important but for reasons, which were not even noticed in their publication because their results show that the mechanism of the so-called Malthusian stagnation does not work. We shall discuss this issue in one of our forthcoming publications.

*6.4. The allegedly characteristic features of the past human history*

> Stage 1 [of the Demographic Transition Theory] presumably characterizing *most of human history*, involves high and relatively equal birth and death rates and little resulting population growth" (Guest & Almgren, 2001; p. 621. Italics added.).
>
> This stage is characterized not by changes in *average* death rates but by a *stagnation of death rates at extremely high levels* for a period of what is believed to be *thousands of years*" (Olshansky & Ault, 1986, p. 357. Italics added.).

Claims:

1. Stage 1 proposed by the Demographic Transition Theory characterised presumably most of human history.
2. During this stage, there were high and relatively equal birth and death rates.
3. During this stage, there was little resulting population growth.
4. This stage was not characterised by changes in the average death rates.
5. This stage was characterised by stagnation of death rates at extremely high levels.
6. This stagnation is believed to have lasted for thousands of years.

It is amazing how firmly the established knowledge is now established if so much can be so easily claimed. The declaration that Stage 1 proposed by the Demographic Transition Theory was "characterized not by changes in *average* death rates but by a *stagnation of death rates at extremely high levels* for a period of what is believed to be *thousands of years*" has to be accepted by faith and by faith alone because we can never expect to have systematic data describing death rates to check its validity. No-one has yet demonstrated the validity of the Demographic Transition Theory. No-one has yet demonstrated the existence of the first two stages of growth, let alone the existence of all stages of growth.

Examples used in support of the Demographic Transition Theory are in fact in its direct contradiction (Nielsen, 2016c). As pointed out earlier (Nielsen, 2016c), the only way to demonstrate the apparent empirical features, which seem to be in agreement with the Demographic Transition Theory, is by a suitable manipulation of data consisting in stitching together the birth and death rates data for Mauritius with the data for Sweden.

It should be also remembered that any scientific theory is acceptable only if it is consistently confirmed by empirical evidence. A single convincingly contradicting evidence questions the validity of an accepted theory. For the Demographic Transition Theory, it is the other way round. There is not a single convincing empirical evidence in support of this theory but there is overwhelming empirical evidence showing that this theory is incorrect. This theory is contradicted by birth and death rates and by the corresponding distributions describing the growth of population (Nielsen, 2016c). Furthermore, within the range of analysable data (Biraben, 1980; Clark,1968; Cook,1960; Durand, 1974; Gallant, 1990; Haub, 1995; Livi-Bacci, 1997; Maddison, 2001, 2010; McEvedy & Jones, 1978; Taeuber & Taeuber, 1949; Thomlinson, 1975; Trager, 1994, United Nations, 1973, 1999, 2013) growth of population was hyperbolic (Kapitza, 2006; Kremer, 1993; Nielsen, 2016b, 2016d; Podlazov, 2002; Shklovskii, 1962, 2002; von Hoerner, 1975, von Foerster, Mora & Amiot, 1960). Stages of growth proposed by the Demographic Transition Theory did not exist.

Birth and death rates may have been high and strongly fluctuating but high and fluctuating birth and death rates do not prove the existence of a stagnant state of growth because, as mentioned earlier, growth is determined by the average *difference* between these two quantities. Furthermore, these two quantities have to behave in a very specific way to produce the stagnant state of growth. Studying just death rates *or* birth rates, or equivalently studying just the fertility rates (Lehr, 2009) cannot be used as the evidence in support of the Demographic Transition Theory. Using scraps of favourable information while ignoring contradicting evidence is strongly misleading and consequently scientifically unacceptable.

*6.5. The allegedly well-documented evidence*

> It is well documented that the fluctuations experienced by the world's population throughout history did not have a regular, cyclical pattern, but were, to a large extent, brought about by randomly determined demographic crises (wars, famines, epidemics, etc.). As McKeown and others have pointed out, the main cause of these fluctuations of the past were mortality crises. There are four kinds of crises: subsistence crises, epidemic crises, combined crises (subsistence/epidemic), and finally crises from other causes, which are mainly exogenous (wars, natural or other catastrophes)
>
> Crises followed by *periods of population decline* during which the nutritional status of the population improved gave rise to fluctuations which testify to the continued existence of the 'Malthusian trap': population would not grow beyond its carrying capacity for long, and when it did, the resulting overshoot was followed by a 'crash' (i.e. the positive checks such as diseases, famines, wars, etc.) (Artzrouni & Komlos 1985, p. 24. Italics added.).

Claims:
1. There were fluctuations in the world's population throughout history.
2. These fluctuations are well documented.
3. It is well documented that these fluctuations did not have a cyclic pattern.
4. It is well documented that these fluctuations were, to a large extent, brought about by randomly determined demographic crises (wars, famines, epidemics, etc.).
5. The main cause of these fluctuations were mortality crises.
6. There are four types of crises.
7. Crises were followed by periods of population decline.
8. Population decline improved nutritional status.
9. Fluctuations testify to the continuing existence of the Malthusian trap.
10. Population was repeatedly reaching its carrying capacity.
11. Population would not grow beyond its carrying capacity for long.
12. Population growing beyond its carrying capacity was reflected in overshoots.
13. Overshoots were followed by crashes.

If all this is so well documented, where is the documentation of this well documented research? It would be interesting to see at least a few references to this important and fundamental research work, to see the *data* showing fluctuations "throughout history," to see a positive proof that the "the fluctuations experienced by the world's population throughout history" are *correlated* with "demographic crises (wars, famines, epidemics, etc.)," that they were "brought about by randomly determined demographic crises." It would be also interesting to see convincing evidence that population was reaching its carrying capacity, that "population would not grow beyond its carrying capacity for long," the convincing evidence of overshoots and crashes, evidence that crashes were associated with "positive checks such as diseases, famines, wars, etc." It would be interesting to see the compelling evidence of the existence of the Malthusian trap, the demonstration of frequent "periods of population decline," the compelling proof that periods of population decline caused by demographic crises were improving nutritional status. All this vital and "well documented" evidence is missing.

What is well documented is the repeated fiction stories, which have to be accepted by faith. We have many publications propagating such stories. The repeatedly related stories of fiction are by now accepted as the undisputable facts. What is well documented is a system of beliefs, doctrines, wished-for explanations, opinions, views, theories, hypotheses, conjectures and speculations, added gradually over a long time until they became the established knowledge, the "well-documented" established knowledge but the knowledge, which is contradicted by science.

In contrast, it is well documented (Biraben, 1980; Clark,1968; Cook,1960; Durand, 1974; Gallant, 1990; Haub, 1995; Livi-Bacci, 1997; Maddison, 2001, 2010; McEvedy & Jones, 1978; Taeuber & Taeuber, 1949; Thomlinson, 1975; Trager, 1994, United Nations, 1973, 1999, 2013) that the growth of human population does *not* show fluctuations or random behaviour. It is well documented that the data show no signs of frequent overshoots and crashes, no signs of growth reaching its carrying capacity, no signs of the "continued existence of the 'Malthusian trap'," no evidence that the "population would not grow beyond its carrying capacity for long," and no repeated "periods of population decline." All these colourful and dramatic descriptions associated with the narrative based on the assumption of the existence of the mythical epoch of the so-called Malthusian stagnation are contradicted by data.

It is obvious, that demographic crises were often causing decline in the size of *local* populations, depending on the scale of these crises and depending on what we understand by a local crisis. Sometimes it might have been just a large death toll in a city, in a part of a country, as for instance in China (Mallory, 1926), or maybe in the whole country or even extending over a few countries. However, a large death toll does not necessarily mean a significant impact on the growth of human population. A large death toll should not be immediately interpreted as a population decline; it could have been just a slower growth over a limited time followed by a more intensified growth, as it happened after AD 1400 for the world population.

All these issues should be closely investigated by examining records of demographic catastrophes. To arrive at any reasonably supported conclusion, we would have to do some hard work. However, data which should be used for such investigations are strongly limited. We have no data showing that local demographic crises were repeatedly causing fluctuations in the growth of regional or global populations. In fact, the data show remarkably stable growth of human population, generally unaffected by demographic crises (Nielsen, 2013a, 2013c, 2016b, 2016d).

The opening paragraph in the above quotation contains two interesting and characteristic elements, the elements occurring repeatedly in the descriptions of the concept of the epoch of the so-called Malthusian stagnation: (1) it makes a highly-questionable but confident declaration about the *existence* of certain features (in this case about the existence of fluctuations) and (2) it equally confidently *explains* them while ignoring empirical evidence. The normal progression in scientific research is *first to observe* certain features and then try to *explain* them. We can also reverse the process: we can first *predict* the existence of certain features. However, to accept the prediction and the associated explanation, we would have to *demonstrate the existence* of the predicted features. This is how science works but for doctrines accepted by faith scientific process of investigation is too tedious and consequently it is readily ignored.

So, in this case, we would have to show first that there were significant fluctuations in the birth and death rates and in the size of human population extending over thousands of years, and then we would also have to explain them convincingly by demonstrating that they were correlated with demographic crises. Alternatively, we would have to predict (using a suitable mathematical model) the existence of fluctuations in birth and death rates and in the size of human population and then we would have to show that our predictions are confirmed by relevant data.

We cannot prove that there were fluctuations "throughout history" in the birth and death rates because we do not have relevant data, but we can prove that there were no fluctuations "throughout history" in the *size* of human population because we have the relevant data (Biraben, 1980; Clark,1968; Cook,1960; Durand, 1974; Gallant, 1990; Haub, 1995; Livi-Bacci, 1997; Maddison, 2001, 2010; McEvedy & Jones, 1978; Taeuber & Taeuber, 1949; Thomlinson, 1975; Trager, 1994, United Nations, 1973, 1999, 2013). There is nothing in the data, which calls for the explanations of fluctuations in the growth of population because there are no fluctuations. What needs to be explained is perhaps the remarkable *absence* of fluctuations, the absence of random behaviour, crashes, overshoots or "periods of population decline." What needs to be explained is why the growth of population was so remarkably stable during the past 12,000 years (Nielsen, 2016d) and why it was hyperbolic. The quoted declarations are in perfect agreement with the established knowledge but they are in conflict with science.

*6.6. The allegedly long-run equilibrium between population size and the food supply*

Referring to three sources (Habakkuk, 1953; Kunitz, 1983; McKeown, 1983), Komlos explains:

> Malthusian positive checks (mortality crises) maintained *a long-run equilibrium between population size and the food supply*. Crises followed by periods when human nutritional status was above the level of subsistence gave rise to *cycles. …the cycles testify to the continued existence of the 'Malthusian population trap': population could not grow beyond an upper bound imposed by the resource and capital constraints* of the economic structure in which it was imbedded. The *'escape' from this trap* occurred only when the aggregate capital stock was large enough and grew fast enough to provide additional sustenance for the population, which thereby overcame the effects of the diminishing returns that had hindered human progress during *the previous millennia*. After escaping from the Malthusian trap, population was able to grow unchecked. In historic terms, this escape corresponds to the industrial and demographic revolutions. Removal of the nutritional constraint, at least for the developed part of the world, resulted in the population explosion (Komlos, 1989, pp. 194, 195. Italics added.).

Claims:

1. There was a long-term equilibrium between population size and the food supply.
2. This equilibrium was maintained by positive checks (mortality crises).
3. Crises were followed by periods when human nutritional status was above the level of subsistence.
4. This process gave rise to cycles.
5. The cycles testify to the continued existence of the 'Malthusian population trap'.
6. Population could not grow beyond an upper bound imposed by the resource and capital constraints of the economic structure in which it was imbedded.
7. Malthusian trap was active for millennia.
8. The escape from the Malthusian trap occurred when the aggregate capital stock was large enough and grew fast enough to provide additional sustenance for the population.
9. The removal of nutritional constrains caused population explosion.

Massive amount of work would have to be done to support all these impressive declarations. We would have to study food supply over millennia and determine how they were correlated with the growth of human population. We would have to prove that there was "a long-run equilibrium between population size and the food supply." We would have to study mortality crises over millennia. We would have to establish a correlation between the growth of human population, food supply and mortality crises. We would also have to investigate upper bounds of "resource and capital constraints" and prove that over millennia the size of the population was repeatedly reaching the limits of these upper bounds.

Conducting scientific research is not easy but results have a high degree of reliability. Writing fictions stories, whose general script is already provided by the established knowledge based largely on faith is much easier, but this is not science.

It is easy to declare so much so quickly and with such a confidence, but it is harder to prove it. It is also hard to accept it, but accept we must if we want to accept the concept of the epoch of the so-called Malthusian stagnation promoted by the established knowledge.

The claimed cycles cannot possibly testify to "the continued existence of the 'Malthusian population trap'" because they did not exist in the growth of population (Nielsen, 2016b, 2016d, von Foerster, Mora & Amiot, 1960). Population growth, global and regional, was remarkably stable and unconstrained. The claim that "population could not grow beyond an upper bound imposed by the resource and capital constraints" is contradicted by the analysis of population data. This claim appears to be based on pure fantasy and on a wished-for mechanism that did not exist. There was no Malthusian trap in the growth of population.

We know nothing about any possible cycles in birth and death rates because we have no relevant data extending over a long time in the past. We do not know how large were these alleged cycles. We do not even know whether they existed. Discussions of these cycles are irrelevant because we know that cycles in birth and death rates are of little or no consequence for explaining the mechanism of growth (Nielsen, 2016c). Even if they were present they did not have any significant influence on the growth of the world population in the past 12,000 years (Nielsen, 2016d). They also had no influence on the growth of regional populations (Nielsen, 2016b). The absence of cycles in the growth of population combined with the evidence of the steadily increasing growth testify that the Malthusian trap did *not* exist. We cannot also claim that there was "'escape' from this trap" because there was no trap in the growth of population. There was also no trap in the economic growth (Nielsen, 2014, 2015a, 2016a, 2016e, 2016f, 2016g, 2016h, 2016i). Again, the quoted declarations are in perfect agreement with the established knowledge but they are in conflict with science.

*6.7. The alleged fluctuation of fertility and mortality rates around zero*

Discussing the first stage of the Demographic Transition Theory, Warf explains:

> Because both fertility and mortality rates are high, the *difference* between them — natural population growth — is relatively low, *fluctuating around zero*" (Warf, 2010, p. 708. Italics added.).

Claims:
1. During the first stage of the demographic transition fertility and mortality rates were high.
2. The difference between them (the natural population growth) was fluctuating around zero because they were high.

Just because fertility and mortality rates were high it does not follow that the difference between them was zero. The difference between them can fluctuate around zero even if they are low. However, this is just a minor issue.

In this quotation, the "natural population growth" is identified as the *difference* between the fertility and mortality rates. It is, therefore, the rate of natural increase or the *growth rate* because, in general, migration rates are relatively small and can be neglected.

We shall recall that while the growth rate fluctuating around a constant value describes exponential growth, the growth rate "fluctuating around zero' describes the constant size of the growing entity, i.e. in our case, the constant size of the population. The claim made by Warf is contradicted by data, which show that for thousands of years the size of human population was not constant but steadily *increasing* (Nielsen, 2016b, 2016d, von Foerster, Mora & Amiot, 1960). Furthermore, the size of population was increasing hyperbolically. The "natural population growth" (growth rate) could not have been "fluctuating around zero" but it must have been increasing hyperbolically because for the hyperbolic growth, the growth rate also increases hyperbolically (Nielsen, 2016h).

*6.8. The alleged roughly constant population*

In line with the accepted interpretations of the first stage of the Demographic Transition Theory, Lagerlöf writes:

> The Malthusian Regime in our model is a stable situation where death and birth rates are both high, and *population roughly constant*. Moreover, mortality is highly volatile, increasing dramatically in periods of big epidemic shocks. In periods with mild shocks population expands. This worsens the impact of the next epidemic, equilibrating population back to its Malthusian state (Lagerlöf, 2003a, p. 756. Italics added.).
>
> In our model, the world can thus be stuck in a *Malthusian equilibrium* for centuries and then suddenly escape, and never contract back. As suggested by a referee, this process could possibly be interpreted in terms of wars, instead of epidemics (Lagerlöf, 2003a, p. 766. Italics added.).
>
> *Throughout human history*, *epidemics, wars and famines have shaped the growth path of population*. Such shocks to mortality are the central theme of the model set up by Lagerlöf, which endogenously generates a long phase of *stagnant population* and living standards, followed by an industrial revolution and a demographic transition (Lagerlöf, 2003b, pp. 434, 435. Italics added.).

Claims:
1. It is assumed that there was a Malthusian regime.
2. It is assumed that Malthusian regime is characterised by high birth and death rates.
3. During the Malthusian regime population is roughly constant.
4. Mortality is highly volatile.
5. Mortality increases dramatically in periods of big epidemic shocks.
6. Population expands when the mortality shocks are mild.

7. Expanding population worsens the impact of the next epidemic and equilibrates population to the Malthusian state.
8. Malthusian equilibrium lasts for centuries.
9. The process of Malthusian equilibrium can be also explained by wars instead of epidemics.
10. Throughout human history, epidemics, wars and famines have shaped the growth path of population.
11. Model based on the assumption of shocks to mortality generates a long phase of stagnant population.
12. The "long phase of stagnant population and living standards" is "followed by an industrial revolution and a demographic transition."

Here again, and quite typically, we have a series of declarations that have to be accepted by faith. However, paradoxically if not ironically, Lagerlöf was on the verge of discovering that doctrines accepted by faith were contradicted by his own model.

He has carried out an interesting and important research work but unfortunately, he did not finish it: he did not compare results of his calculations with data (Maddison, 2001), which were available to him before publication of his work. He did not take the final and the most essential step. If he did, he would have discovered that the mechanism of the so-called Malthusian stagnation incorporated in his model did not produce fluctuations in the model-generated growth of population, that model-generated growth of population was not stagnant and it did not fit the relevant data. He would have found that contrary to what he claims in his publication, his model generated population was *not* "roughly constant." If he cared to consult data (Maddison, 2001) he would have also found the population reported by Maddison was also not "roughly constant." We shall discuss these issues in a separate publication.

Lagerlöf presents a plot of the growth rate and calls it erroneously "Population growth" (Lagerlöf, 2003b, p. 436). He fails to take the most essential step in this type of work and to use his model-generated growth rate to calculate model-generated distribution describing the growth of population. He ignores data (Maddison, 2001) and yet his unfinished work is accepted for publication maybe because it proclaims loud and clear the doctrines of the established knowledge. Science appears to be of no importance.

*6.9 Incorrect claims about the growth rate*

> In our model, this leads to a *constant rate* of population growth prior to the adoption of the Solow technology. This result is consistent with population data from Michael Kremer (1993), where *the growth rate of population fluctuates around a small constant* throughout most of the Malthusian period (from 4000 B.C. to A.D. 1650) (Hansen & Prescott (2002, p. 1205. Italics added.).

Claims:
1. Growth rate of population fluctuates around small constant during the Malthusian period (i.e. prior to the adoption of Solow technology).
2. Small and roughly constant growth rate is consistent with population data from Michael Kremer (1993).

*First*, it appears that Hansen and Prescott might be confusing constant growth rate with constant population. It might be the same mistake as it appears to have been made by Lagerlöf (2003b). A constant (non-zero) growth rate does not produce a constant (non-zero) size of population. A constant (non-zero) growth rate produces *exponential* growth.

*Second*, this declaration appears to contain conflicting information. It is hard to imagine that random forces characterising the mythical Malthusian period would produce a steadily increasing exponential growth. Steadily-increasing growth suggests the presence of a dominating constant force, overruling any random forces.

*Third*, fluctuations in the growth rate are not readily reflected as fluctuations in the growth of population (Nielsen, 2016c). We can demonstrate it even without carrying mathematical analysis of the fluctuating growth rate. Data alone (Lehmeyer, 2004; Mauritius, 2015; Statistics Mauritius, 2014; Statistics Sweden, 1999; Wrigley & Schofield, 1981) show clearly that fluctuating growth rates do not produce significant fluctuations in the growth of population and that they have no impact on the mechanism of growth because they do not alter growth trajectories.

*Fourth*, we would have to show convincingly that the growth rate was indeed fluctuating around a small constant value as claimed by Hansen and Prescott (2002). There is no such proof because we do not have the data for the growth rate extending over thousands of years. However, there is a proof that the growth rate during the AD and BC eras was *not* fluctuating around a small constant value but that it was increasing hyperbolically because the growth of the population was hyperbolic (Kapitza, 2006; Nielsen, 2016b, 2026d; Kremer, 1993; Podlazov, 2002; Shklovskii, 1962, 2002; von Hoerner, 1975, von Foerster, Mora & Amiot, 1960). For the hyperbolic growth, the growth rate increases hyperbolically with time or in the direct proportion to the size of population (Nielsen, 2016h), as observed also by Kremer (1993).

*Fifth*, Kremer (1963) did not carry out an extensive study of the growth rate. He has presented rough calculations of this quantity using strongly varying local gradients, which do not represent the real gradient of growth. His calculations are strongly inaccurate for the BC era when individual data values are separated by large time intervals. It is scientifically unjustifiable to use such calculations and claim fluctuations around a constant value.

*Sixth*, for the hyperbolic growth, growth rate is small over a long time because it is also hyperbolic. Growth rate might appear to vary around a small constant but such interpretation is incorrect. Growth rate should be preferably calculated using interpolated gradients to avoid spurious effects of strongly-varying local gradients between adjacent data values. It is also useful to display growth rate using various types of displays to help in its interpretation. Using the approximate calculations of Kremer (1993) and claiming that growth rate was varying around small constant is self-misleading and scientifically unjustified.

This example illustrates that in science it is essential to carry out methodical analysis of data. In economic and demographic research, this is particularly important because historical economic growth and historical growth of population were increasing hyperbolically. Hyperbolic distributions are strongly misleading and can easily lead to their misinterpretations. Furthermore, for hyperbolic distributions, the growth rate and the gradient increase in a similar fashion. The growth rate increases hyperbolically and the gradient follows the second-order hyperbolic distribution, both of them containing the same confusing features of a slow growth over a long time and a fast growth over a short time, but both increasing *monotonically* over the entire range of time. Hyperbolic growth of the GDP and population as well the monotonically-increasing growth rates and gradients cannot be divided into two or three distinctly different sections. They all have to be analysed and interpreted as a whole. The same mechanism has to be applied to the slow and to the fast growth because slow and fast growth belongs to the same, monotonically-increasing distributions.

*6.10. The alleged density-dependent variations in mortality*

> If population *density* increases the mortality rate rises, equilibrating population back to the Malthusian trap (Lagerlöf, 2003a, p. 765. Italics added.).

This statement has to be also accepted by faith because there is no convincing research supporting such declaration. Creative imagination appears to be taking full control in the established knowledge.

Here we have an example of an interesting *detail* added to the concept of the epoch of the so-called Malthusian stagnation, illustrating how one fantasy can lead easily to a new fantasy and how such gradual additions reinforce the established knowledge. This statement claims the dependence of mortality rate on the *density* of human population. It offers an *explanation* how the phantom Malthusian trap regulates the growth of human population. It describes some kind of a general rule that the Malthusian trap is activated when the population *density*, not its size, reaches a certain limiting value.

There is no research confirming the described mechanism; no research showing how the growth of human population depends on its *density*. Even if we could show some isolated examples of the density-dependent growth we would have to demonstrate that such mechanism applies also to regional and global populations. The best data available to us show the *time*-dependence of the size of human population and there is nothing in them to suggest any form of *density*-dependence, let alone the existence of the Malthusian trap triggered by the density of population.

This statement is yet another example of the leaps of faith, of confident declarations requiring a huge amount of work to be accepted as a reliable contribution to science. The descriptions of the epoch of the so-called Malthusian stagnation are full of such unscientific declarations. Indeed, they are made of them.

Other terms used to describe the alleged stagnant and fluctuating state of growth during this mythical epoch of Malthusian stagnation are "equilibrium trap" or "population trap" (Leibenstein, 1957; Nelson, 1956), "multiple equilibria" or "poverty trap" (Wang, 2005).

The belief in the stagnant and fluctuating growth is so strong that mathematical models are deemed successful if they can generate the desired oscillations during this mythical epoch of Malthusian stagnation, and no-one seems to care to take the next and the most essential step and to compare model calculations with population data. As long as oscillations of some kind are generated by a mathematical model, they are taken as the proof of the existence of the epoch of the so-called Malthusian stagnation. This line of reasoning shows that the primary, if not the exclusive, aim of such mathematical exercises is to translate a story into a mathematical language and when the translation is done properly, when mathematical formulae generate *any kind of oscillations*, large or small, significant or negligible, these formulae are then taken as a proof of the existence of the so-called Malthusian stagnation.

*6.11. The alleged Age of Pestilence and Famine*

The so-called epoch of Malthusian stagnation is also described as the Age of Pestilence and Famine (Omran 1971, 1983, 1998).

> In this stage, the major determinants of death are the Malthusian positive checks, namely epidemics, famines and wars (Omran, 1983, p. 306; Omran, 2005, p. 737).
> 
> Even if fertility approached its biologic maximum, depopulation could and did occur as a result of epidemics, wars and famines, which repeatedly pushed mortality levels to high peaks (Omran, 2005, p. 733).
> 
> The pattern of growth [of human population] until about 1650 is cyclic (Omran, 1971, Table 4, p. 533).

Claims:
1. During the Age of Pestilence and Famine (i.e. during the so-called epoch of Malthusian stagnation) major determinants of death are the Malthusian positive checks (epidemics, famines and wars).
2. Depopulation was occurring even when fertility was approaching its biological maximum because epidemics, wars and famines were repeatedly pushing mortality levels to high peaks.
3. Growth of population before AD 1650 was cyclic.

To justify the first claim, we would have to have reliable records of the *causes of death* over thousands of years. We would then have to show convincingly that indeed the major causes of death were epidemics, famines and wars. We would also have to show that there was a clear change in the causes of death when the so-called epoch of Malthusian stagnation ceased to exist. We cannot present such proofs because we do not have the supporting data. In principle, therefore, this claim is not scientific because we cannot check it by data. It has to be accepted by faith.

To justify the second claim, we would have to have reliable records of fertility and mortality over thousands of years. We would then have to demonstrate that fertility was approaching biological limits, that such events were coinciding with high mortality peaks and that these high mortality peaks were caused by epidemics, wars and famines. We do not have relevant data to check whether these descriptions are true. They are therefore also unscientific and they have to be accepted by faith.

The growth of population, global and regional, before AD 1650 was *not* cyclic (Nielsen, 2016b, 2016d). This statement is contradicted by data (Biraben, 1980; Clark,1968; Cook,1960; Durand, 1974; Gallant, 1990; Haub, 1995; Livi-Bacci, 1997; Maddison, 2001, 2010; McEvedy & Jones, 1978; Taeuber & Taeuber, 1949; Thomlinson, 1975; Trager, 1994, United Nations, 1973, 1999, 2013).

*6.12. The alleged main cause of mortality*

> During the first stage, *mortality vacillated at high levels*, with infectious disease as the main cause of death plus a large proportion due to wars and famines (Robine, 2001, p. 191. Italics added.).

Claims:
1. During the first stage of demographic transitions mortality vacillated at high levels.
2. The main causes of death were infectious diseases.
3. Large proportion of death were caused by wars and famines.

We cannot prove that "mortality vacillated at high levels" because we have no relevant data for the so-called "first stage" to carry out such a study, the stage that is assumed to have lasted for thousands of years. We cannot prove that these imagined and wished-for vacillations were correlated with infectious disease, wars and famines. We cannot prove that the *main* causes of deaths were infectious diseases. We cannot prove that a *large propor*tion of death was due to wars and famines. We do not have sufficiently extensive records of causes of death extending over thousands of years. We do not know how the causes of death were changing over time. We do not have the records to help us to distinguish between the major and minor causes. We do not know whether the main cause of death was the same over thousands of years. The concept of the epoch of Malthusian Stagnation and all these claims have to be accepted by faith.

*6.13. The alleged unsustained growth of population*

The first transition phase, called the 'Age of Pestilence and Famine,' is characterized by *high and fluctuating mortality rates*, variable life expectancy with low average life span, and *periods of population growth that are not sustained* (McKeown, 2009, p. 20S. Italics added.).

Claims:
1. During the Age of Pestilence and Famine (i.e. during the hypothetical but non-existent epoch of the so-called Malthusian stagnation) mortality rates were high and fluctuating.
2. Average life span was low.
3. There were periods when the population growth was not sustained.

Mortality rates might have been high and fluctuating but we have no data extending over thousands of years to prove it. Furthermore, we would yet have to show that these hypothetical high and fluctuating mortality rates could have been responsible for creating stagnation. What we know is that strongly-fluctuating mortality rates do not change the growth of population (Lehmeyer, 2004; Mauritius, 2015; Nielsen, 2016c; Statistics Mauritius, 2014; Statistics Sweden, 1999; Wrigley & Schofield, 1981). There is also nothing in the data and in their analysis to show that "low average life span" was affecting the growth of population. As for the "periods of population growth that are not sustained" this claim is contradicted by the analysis of data (Nielsen, 2016b, 2016d).

*6.14. Positive forces were allegedly balanced by negative forces*

The positive forces of growth had existed all along. However, they had been counterbalanced by the negative forces of malnutrition and disease (Komlos & Baten, 2003, p. 19).

We have no reliable empirical evidence to support this claim, no study of positive and negative forces, no study of their balancing, and no study of their influence on the growth of human population. This is not science but story-writing prompted and approved by the established knowledge.

How do we know that the so-called positive forces were balanced by forces of malnutrition and disease? They obviously were not because economic growth and the growth of population were hyperbolic and remarkably stable (Nielsen, 2016a, 2016b, 2016d). Such a strong and stable growth could have been only generated by a strong and dominating force.

Here again, authors of this declaration take an easy way out. They have made no attempt to consult data available to them at the time of the publication of their paper (Biraben, 1980; Clark,1968; Cook,1960; Durand, 1974; Gallant, 1990; Haub, 1995; Livi-Bacci, 1997; Maddison, 2001; McEvedy & Jones, 1978; Taeuber & Taeuber, 1949; Thomlinson, 1975; Trager, 1994, United Nations, 1973, 1999). They have made no attempt to reconcile their interpretations with the already documented evidence of hyperbolic growth (Kapitza, 2006; Kremer, 1993; Podlazov, 2002; Shklovskii, 1962, 2002; von Hoerner, 1975, von Foerster, Mora & Amiot, 1960). Again, this declaration is in perfect agreement with the established knowledge but is in conflict with science.

*6.15. The continuing misinformation*

The established knowledge is by now so strongly established that it will be difficult to change it. It continues to be supported by the scientifically-unsubstantiated claims and descriptions. It would take volumes to list and discuss all such examples and to show that these repeatedly propagated doctrines, explanations and interpretations have to be accepted by faith.

The current established knowledge based on the assumption of the so-called Malthusian stagnation followed by explosion and reinforced by many complicated explanations is similar to the established knowledge about the dynamics of celestial bodies, interpretations which were established for about two millennia before they were eventually abandoned. Describing the work of mathematicians of his time, Osiander wrote:

> With them it is as though an artist were to gather the hands, feet, head and other members from his images from divers models, each part excellently drawn, but not related to a single body, and since they in no way match each other, the result would be monster rather than man (Copernicus, 1995).[*]

Historical economic growth and historical growth of population can be expected to be described by a simple mechanism because hyperbolic growth is simple. This issue will be discussed in a separate publication, where a simple explanation of the mechanism of hyperbolic growth will be also presented. Hyperbolic growth prevailed for at least 12,000 years for the growth of population (Nielsen, 2016d) and for hundreds of years for the economic growth (Nielsen, 2016a). The established knowledge in demography and in economic research offers complicated explanations, which have to be accepted by faith. Hopefully we shall not have to wait for two thousand years to abandon these erroneous doctrines and replace them by science.

## 7. Summary and conclusions

Established knowledge in demography and in economic research is based on a series of doctrines and explanations revolving around the concept of the so-called Malthusian stagnation and around the concept of the escape from the so-called Malthusian trap described as explosion, takeoff, sprint or spurt. It is a system of interpretations, which have to be accepted by faith.

It is easy to understand why these concepts are so attractive because the growth of population and economic growth were increasing hyperbolically and hyperbolic growth creates an illusion of stagnation followed by explosion.

It is essential to understand that hyperbolic distributions should be analysed and interpreted *as a whole*. If we take just a few examples along the hyperbolic growth, we can easily make a mistake and arrive at incorrect conclusions. If hyperbolic distributions are already difficult to understand without their methodical analysis, linearly-modulated hyperbolic distributions (Nielsen, 2015a) describing income per capita are even more difficult to understand because they create even stronger illusion of stagnation followed by a sudden explosion. Here again, just taking a few examples along these distributions is bound to lead to incorrect conclusions. These distributions have to be also analysed with care. Careful and methodical mathematical analysis of data describing historical economic growth and the growth of population is unavoidable.

Distributions describing income per capita are generated by a division of two hyperbolic distributions. The characteristic feature of this ratio is that for a long time the growth of income per capita was not just slow, as for hyperbolic distributions, but nearly constant. This feature characterises the division of any hyperbolic distributions, not just the division of the GDP and population (Nielsen, 2015a). It is a purely mathematical property, which has nothing to do with specific properties of economic growth,

The nearly constant income per capita should never be interpreted automatically as stagnation. The only way to claim stagnation for this nearly-constant income per capita is to *analyse* the GDP and population data separately and to *prove* that these distributions are not hyperbolic but stagnant.

It is incorrect to take a few values of income per capita, show that they are nearly constant and claim stagnation. If the GDP and population increase hyperbolically, then income per capita increases by following the *monotonically*-increasing linearly-modulated hyperbolic destitution and it is incorrect to try to divide such a monotonically-increasing distribution into two different sections, slow and fast. Mathematically, it is impossible to make such a division. It is impossible to identify a point or a range of points and claim them as marking the place of transition.

Even though the ratio of two hyperbolic distributions is nearly constant over a long time and nearly vertical over a short time, the transition from the nearly constant to the nearly vertical patterns occurs all the time along the entire range of such distributions. Linearly-modulated hyperbolic distributions representing income per capita should be also interpreted as a whole. The same mechanism should be applied to the nearly constant and to the nearly vertical growth, unless we can prove that the GDP and population were not following hyperbolic distributions but were stagnant.

---

[*] This quotation comes from a letter written by Andreas Osiander, Lutheran theologian and a friend of Copernicus, a letter addressed to the chief editor, Pope Paul III. Osiander argues in favour of the mathematically simple and elegant heliocentric system as opposed to the complicated geocentric descriptions. This letter was later used as an unsigned introduction to the book *De revolutionibus orbium coelestium*, and was mistakenly attributed to Copernicus.

We have presented many examples of claims revolving around the concepts of stagnation followed by explosion. We have shown why such claims are scientifically unacceptable.

The origin of the fundamental concepts of the established knowledge can be traced, perhaps not entirely correctly, to Malthus (1798). He has presented an important pioneering work but unfortunately the ensuing studies of economic growth and of the growth of population have taken a wrong turn at a certain time in the past, perhaps because relevant data were not available.

By the time the relevant data (Biraben, 1980; Clark,1968; Cook,1960; Durand, 1974; Gallant, 1990; Haub, 1995; Livi-Bacci, 1997; Maddison, 2001, 2010; McEvedy & Jones, 1978; Taeuber & Taeuber, 1949; Thomlinson, 1975; Trager, 1994, United Nations, 1973, 1999, 2013) became available, they were ignored. More recently, some of them (Maddison, 2001) were manipulated to support the established knowledge (Ashraf, 2009; Galor, 2005a, 2005b, 2007, 2008a, 2008b, 2008c, 2010, 2011, 2012a, 2012b, 2012c; Galor & Moav, 2002; Snowdon & Galor, 2008). Earlier analyses of data (Kapitza, 2006; Kremer, 1993; Podlazov, 2002; Shklovskii, 1962, 2002; von Hoerner, 1975, von Foerster, Mora & Amiot, 1960) showing that the growth of population was hyperbolic were also ignored. By now, the established knowledge is so well established that anything being in its conflict is methodically ignored, rejected or suppressed. This is not science.

Recent analyses of data (Nielsen, 2014, 2015a, 2016a, 2016b, 2016d, 2016e, 2016f, 2016g, 2016h, 2016i) confirmed the earlier studies (Kapitza, 2006; Kremer, 1993; Podlazov, 2002; Shklovskii, 1962, 2002; von Hoerner, 1975, von Foerster, Mora & Amiot, 1960) and demonstrated that the historical growth of population and the historical economic growth were hyperbolic. The established knowledge based on the scientifically-contradicted concepts of stagnation followed by explosion (takeoff or the escape from the Malthusian trap) has to be replaced by explanations based on accepting hyperbolic growth.

It is incorrect to interpret the past harsh living conditions as a proof of the existence of the so-called Malthusian stagnation. Whatever harsh living conditions might have been present in the past, their effects are generally not reflected in growth trajectories. The only known example is for the growth of global population between AD 1200 and 1400 coinciding with the convergence of *five* major demographic catastrophes (Nielsen, 2016d). However, even then, the recorded effect is small.

Negative effects of the Malthusian positive checks should be never used robotically to describe the past growth of population or the economic growth. If we want to claim that positive checks were shaping the growth of population or economic growth, we have to prove it. If we want to claim that the so-called Law of Population was shaping growth trajectories, we have to prove it. If we want to claims that demographic catastrophes were shaping the growth of population, we have to prove it. If we want to claim that Industrial Revolution was shaping growth trajectories, we have to prove it. We cannot take shelter in the established knowledge because in this respect established knowledge is repeatedly contradicted by data. Any data we might have, should to be methodically analysed to prove the negative effects of Malthusian positive checks but whatever we would prove would be just an exception from the general and well-demonstrated pattern that the historical growth of population and historical economic growth were not only hyperbolic but that they also remarkably stable.

Interpretations based on the concepts of the so-called Malthusian stagnation and on the claims of the escape from the so-called Malthusian trap are not only incorrect but also dangerously misleading. They suggest that after the endless epoch of stagnation we have now entered the sustained growth regime (Galor, 2005a, 2011). This hypothesis creates a sense of security. In contrast, analysis of data shows that the past growth was sustainable but now for the first time in human history it is unsustainable and insecure (Nielsen, 2015b). While in the past, economic growth and the growth of population, global and regional, were following the slowly increasing hyperbolic trajectories (Nielsen, 2014, 2015a, 2016a, 2016b, 2016d, 2016e, 2016f, 2016g, 2016h. 2016i) indicating the unconstrained and secure growth, now the growth is at the stage of the dangerously fast increase (see Figures 1 and 2). The growth is no longer hyperbolic but the current growth increases close to the historical hyperbolic trajectories. For the first time in human history, these growth trajectories are clearly unsustainable because such a fast increase cannot be possibly tolerated for much longer.

The established knowledge is not only in conflict with data describing the past economic growth and the growth of human population but also in conflict with the general knowledge about the current mounting problems threatening our future. The established knowledge in demography and in economic research created its own world of fiction divorced from the real world.

We have not escaped the Malthusian trap because there was no trap in the economic growth or in the growth of population. The past growth was unconstrained and sustainable as demonstrated by the undisturbed hyperbolic distributions. However, now we are in the trap. For the first time in human history we are in the trap of numerous critical problems, which threaten our global security and our survival (Nielsen, 2006). For the first time in human history our combined ecological footprint is larger than the ecological capacity and it continues to increase (WWF, 2010). For the first time in human history our growth is supported by the increasing ecological deficit.

In order to understand the past and present economic growth, erroneous interpretations revolving around the concept of the so-called Malthusian stagnation have to be abandoned and replaced by scientifically acceptable interpretations. What needs to be explained is why the past economic growth and the growth of population were hyperbolic. Why was the growth so remarkably stable? Why was it not influenced by many random forces, which might have been present? Why did the growth start to divert to slower trajectories? Why does it continue so closely to the dangerously fast hyperbolic trajectories? And the most important questions of all: How to slow down the current growth? How to control growth?

Examples presented here suggest that there is a problem not just with certain interpretations adopted and protected by the established knowledge in the demographic and economic research but with the way research is carried out in these two fields. It is not just the problem with one or two theories, such as the Demographic Transition Theory or the Unified Growth Theory, which need to be corrected or most likely replaced. It is not even just the problem with the accepted paradigm based on the concept of the so-called Malthusian stagnation, which needs to be abandoned. It is a *systemic* problem. It is a problem, with the way research is conducted in these two fields. It is a problem with creating stories and interpretations, which have to be accepted by faith. It is a problem with a selective use of data. It is a problem of ignoring contradicting evidence, such as the contradicting evidence published over 50 years ago by von Foerster, Mora and Amiot (1960). It is a problem with manipulating and distorting data to fit the preconceived ideas, as it has been done repeatedly in the Unified Growth Theory and in other related publications (Ashraf, 2009; Galor, 2005a, 2005b, 2007, 2008a, 2008b, 2008c, 2010, 2011, 2012a, 2012b, 2012c; Galor & Moav, 2002; Snowdon & Galor, 2008). It is a problem of testing data by a theory rather than testing theory by data. It is a problem with protecting a system of doctrines, which are accepted on faith.

As outlined briefly elsewhere (Nielsen, 2016i), there are two ways of conducting research: (1) the dynamic scientific method, which is used in the self-correcting disciplines of science and (2) the stale method, which is used routinely in the usually emotional and dishonest defence of doctrines accepted by faith. It is unfortunate, that as pointed out earlier (Nielsen, 2013a, 2013b, 2013c, 2014, 2015a, 2016a, 2016b, 2016c, 2016d, 2016e, 2016f, 2016g, 2016h, 2016i), demographic and economic research appears to gravitate strongly towards the unscientific method.

The established knowledge revolving around the concept of the so-called Malthusian stagnation has to be changed because there was no stagnation in the historical economic growth and in the historical growth of population. There was also no escape from the Malthusian trap because there was no trap. This paradigm has to be changed because historical economic growth and the historical growth of population were hyperbolic. However, in order to make the demographic and economic research scientifically acceptable, the systemic problem has to be also solved. Scientific research can be based only on the well-known and generally recognised scientific rules of investigation. Anything else is not science.

# Changing the direction of the economic and demographic research

*By* Ron W. NIELSEN †

**Abstract.** A simple but useful method of reciprocal values is introduced, explained and illustrated. This method simplifies the analysis of hyperbolic distributions, which are causing serious problems in the demographic and economic research. It allows for a unique identification of hyperbolic distributions and for unravelling components of more complicated trajectories. This method is illustrated by a few examples: growth of the world population during the AD era; growth of population in Africa; economic growth in Western Europe; and the world economic growth. They show that fundamental postulates of the demographic and economic research are contradicted by data, even by precisely the same data, which are used in this research. The generally accepted postulates are based on the incorrect understanding of hyperbolic distributions, which characterise the historical growth of population and the historical economic growth. In particular, data used, but never analysed, during the formulation of the Unified Growth Theory show that this theory is based on fundamentally incorrect premises and thus is fundamentally defective. In this theory, distorted representations of data are used to support preconceived and incorrect ideas. Precisely the same data, when properly analysed, show that the theory is incorrect. Application of this simple method of analysis points to new directions in the demographic and economic research. It suggests simpler interpretations of the mechanism of growth. The concept or the evidence of the past primitive and difficult living conditions, which might be perhaps described as some kind of stagnation, is not questioned or disputed. It is only demonstrated that trajectories of the past economic growth and of the growth of population were not reflecting any form of stagnation and thus that they were not shaped by these primitive and difficult living conditions. The concept or evidence of an explosion in technology, medicine, education and in the improved living conditions is not questioned or disputed. It is only demonstrated that this possible explosion is not reflected in the economic growth and in the growth of population. Growth trajectories were increasing monotonically during the generally claimed epoch of stagnation and during the claimed explosion.
**Keywords.** Hyperbolic Distributions; Reciprocal Values; Economic Growth; Growth of Human Population; Industrial Revolution; Unified Growth Theory; Growth Regimes; Gross Domestic Product; GDP

## 1. Introduction

What we are going to see will change the fundamental postulates in the demographic and economic research. It will change radically the way the mechanism of economic growth and of the growth of population is interpreted. Maybe the change will not come immediately because it is usually difficult to change the well-established interpretations and explanations but the change will come because this is the way science works. Incorrect interpretations are not tolerated for too long and it does not matter who are their advocates.

It might be expected that a complicated proof would be required to achieve such a radical change of direction in the economic and demographic research, that perhaps some new and complicated description of the mechanism of growth would have to be proposed. However, the proof turns out to be exceptionally simple. No complicated mathematics is required but only the way we describe data using the simplest mathematical representation: the straight line.

George Pólya, Hungarian mathematician, observed that when a proof is too simple, "youngsters" will be unimpressed (Pólya, 1981), but mathematics does not have to be complicated to be useful. He

† AKA Jan Nurzynski, Griffith University, Environmental Futures Research Institute, Gold Coast Campus, Qld, 4222, Australia.

☏. +61407201175

✉. ronwnielsen@gmail.com

also pointed out that solving problems is a quintessential human activity and the aim is always to find *the simplest solutions*.

We are going to present here a proof so simple that it might look trivial. We are going to show how to change the confusing and complicated distributions describing the historical economic growth and the historical growth of human population into the simplest representations. We are going to show how the distributions, which suggest complicated explanations of the mechanism of growth are in fact so simple that they suggest also a simple mechanism.

Analysis of data describing the historical economic growth and the historical growth of population might look complicated but it is exceptionally simple. Anyone can do it. However, there is more to the analysis of data then just looking for their mathematical descriptions. We are going to demonstrate that this simple method of analysis makes a significant contribution to a better understanding of the mechanism of the historical growth of population and of the economic growth. It also demonstrates that there is a need to replace the traditionally used postulates based largely on impressions and conjectures by postulates based on the mathematical analysis of data.

## 2. The common problem

Hyperbolic processes appear to be causing a serious problem in the economic and demographic research. They create such a strong illusion that it deceives even the most experienced and respected researchers. The common mistake is to see them as being made of two distinctly different components, slow and fast, with a clear transition between them (Ashraf, 2009; Artzrouni & Komlos, 1985; Baldwin, Martin & Ottaviano, 2001; Becker, Cinnirella & Woessmann, 2010; Clark 2003, 2005; Currais, Rivera & Rungo 2009; Dalton, Coats & Asrabadi, 2005; Desment & Parente, 2012; Doepke, 2004; Ehrlich, 1998; Elgin, 2012; Galor 2005a, 2005b, 2007, 2008a, 2008b, 2010, 2011, 2012a, 2012b; Galor & Michalopoulos, 2012; Galor & Moav 2001, 2002; Galor & Mountford, 2003, 2006, 2008; Galor & Weil, 1999, 2000; Goodfriend & McDermott 1995; Hansen & Prescott 2002; Jones, 2001; Johnson & Brook 2011; Kelly, 2001; Khan 2008; Klasen & Nestmann 2006; Kögel & Prskawetz 2001; Komlos 1989, 2000, 2003; Komlos & Artzrouni 1990; Lagerlöf 2003a, 2003b, 2006, 2010; Lee, 2003, 2011; Mataré, 2009; McFalls, 2007; McKeown, 2009; McNeill 2000; Møller & Sharp, 2013; Mongomery, n.d.; Nelson, 1956; Omran 1971, 1983, 1986, 1998, 2005; Robine 2001; Smil 1999; Snowdon & Galor, 2008; Steinmann, Prskawetz & Feichtinger, 1998; Strulik, 1997; Tamura 2002; Thomlinson 1965; van de Kaa 2008; Voigtländer & Voth, 2005; Vollrath, 2011; Wang 2005, Warf 2010; Weisdorf 2004; Weiss 2007). The next step is then to try to explain these two perceived stages of growth and the associated but non-existent transition by proposing distinctly different mechanisms for each of these imagined components rather than seeing them as representing a *single*, *monotonically increasing* distribution governed by a *single mechanism of growth*.

This step leads progressively further away from the correct understanding of studied processes because all efforts are now concentrated on explaining the non-existing features. An increasing number of scholars are being involved. They do not analyse the relevant data but only describe their impressions created by hyperbolic illusions. The participating researchers do not question the existence of the distinctly different stages of growth or of the postulated transition – they take them for granted and concentrate their attention only on the explanation of these phantom features, proposing new mechanisms, theories and mathematical descriptions without realizing that the apparent distinctly different two stages of growth do not exist and that there is no transition but a monotonically increasing hyperbolic distribution. Their mathematical descriptions, complicated and elaborate as they might be, are not the descriptions of the studied processes but rather the descriptions of phantom impressions created by hyperbolic illusions.

The perceived two stages of growth are commonly described as stagnation and sustained growth, while the perceived but non-existent transition as an escape, sprint, sudden spurt, intensification, acceleration, explosion or by some other similar terms all emphasizing a clear and dramatic change in the pattern of growth at a certain time. Variety of forces and mechanisms are then proposed to explain the phantom stages of growth and of the associated but non-existent transition. Efforts are also made to determine the precise time of the non-existent transition, often placing it around the Industrial Revolution but sometimes around 1950, without realizing that the determination of this time is impossible because there was no unusual acceleration at any particular time or over a certain range of time.

Hyperbolic processes are prone to misinterpretations and consequently they have to be analysed with care. Fortunately, their analysis is exceptionally simple. To show how to avoid being guided by hyperbolic illusions we shall describe the simple method of their analysis and illustrate it by a few examples.

## 2. The method of reciprocal values

Hyperbolic processes can be easily analysed using the method of reciprocal values. This method is so simple that it can be explained by using just two elementary equations, and yet so powerful that it can turn around and revolutionize such fields of research as the economic growth and the growth of human population, the important fields of study because for the first time in human existence we have now reached ecological limits of our planet and the correct understanding of these two processes is essential to avoid the undesirable unsustainable developments. We have to know how these processes work and how to control them. Incorrect interpretations are potentially dangerous and cannot be tolerated. Every effort has to be made to identify and eliminate any incorrect and misleading explanations.

The first-order hyperbolic distribution is described by the following simple equation:

$$S(t) = (a_0 + a_1 t)^{-1}, \qquad (1)$$

where $S(t)$ is the size of a growing entity, while $a_0$ and $a_1$ are constants. For the hyperbolic growth, $a_1 < 0$.

Example of hyperbolic growth is shown in Figure 1. It represents the growth of the world population during the AD era. We can see that hyperbolic distribution describes well the growth of population during the entire range of data.

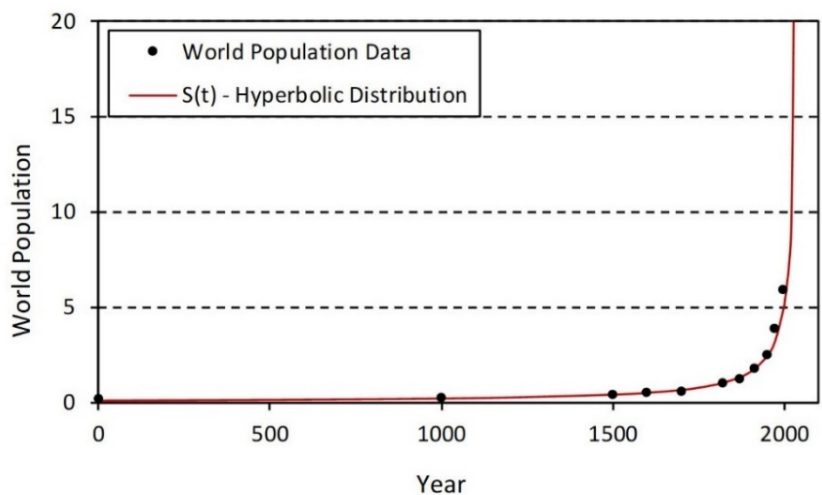

**Figure 1.** Example of hyperbolic growth. Population data (Maddison, 2001) taken from the same source as used by Galor in his Unified Growth Theory (Galor, 2005a, 2011) are compared with hyperbolic distribution.

Data have to be analysed but in general they are not. Meticulous analysis of data is particularly important in the study of hyperbolic processes because they may be strongly misleading. They easily create an illusion of stagnation followed by explosion. Unfortunately, on seldom occasions when data are used and displayed, they are displayed in a grossly distorted and self-misleading way (Ashraf, 2009; Galor 2005a, 2005b, 2007, 2008a, 2008b, 2010, 2011, 2012a, 2012b; Galor & Moav, 2002; Snowdon & Galor, 2008) as shown in Figure 2.

Figure 2 was repreduced from Galor's publication (Galor, 2005a, p. 181). His figure was based on precisely the same source of data (Maddison, 2001) as used in Figure 1 but in this distorted way they show no resemblance to the the original data. Such distortions were used repeatedly during the development of the Unified Growth Theory (Galor, 2005a, 2011) making it scientifically unacceptable, incorrect and unreliable. This Figure shows incorrectly that there was a long epoch of stagnation followed by a takeoff to a fast growth.

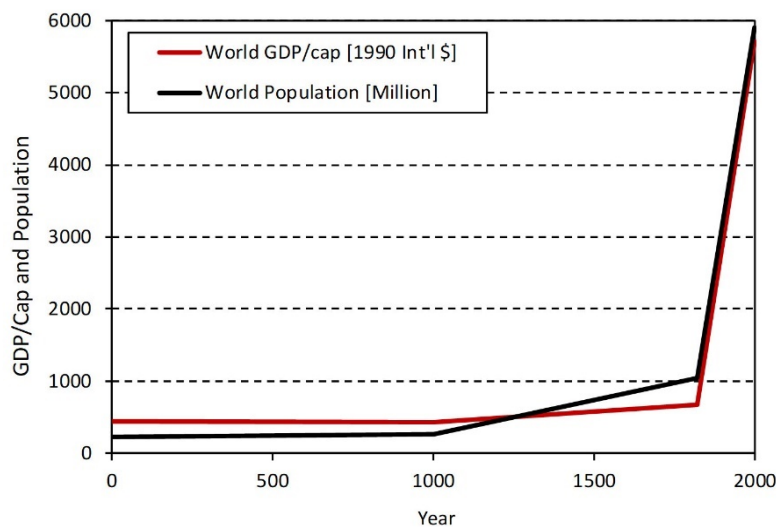

**Figure 2.** Example of a distorted presentation of data used in academic literature (Ashraf, 2009; Galor 2005a, 2005b, 2007, 2008a, 2008b, 2010, 2011, 2012a, 2012b; Galor & Moav, 2002; Snowdon & Galor, 2008). Data presented in this figure come from precisely the same source (Maddison, 2001) as the data presented in Figure 1 but in this distorted way they bear no resemblance to the original data and they suggest incorrect interpretation of the mechanism of growth. Population is in millions and the GDP/cap in the 1990 International Geary-Khamis dollars

In discussions of the growth of population or of the economic growth it is easy to use some selected numbers and show that the growth was slow over a long time and fast over a short time. The slow growth is then interpreted as stagnation controlled by random forces of growth usually associated with Malthusian positive checks. The fast growth is interepreted as explosion controlled by distinctly different forces. The triggering mechanism of the alleged explosion is usually claimed to have been associated with the Industrial Revolution and Galor conveniently locates this alleged takeoff time around the time of the Industrial Revolution. Of course the growth was slow over a long time and fast over a short time because it was hyperbolic. It was not because there was stagnation followed by a takeoff or explosion leading to a new type of growth.

Hyperbolic distribution shown in Figure 1 is described by the eqn (1) with the following parameters $a_0 = 8.724$ and $a_1 = -4.267 \times 10^{-3}$. The fit to the data is remarkably good. Details of analysis are described in a separate publication (Nielsen, 2016a). They show that there was a major transition from a fast hyperbolic growth to a slow hyperbolic growth around AD 1 and that there was a minor disturbance around AD 1300. However, these details are of no concern to us in our present discussion. What is important to notice is that the growth of human population was indeed slow over a long time and fast over a short time but that these features are described remarkably well by a *single* hyperbolic distribution. These features represent nothing more than mathematical properties of hyperbolic distribution. They represent a *single* mechanism of growth.

It is important to point out that hyperbolic distribution increases monotonically. It makes no sense to divide it into two or three components and assign different mechanisms of growth to each perceived component. Hyperbolic distribution cannot and should not be divided into separate components and the best way to see it is to plot their reciprocal values $[S(t)]^{-1}$ because they convert hyperbolic distribution to a straight line:

$$[S(t)]^{-1} = a_0 + a_1 t \quad (2)$$

Reciprocal values of hyperbolic distribution shown in Figure 1 are plotted in Figure 3. It is precisely the same distribution as shown in Figure 1 but it is presented in a different way. The confusing features such as the apparent stagnation followed by a takeoff to a fast growth increasing to infinity are replace by a clear straight line, which is easy to understand. It is obvious now that it would make no sense to divide such a straight line into distinctly different components and to claim distinctly different mechanisms of growth. It is also clear that it is impossible to identify a transition from a slow to a fast growth for hyperbolic distributions. There is no transition at any time. The transition occurs gradually over the entire range of growth. It is impossible to identify a takeoff time because there was no takeoff.

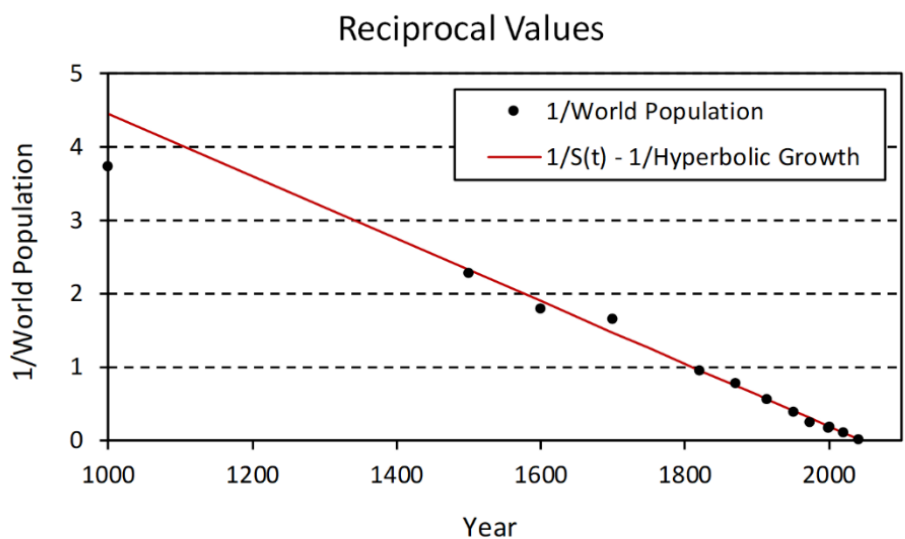

**Figure 3.** Reciprocal values of the hyperbolic distribution presented in Figure 1 together with the reciprocal values of the size of population. Complicated hyperbolic distribution is now represented by a simple straight line, which is easy to understand. The takeoff around 1800 shown in Figure 2 did not happen. The straight line cannot be divided into two distinctly different components making it clear that hyperbolic distribution shown in Figure 1 is also made of a single component. The slow and the fast growth shown in Figure 1 follow the same, monotonically-increasing distribution.

The display in Figure 3 is from AD 1000 for two reasons. (1) There is a large gap between AD 1 and 1000 so the display from AD 1000 shows better the agreement of the fitted hyperbolic distribution with data. (2) Detailed analysis of data for the AD and BC eras shows clearly that between around 500 BC and AD 500 there was a massive transition from a fast hyperbolic growth during the BC era to a significantly slower hyperbolic growth during the AD era (Nielsen, 2016a). The point at AD 1 is right in the middle of this transition and belongs to an entirely different distribution, the distribution describing the process of transition.

It should be also noticed that the point at AD 1000 in Figure 3 appears to be much further away from the fitted distribution then the point in Figure 1. The distributions are precisely the same but the display of reciprocal values magnifies the discrepancies between data and the calculated curve for small values (Nielsen, 2016b). The smaller are the values of the data and of the calculated distribution the larger is the magnification.

Reciprocal values allow for a unique identification of the first-order hyperbolic distributions because only these distributions are represented then by straight lines. This representation allows also for an easy study of departures from hyperbolic growth because deviations from a straight line are easy to notice.

Properties of growth do not change by changing the display of data but certain features, which are difficult or even impossible to recognize in one display can be easily identified in another. It is essential to remember that in the display of reciprocal values effects are reversed. Thus, for instance, a deviation to a slower trajectory will be indicated by an *upward* bending and deviation to a faster trajectory by a *downward* bending. An increasing growth is represented by a decreasing trajectory of the reciprocal values.

When hyperbolic growth is represented by a mathematically generated and gradually changing curve, such as shown in Figure 1, it might be clear that there was no particular time when the growth changed from being nearly horizontal to nearly vertical, but when data represented by discrete points are displayed, such a conclusion might be less obvious. The illusion becomes particularly strong when only a few strategically located points are selected (Ashraf, 2009; Galor 2005a, 2005b, 2007, 2008a, 2008b, 2010, 2011, 2012a, 2012b; Galor & Moav, 2002; Snowdon & Galor, 2008) from a significantly

larger set of data as if to make the deception even more pronounced. Even if the enforcement of the perceived illusion is unintended, such crude displays of data lead readily to grossly incorrect interpretations.

However, if reciprocal values of data are displayed, their analysis is immediately made significantly simpler because if the data follow a simple, first-order hyperbolic distribution, their reciprocal values will be clearly aligned along a decreasing straight line. It is then obvious that dividing such a straight line into two sections and claiming two distinctly different regimes of growth governed by two distinctly different mechanisms simply makes no sense. It also makes no sense to try to locate a point on the decreasing straight line and claim a transition to a new trajectory because there is obviously no transition to a new trajectory on a decreasing straight line.

It should be stressed that in this representation only the first-order hyperbolic distributions describing growth will follow the decreasing straight-line trajectories. It is for this reason that this simple method is so useful in identifying the first-order hyperbolic distributions. It is a simple and yet powerful method, which can be used successfully in the analysis of data describing the historical economic growth and the growth of human population, global, regional or local, because in general they follow simple, first-order hyperbolic trajectories. Any deviations from such trajectories can be easily investigated. Higher-order hyperbolic distributions describing growth will be represented by gradually decreasing trajectories, which could be fitted using higher-order polynomial functions intercepting the horizontal axis, while the exponential growth will be represented by a decreasing exponential function.

This method might have a more general application but its specifically intended application described in this publication is to help to avoid being guided by hyperbolic illusions, the unfortunate common mistake, which often leads to seriously incorrect conclusions as we shall demonstrate in the examples 2 and 3.

Going beyond the intended application, the first-order decreasing hyperbolic distributions will be represented by the increasing straight lines. Again, in this representation, any deviation from the decreasing hyperbolic distributions can be easily detected and investigated. Pareto distributions, which resemble the decreasing hyperbolic distributions, will be represented by gradually increasing functions, which in this representation might be also easier to investigate.

We shall now illustrate the application of the method of reciprocal values by using three additional examples: the growth of human population in Africa, the economic growth in Western Europe and the world economic growth.

## 3. Further examples
### 3.1. Growth of population in Africa

The method of reciprocal values can be used to study fine details of growth trajectories, the study which can then be used not only to improve the fit to data but also to understand the mechanism of growth. Some distributions might be made of different components, which could be difficult or even impossible to see in the direct display of data but they could be easily revealed by displaying their reciprocal values. An excellent example is the growth of human population in Africa shown in Figure 4, constructed using Maddison's data (Maddison, 2010). These Figure illustrates the added advantage of using the reciprocal values of data.

The top panel in Figure 4 contains the direct display of data for the growth of human population in Africa. The displayed shape suggests hyperbolic growth because it is slow over a long time and fast over a short time.

However, the reciprocal values of data presented in the lower panel reveal that the growth trajectory is in fact made of two major components: a slow hyperbolic distribution until around 1870 and a fast hyperbolic distribution after that year. Parameters describing the two hyperbolic components are $a_0 = 5.105 \times 10^1$, $a_1 = -2.036 \times 10^{-2}$ for the slow component and $a_0 = 1.705 \times 10^2$, $a_1 = -8.515 \times 10^{-2}$ for the fast component.

Figure 4 shows also that at a later stage, the fast hyperbolic growth started to be diverted to a slower trajectory as indicated by the upward bending of the trajectory representing the reciprocal values. Furthermore, it is now clear that the growth of population in Africa was never stagnant and that there was never a transition from stagnation to growth. The first stage of growth was hyperbolic and the transition around 1870 was a transition from hyperbolic growth to another hyperbolic growth. All these features, which are unrecognisable in the direct display of data are clearly seen in the display of the reciprocal values.

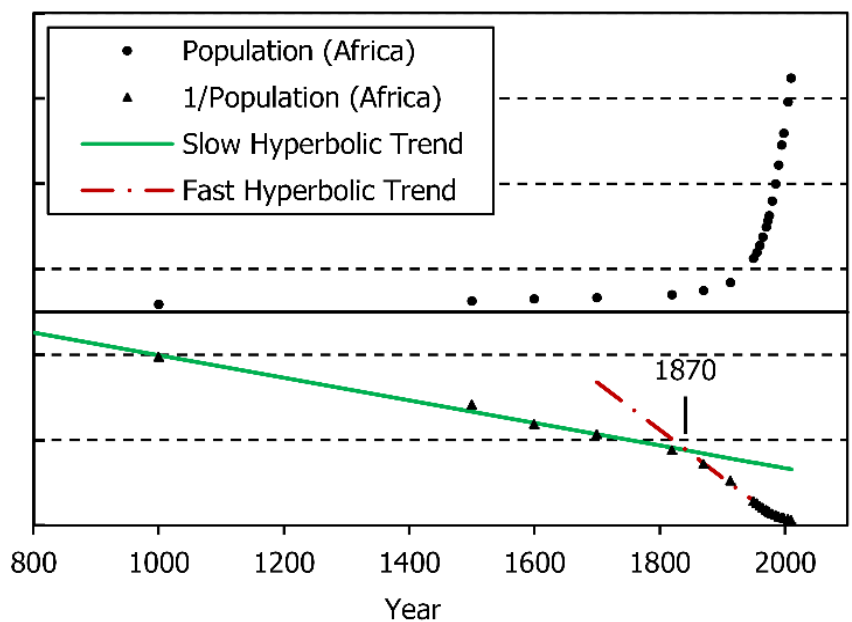

**Figure 4.** Growth of human population in Africa (Maddison, 2010) illustrates how the method of reciprocal values can serve as an excellent tool in revealing hidden features of studied distributions.

The pattern revealed by data contradicts the traditional interpretations of the mechanism growth of human population. There was no escape from the Malthusian trap because there was obviously no trap in the growth of population. The growth was slow but it was increasing monotonically with no signs of restrictions imposed by a mythical trap.

The transition from a slow to a fast hyperbolic growth in Africa occurred around the time of the Industrial Revolution but it was not a transition from a stagnant growth to a new, so-called sustained growth regime (Galor, 2005a, 2011) but from a hyperbolic growth to another but faster hyperbolic growth. It was the boosting that coincides with the intensified colonisation of Africa (Duignan & Gunn, 1973; McKay, Hill, Buckler, Ebrey, Beck, Crowston, & Wiesner-Hanks, 2012; Pakenham, 1992).

Contrary to the commonly accepted interpretations, the boosting in the growth of population was *not* triggered by a dramatically *decreased* intensity of Malthusian positive checks but by their dramatic *escalation*. It is clear that the accepted interpretations of the effects of Malthusian positive checks are incorrect. Their increased intensity does not lead to stagnation but to a more intensified growth (Malthus, 1798; Nielsen, 2016c). The increased intensity of Malthusian positive checks increases the mortality rate but it also increases the fertility rate with the net result of increasing the rate of natural increase or the growth rate. This correlation is also clearly demonstrated even now by the growth of population in poor countries. The poorer they are the faster is the growth of their populations. Thus, this simple analysis of data assisted by using the reciprocal values already questions the commonly accepted interpretations of the mechanism of growth of human population.

As shown in Figure 4, reciprocal values of data reveal the details of the mechanism of growth, which were impossible to identify by the direct display of data. Even if we cannot yet fully explain these details, we can already see that the growth of the populations in Africa was following a slow hyperbolic trend until around 1870. Around that year, the growth of human population in Africa experienced an unprecedented 4-fold acceleration, which diverted the growth into a significantly faster hyperbolic trajectory. The fast-hyperbolic growth continued until around 1975 when it started to be diverted to a new but slower trend.

It is this pattern of growth that we have to explain. It is for this pattern of growth that we have to propose the mechanism of growth. It is not the imaginary pattern of stagnation followed by explosion. It is not the fictitious Malthusian regime followed by the mythical takeoff from stagnation to an imagined sustained growth regime (Galor, 2005a, 2011). It is an entirely different pattern, the pattern indicated by the close analysis of data rather than by the pure fantasy. The aim of scientific investigation is not to explain figments of imagination but the evidence presented by data.

Data are essential in scientific investigations. Assisted by data we shall not be guided by the erroneous concept of stagnation but by the clear evidence of hyperbolic growth. We shall also not be guided by the erroneous concept of a takeoff from stagnation to a sustained growth regime but by the clear evidence of a transition from a hyperbolic growth to another hyperbolic growth. We shall also be guided by an observation that at a certain stage, around 1975, the long-lasting pattern of hyperbolic growth has been eventually abandoned and the growth was diverted to an entirely different trajectory.

### 3.2. Economic growth in Western Europe

Economic growth is measured using the Gross Domestic Product (GDP) or the GDP per capita (GDP/cap). Galor and Moav (2002) studied economic growth in Western Europe using the data of Maddison (Maddison, 2001). They have selected a few, strategically located points from a larger set of data, joined them by straight lines and concluded that there were two distinctly different regimes of growth: the "Malthusian regime" (also labelled as the "epoch of stagnation," "Malthusian era," "Malthusian epoch," "Malthusian steady-state equilibrium," "Malthusian stagnation" or "Malthusian trap") and the "sustained economic growth" (described also as the "Modern Growth Regime," "sustained economic growth" and "sustained growth regime"). Their distorted representation of Maddison's data is shown in Figure 5.

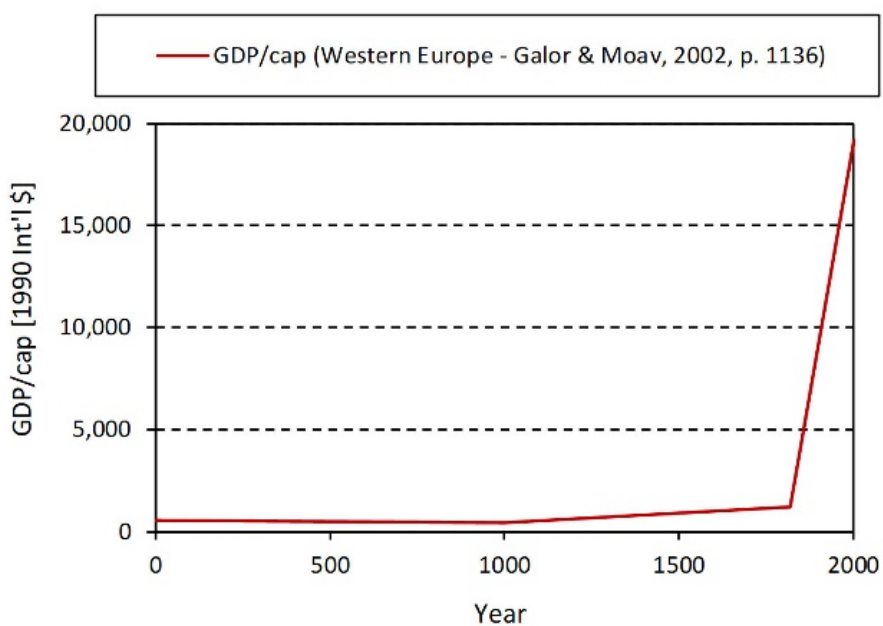

**Figure 5.** A typically distorted and self-misleading representation (Galor & Moav, 2002, p. 1136) of Maddison's data (Maddison, 2001). Compare it with exactly the same data, but not distorted, presented in Figure 7.

Referring to this crude display of data they also concluded that the Industrial Revolution had a strong impact on the economic growth causing a dramatic takeoff from stagnation to a fast growth. They made no attempt to analyse mathematically Maddison's data (Maddison, 2001) but presented a series of mathematical equations describing their imaginations, which were neither related to nor supported by the source of data they have used.

It is remarkable that data coming from *precisely the same source* as they have used contradict their claims and their interpretations of growth. Extensive analysis of the GDP/cap data, global and regional, is presented in a separate publication (Nielsen, 2016d). It is shown there that GDP/cap data follow the *monotonically increasing* trajectories. They are just the linearly modulated hyperbolic trajectories (Nielsen, 2017a), i.e. hyperbolic trajectories modulated by the linear time-dependence of the reciprocal values of the size of population. There is no stagnation and no takeoff to a distinctly different regime of growth. Both, the GDP and the population increase hyperbolically (Nielsen, 2016b, 2016e, 2016f) and thus monotonically. Consequently, their ratios increase also monotonically.

Figure 6 presents the reciprocal values of the Gross Domestic Product (GDP) for Western Europe (Maddison, 2001) in the vicinity of the alleged takeoff. The data are well aligned along a decreasing straight line, which means that they were following the simplest, first-order, hyperbolic distribution given by the eqn (1).

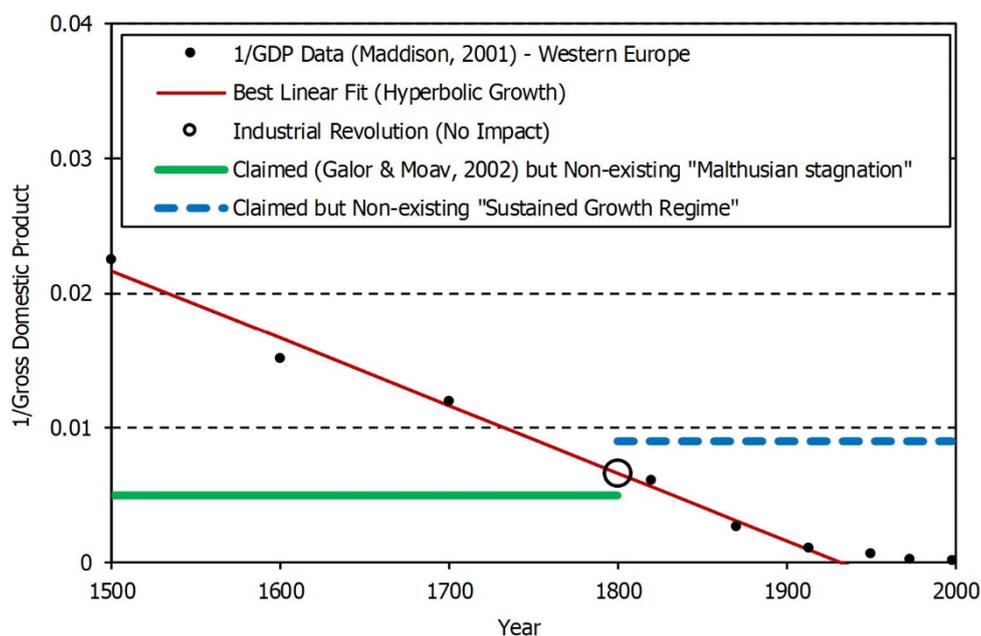

**Figure 6.** Reciprocal values of data describing the Gross Domestic Product (GDP) in Western Europe (Maddison, 2001) in the vicinity of the Industrial Revolution. This is precisely the same source of data as used by Galor and Moav (2002) to construct their distorted representation shown in Figure 5. Contrary to their claim, Industrial Revolution had no effect on shaping the economic growth trajectory in Western Europe, the centre of this revolution. The two regimes of growth claimed by them also did not exist. The GDP is in billions of the 1990 International Geary-Khamis dollars.

Industrial Revolution was between 1760 and 1840 (Floud & McCloskey, 1994), or around 1800 as shown in Figure 6. This figure demonstrates clearly and convincingly that the claimed takeoff around the time of the Industrial Revolution did not happen because the reciprocal values of the GDP data follow an undisturbed straight-line trajectory representing an undisturbed hyperbolic growth. It is now clear that there was no takeoff and no escape, great or small, from the hypothetical but non-existing Malthusian trap, at least from the alleged trap in the economic growth. Maybe there were some other traps but maybe they are just figments of imagination. It is clear, however, that Industrial Revolution had absolutely no impact on shaping the economic growth trajectory in Western Europe, the centre of this revolution.

Industrial Revolution had, no doubt, many other impacts but they are not reflected in the economic growth trajectory. Their study could be important and interesting but they will not explain the growth of the GDP. The mechanism of growth was immune to the changes introduced by the Industrial Revolution. Whatever dramatic changes the Industrial Revolution might have introduced to the general style of living, to technology and even to the economic marked, these changes obviously were not shaping the economic growth trajectory.

The absence of a takeoff eliminates also the need for assuming the existence of two distinctly different regimes of growth. It obviously makes no sense to divide the straight line into two arbitrarily selected sections and claim distinctly different trajectories governed by distinctly different mechanisms of growth. What might not have been clear in the direct display of data, is now perfectly obvious if we display the reciprocal values of data. This display abolishes all elaborate theories and untidy explanations incorporating such concepts as traps, escapes, takeoffs and stagnation and replaces them by a simple interpretation of the mechanism of growth suggested by the simple equation describing hyperbolic growth. This conclusion is in agreement with the general observation that natural phenomena can be usually explained by using simple descriptions.

In Figure 7, the hyperbolic trajectory corresponding to the straight line shown in Figure 6 is extended to AD 1. The economic growth in Western Europe is well described by a simple, first-order, hyperbolic distribution. The corresponding parameters are: $a_0 = 9.697 \times 10^{-2}$ and $a_1 = -5.020 \times 10^{-5}$. The point at 1950 is not fitted by the hyperbolic trend because from the early 1900s the economic growth in Western Europe started to be diverted to a *slower* trajectory, which is again contrary to the claimed boosting or a transition from stagnation to growth. There was a transition but it was a transition from a *monotonically increasing* hyperbolic growth to a *slower* trajectory.

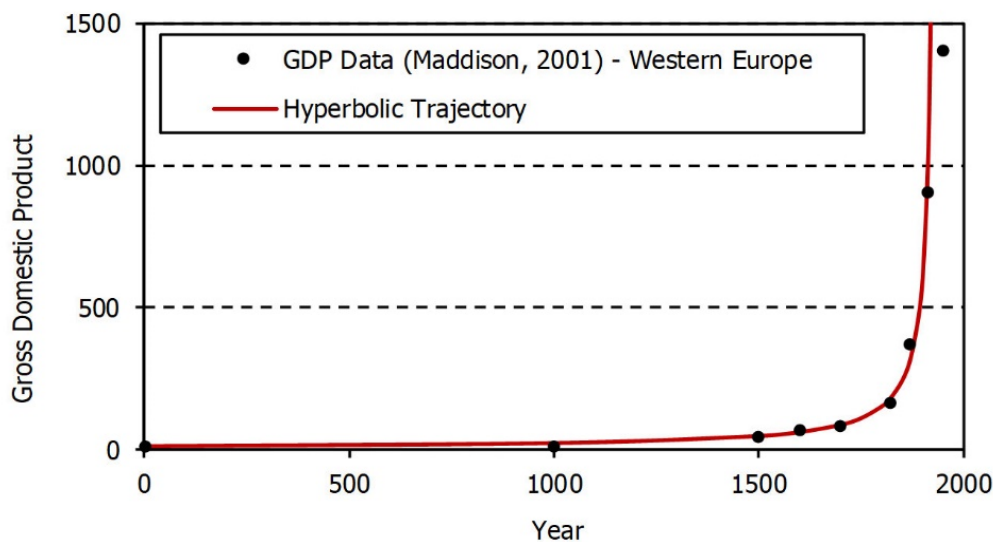

**Figure 7.** Data for the Gross Domestic Product (GDP) in Western Europe (Maddison, 2001) compared with the monotonically increasing hyperbolic distribution. The GDP is in billions of the 1990 International Geary-Khamis dollars.

We cannot claim that the growth was sustained only after the Industrial Revolution because it was sustained equally strongly during the postulated but non-existent "epoch of stagnation." Figure 6 and Figure 7 show clearly that the concept of two stages of growth is unsupported by data. When stripped of the hyperbolic illusions, the economic growth is revealed as a simple process, which can be described using just one, simple mathematical trajectory until the early 1900s when it started to be diverted to a slower, non-hyperbolic, trajectory. There is no compelling need to make this simple description complicated.

Growth of the GDP was slow in the past because it was hyperbolic. However, while being slow it was not stagnant. The growth was fast in recent years because it was hyperbolic. It followed the same undisturbed hyperbolic distribution as in the past.

We now have a completely different understanding of the economic growth in Western Europe, an important turnaround in the economic research. Rather than wasting the valuable time, energy and financial resources on trying to explain the phantom features created by hyperbolic illusions and magnified by the customary crude representation of data (Ashraf, 2009; Galor 2005a, 2005b, 2007, 2008a, 2008b, 2010, 2011, 2012a, 2012b; Galor & Moav, 2002; Snowdon & Galor, 2008) we can now focus our attention on the relevant task of trying to explain why the economic growth was so stable over such a long time and why it was hyperbolic. Rather than writing numerous articles based on impressions and publishing them in peer-reviewed scientific journals and in academic books we can now concentrate our attention on the understanding of *the science* of economic growth. In our investigations, we shall not be guided by impressions, we shall not be guided by the customary crude representations of data but by their rigorous mathematical analysis

### 3.3. *Global economic growth*

Another example of the application of the method of reciprocal values is the global economic growth. It is an important example because it questions Galor's Unified Growth Theory (Galor, 2005a, 2011) representing the culmination of his work extending over 20 years (Baum, 2011). His theory is based on an uncritical acceptance of the common interpretations, descriptions and explanations used in the economic and demographic research. In this sense, his theory offers no new insights.

The fundamental postulate of this theory is again the existence of three regimes of growth: the slow and stagnant Malthusian Regime, the short and intermediary Post-Malthusian Regime and the fast, Sustained Growth Regime. Galor also accepts that Industrial Revolution played a crucial role in the alleged dramatic takeoff from a prolonged stagnation into a rapid and sustained growth.

The welcome initiative in his theory is that he makes an attempt of using repeatedly Maddison's data (Maddison, 2001). However, he makes not even a single attempt to test his theory by the rigorous analysis of data. This is a serious omission. The usual practice in any scientific theory is to test it by data or at least to suggest how it can be tested by data. Galor does not follow this accepted practice. He does not test his mathematical descriptions by data. Data are used repeatedly but they are never analysed. They are presented in a typically distorted way, as illustrated in Figures 2 and 5, and in this distorted way they seem to support the preconceived ideas. His work is based on prejudice and no attempt is made to check its validity.

When data are used but manipulated to confirm preconceived ideas we are not dealing with science. We also make no progress and we are not learning anything new or useful.

We shall now use exactly *the same source of data* and show that the Unified Growth Theory is scientifically unsustainable. For more extensive discussion of these issues see other publications (Nielsen, 2016a, 2016b, 2016d, 2016e, 2016f, 2016g, 2016h, 2016i, 2016j, 2017a).

It is hard to see how much can be rescued from Galor's Unified Growth Theory. It is hard to see how many of his descriptions and explanations are based on pure and unsubstantiated speculations. His theory would have to be minutely analysed. However, its major premises are untenable. All his "mind boggling" "mysteries of the growth process" (Galor, 2005a, p 220), for instance, can be easily explained (Nielsen, 2016a, 2016d, 2016g, 2016i) – *there are no mysteries*. All his mysteries were created by his repeatedly distorted presentations of data coming from a reputable source (Maddison, 2001), the data used during the formulation of his theory but never properly analysed.

His theory certainly does not explain the mechanism of growth because it revolves around the descriptions of hyperbolic illusions. It does not even describe economic growth. His descriptions are incorrect because again they are based on the distorted presentations of data and on the unsubstantiated prejudice.

Theories come and go. Scientific integrity is not tarnished by proposing incorrect explanations and interpretations but by refusing to correct them or to reject them when they are contradicted by reliable data.

Reciprocal values of data for the world Gross Domestic Product (GDP) (Maddison, 2001) are shown in Figure 8. They follow closely a decreasing straight line, which means that the economic growth was increasing hyperbolically. It is clear that there was no takeoff of any kind, large or small, around the time of the Industrial Revolution and no repeatedly claimed great escape from the postulated but non-existing Malthusian trap. The data do not support the existence of the three regimes of growth and thus contradict the fundamental postulates of the Unified Growth Theory.

The last point of the data shown in Figure 8 is not fitted by the straight line, suggesting a possible diversion to a slower trajectory. This region can be studied more closely using the extended compilation of the economic growth data (Maddison, 2010). Their reciprocal values between 1700 and 2003 are shown in Figure 9 demonstrating clearly that while the Unified Growth Theory claims an unusually accelerated growth after the alleged but non-existent epoch of stagnation, the data show the opposite behaviour: a diversion to a *slower* trajectory after the earlier vigorous, well-sustained and secure economic growth. Rather than being boosted by the Industrial Revolution, the economic growth continued along the *undisturbed* hyperbolic trajectory for about one hundred years after this revolution and then started to be diverted to a *slower* trajectory.

Figure 9 illustrates again how the method of reciprocal values can unravel useful details about a studied process. Not only does it help in an unambiguous and easy identification of hyperbolic distributions but also it helps in an easy detection of deviations from such distributions. The world economic growth continues to increase but from the early 1900s it started to be diverted away from the faster accelerating historical hyperbolic trajectory to a slower trend.

The point of intersection of the reciprocal values with the horizontal axis is the point of singularity when the growth escapes to infinity. No growth can go beyond this point and any growth close to it may become unstable, unsustainable and catastrophic. Figures 8 and 9 show how close we are now to the point of the potential global economic instability and unsustainability.

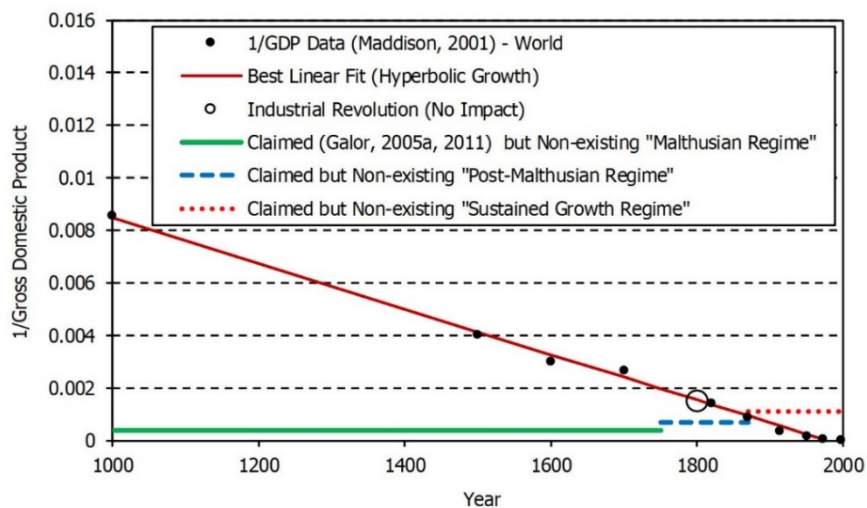

**Figure 8.** Fundamental concepts of the Unified Growth Theory are contradicted by *precisely the same data* (Maddison, 2001), which were used (but never analysed) during its development. Reciprocal values of data follow closely a decreasing linear distribution representing a monotonically increasing hyperbolic growth. The three regimes of growth claimed by Galor (2005a, 2011) did not exist. There was no takeoff around the time of the Industrial Revolution or around any other time. The monotonically increasing hyperbolic growth remained undisturbed until the 1990s. The GDP is in billions of the 1990 International Geary-Khamis dollars.

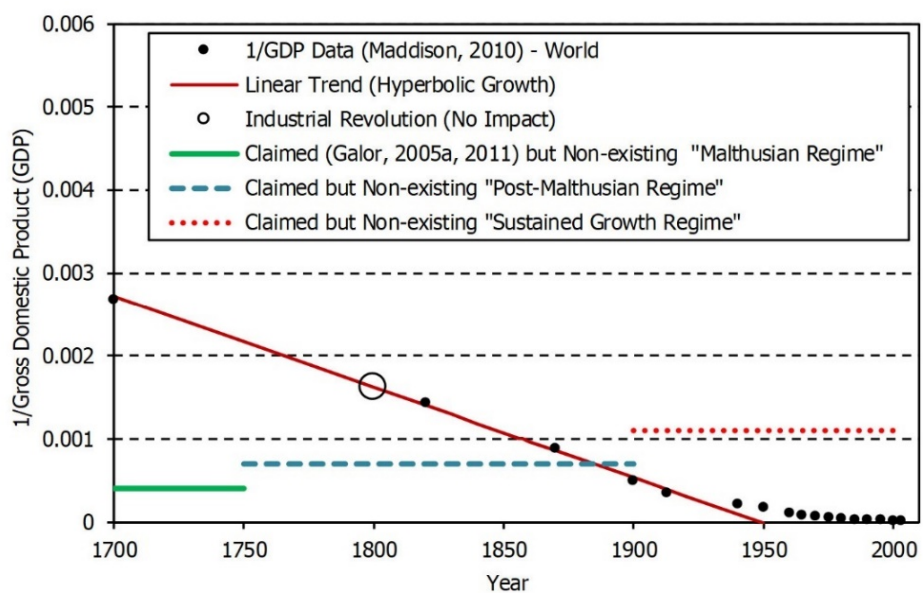

**Figure 9.** Maddison's data (Maddison, 2010) show clearly that while the Unified Growth Theory (Galor, 2005a, 2011) claims a transition from stagnation to a vigorous growth, the data show the opposite behaviour: a transition from a vigorous hyperbolic growth to a *slower* trajectory, as indicated by the upward bending of the growth trajectory of the reciprocal values during the 1990s and 2000s. There was no stagnation and no boosting in the economic growth at any time. The claimed (Galor, 2005a, 2011) but non-existent three regimes of growth are also shown. Their existence is contradicted by data. The GDP is in billions of the 1990 International Geary-Khamis dollars.

Unified Growth Theory claims that after a long epoch of stagnation we have now reached an era of "sustained economic growth," the term repeated 82 times in the first detailed formulation of this theory (Galor, 2005a), the potentially misleading description because while it is true that the current economic growth is still sustained the past economic growth was not only sustained but also it was increasing along a more secure trajectory, far away from the point of singularity. Even though the growth is now diverted to a slower trajectory any further increase can be potentially dangerous.

Reciprocal values of data show that for the first time during the AD era, and probably for the first time in human existence, we are now trapped between the already high level of economic growth and a point of no return, or equivalently between the very small reciprocal values of the GDP and zero. Any intrusion into this narrow gap has to be closely monitored. Even if the trend of the reciprocal values of the GDP data does not cross the horizontal axis any close approach to this axis could be dangerous, because it could trigger global economic instability and even a possible global economic collapse.

This simple analysis of data shows how dangerous are the generally accepted postulates presented in the Unified Growth Theory. The concept of a transition from stagnation to the so-called sustained growth regimes suggests that now for the first time in human history we can enjoy the sustained economic growth. Data, however, reveal a diametrically different pattern of growth. It was in the past that the economic growth was sustainable because it was following a stable hyperbolic distribution, which was still far away from the point of singularity. Now, however, the reciprocal values of the GDP are so close to zero, i.e. to the point when the GDP escapes to infinity, that the economic growth is no longer easily sustainable. The possibility of a serious economic instability is real. Now, the economic growth has to be closely monitored and controlled. The claim that we are now in the regime of sustained economic growth is dangerously inaccurate and misleading.

The method of reciprocal values can be also used to demonstrate that two other postulates of the Unified Growth Theory, the postulate of the differential takeoffs and the postulate of the great divergence, are contradicted by the mathematical analysis of data coming from *the same source,* which was used during the formulation of this theory (Nielsen, 2016b, 2016e, 2016i). Takeoffs never happened and consequently it makes no sense to claim that they happened at different times for developed and developing regions. The so called great divergence also never happened. Different regions are on different levels of development but they follow closely similar trajectories. They are like athletes running along similar tracks. They do not run in distinctly different directions as incorrectly claimed in the Unified Growth Theory but in the same direction.

If the economic growth continued along the historical hyperbolic trajectory it would have already reached a point of no return as indicated by the fitted straight line crossing the horizontal axis. To use the colourful description of von Foerster, Mora and Amiot (1960), we have been saved from experiencing a doomsday in the global economic growth. However, the danger of an excessive and unsustainable growth is still not averted.

Under a suitable control, the economic growth can continue for a long time, but this is precisely the important point: from now on the economic growth has to be closely monitored and controlled because it can easily become unsustainable.

Data between 1965 and 2003 follow closely exponential trajectory. Exponential growth does not increase to infinity at a fixed time but this is hardly any consolation because eventually such a growth also becomes unsustainable.

Any other continually increasing growth can be unsustainable unless it is increasing to a certain constant asymptotic value. However, it is extremely difficult to control such a growth because the growth rate would have to finely tuned to decrease slowly to zero. A constant growth rate, even if small, would represent the undesirable exponential growth. A growth rate fluctuating around zero would be safe but our general tendency is to try to increase the growth rate or at least to keep it constant, both options leading to unsustainable economic growth.

Data describing the world economic growth (Maddison, 2001) are compared in Figure 10 with the hyperbolic trajectory calculated using the straight-line fitted to the reciprocal values shown in Figure 8. Parameters describing the historical hyperbolic growth of the world GDP are: $a_0 = 1.716 \times 10^{-2}$ and $a_1 = -8.671 \times 10^{-6}$.

Now the puzzling features of the economic growth, the features that prompted so many discussions in numerous peer-reviewed scientific journals culminating in the formulation of the Unified Growth Theory, are manifestly clear, and their explanation is surprisingly simple. Over hundreds of years, the world economic growth was slow because it was hyperbolic. Over a short time, until the early 1900s, the economic growth was fast because it was hyperbolic – it followed *the same* undisturbed hyperbolic trajectory as in the past. The apparent transition from a slow to a fast growth is just an illusion created by the hyperbolic distribution. There was no unusually accelerated transition from the slow to the fast economic growth. The acceleration was gradual over the entire range of time.

The study presented here shows how important it is to have a clear understanding of the economic growth and how the simple method of reciprocal values can assist in such studies. Application of this method can not only assist in unravelling different components of growth trajectories but also to avoid being guided by hyperbolic illusions, which are the source of numerous misinterpretations of economic growth and of the growth of population culminating in the formulation of the fundamentally flawed and strongly misleading Unified Growth Theory (Galor, 2005a, 2011).

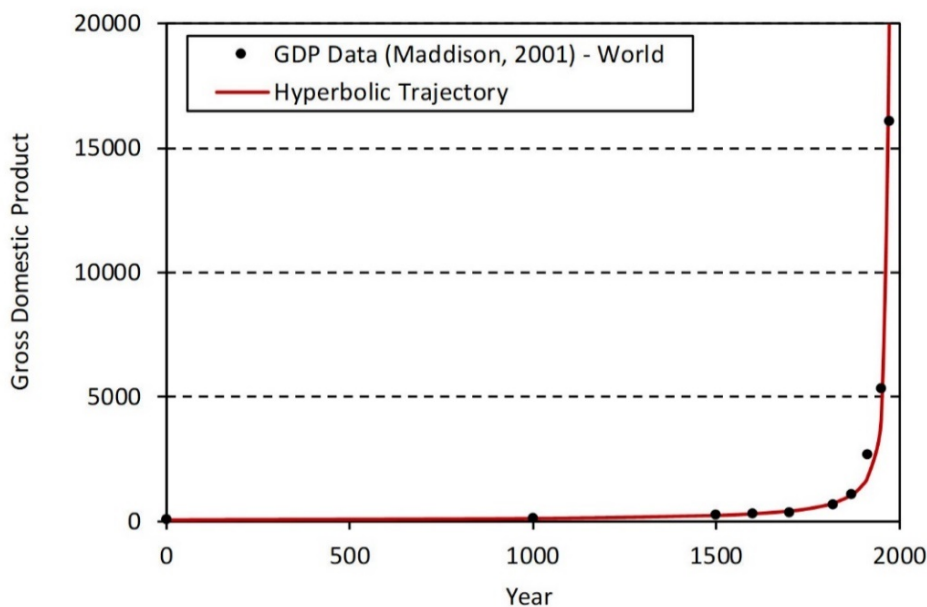

**Figure 10.** The data for the world Gross Domestic Product (GDP) (Maddison, 2001) follow closely the first-order hyperbolic distribution. The claimed three regimes of growth (Galor, 2005a, 2011) did not exist. They are replaced by an uninterrupted and monotonically increasing hyperbolic growth. The GDP is in billions of the 1990 International Geary-Khamis dollars.

## 4. Summary and conclusions

We have described a simple but effective method of analysis of hyperbolic distributions and we have explained its application by using the growth of the world population during the AD era. We have then demonstrated the flexibility of this method by using an example of the growth of human population in Africa. This method can be used to identify uniquely the first order hyperbolic distributions, to reveal hidden components of growth trajectories and to remove hyperbolic illusions, which are the source of numerous misinterpretations of economic growth and of the growth of population, misinterpretations prevailing over a long time in academic literature. This simple method redirects the economic and demographic research from explanations of phantom features created by hyperbolic illusions, to explanations based on the scientific analysis of data.

    We have presented two other examples of analysis of data: the economic growth in Western Europe and the global economic growth. All these four examples show that the rigorous analysis of data contradicts the established knowledge in demography and in the economic research, and in particular, that it contradicts the fundamental postulates of the Unified Growth Theory (Galor, 2005a, 2011). However, what we have presented here is just a tip of an iceberg. An entirely new world is opened when we analyse more data (Nielsen, 2016a, 2016b, 2016c, 2016d, 2016e, 2016f, 2016g, 2016h, 2016i, 2016j, 2016k, 2016l, 2016m, 2016n, 2016o, 2016p, 2017a, 2017b, 2017c, 2017d), the world without stagnation in the economic growth and in the growth of population, without takeoffs from the alleged stagnation to growth, the world without complicated and untidy explanations of the mechanism of growth but the elegant world where data can be described by simple mathematical distributions, the world, which opens up new vistas for the demographic and economic research.

Impressions can be strongly deceptive and persuasive. "It is clear that the earth does not move, and that it does not lie elsewhere than at the centre" declared Aristotle. Fortunately, however, in science, incorrect interpretations are sooner or later corrected.

1. *Stagnation*. Research based on impressions and reinforced by the customary crude and self-misleading representations of data (Ashraf, 2009; Galor 2005a, 2005b, 2007, 2008a, 2008b, 2010, 2011, 2012a, 2012b; Galor & Moav, 2002; Snowdon & Galor, 2008), such as shown in Figures 2 and 5, seems to confirm the generally accepted belief that there was an epoch of stagnation in the economic growth and in the growth of population. Scientific analysis of *precisely the same* (but undistorted) data demonstrates that there was no stagnation and that the economic growth and the growth of population followed *monotonically increasing hyperbolic distributions*.

2. *Takeoffs*. Research based on impressions seems to indicate that there was a transition from stagnation to growth described usually as a takeoff or explosion. Scientific analysis of precisely the same (but undistorted) data demonstrates that there was no takeoff or explosion and that economic growth and the growth of population continued to follow the monotonically increasing hyperbolic distributions. *What appears as a takeoff or explosion is in fact the natural continuation of hyperbolic growth.*

3. *Industrial Revolution*. Research based on impressions seems to indicate that Industrial Revolution played a crucial role in the economic growth and in the growth of population causing a dramatic acceleration (boosting) in the growth trajectories, described as takeoffs. Scientific analysis of precisely the same (but undistorted) data demonstrates that *Industrial Revolution had absolutely no impact on shaping growth trajectories*. Industrial Revolution can be linked to other impacts but not to shaping the population or the economic growth trajectories. This might be surprising but the evidence in data is undisputable and we have to accept it.

4. *Regimes of growth*. Research based on impressions seems to suggest that there were two or maybe even three distinctly different regimes of growth governed by distinctly different mechanisms (Galor, 2005a, 2011). Scientific analysis of precisely the same (but undistorted) data demonstrates that these two or three distinctly *different regimes of growth did not exist*. The growth was hyperbolic until recently when it started to be diverted to *slower* trajectories.

5. *Mysteries*. Research based on impressions resulted in claiming a series of "mind-boggling" and "perplexing" "mysteries of the growth process" (Galor, 2005a, pp. 177, 220). Scientific analysis of precisely the same data demonstrates that all these mysteries belong to the world of fiction created by a good dose of fantasy guided by the misleading impressions and reinforced by the customarily distorted presentations of data (Ashraf, 2009; Galor 2005a, 2005b, 2007, 2008a, 2008b, 2010, 2011, 2012a, 2012b; Galor & Moav, 2002; Snowdon & Galor, 2008) such as shown in Figures 2 and 5. Science is supported by a methodical analysis of data. There are no mysteries when precisely the same data are properly analysed.

   In particular, the mystery of the great divergence is explained: there was no great divergence (Nielsen, 2016i). Various regions are on different levels of economic growth but they all follow closely similar trajectories. Their economic growth did not diverge into distinctly different trajectories as incorrectly suggested by the crude representations of data.

   The mystery of the alleged sudden spike in the growth rate of income per capita has been explained: there was no sudden spike (Nielsen, 2016g). The growth rate of income per capita followed a monotonically increasing trajectory, which is readily represented by a mathematical distribution derived using hyperbolic growth for the growth of the GDP and for the growth of population.

   The mystery of the puzzling features of income per capita has been explained (Nielsen, 2017a). The distribution representing income per capita is nothing more than just a linearly modulated hyperbolic distribution. It reflects nothing more than the purely mathematical property of dividing two hyperbolic distributions.

   Other questions listed by Galor as representing the mysteries of the growth process can be easily answered. They refer to features that do not exist, features based on impressions reinforced by ineffectual handling of empirical evidence. They are in the same category as the question "Why does the sun revolve around the earth?"

6. *Mechanism*. Research based on impressions leads to proposing numerous complicated mechanisms of growth. Scientific analysis of data shows that the mechanism of growth is exceptionally *simple* (Nielsen, 2016p), which is hardly surprising because hyperbolic distributions are described by an exceptionally simple equation [see eqn (1)].

7. *Unified Growth Theory.* Research based on impressions prompted the development of a Unified Growth Theory (Galor, 2005a, 2011). Mathematical analysis shows that the fundamental postulates of this theory are *contradicted by the same data,* which were used during its development. Galor could have saved 20 years of his life and could have directed his academic skills to developing a useful theory if he did what any scientist is supposed to do: if he based his deductions and explanations on a scientific analysis of data. He had access to excellent data but he did not analyse them. He was guided by preconceived ideas and he supported them by distorted presentations of data.

The analysis data suggests new lines of research. Thus, for instance, the relevant question is not why the historical economic growth was so unstable in the past or what caused the perceived transition from alleged stagnation to growth but *why the economic growth was so remarkably stable* in the past. The same question applies to the growth of population but it was already answered (Nielsen, 2016c, 2017d). The growth of population was remarkably stable because of the combination of the generally low impacts of demographic catastrophes (at least on the global and regional scales) and the high level of human resilience expressed in the efficient process of regeneration (Malthus, 1798; Nielsen, 2016c). If we accept that there is a close relationship between the growth of population and the economic growth, then the question about the stability of the historical economic growth has been also already answered. However, it is possible that some new insights could be still added to this explanation.

The relevant question is not why the Industrial Revolution and the unprecedented technological development boosted the economic growth because they did not. The relevant question is *why the Industrial Revolution and the unprecedented technological development did not boost the economic growth*. Why these apparently strong technological and socio-economic forces had no impact on shaping the economic growth trajectories.

The relevant question is not why the economic growth increased so fast in modern time, because we have shown that this fast increase was just the natural continuation of the monotonically increasing hyperbolic growth until in recent years it started to be diverted to a *slower* but still fast-increasing trajectory. The relevant question is *why the economic growth was diverted to a slower trajectory*. What new force or forces were so strong that they were able to overpower the historically strong force of growth. Another relevant question is also whether this new trajectory is likely to develop into a historically preferable and potentially catastrophic, hyperbolic growth. Furthermore, the relevant question is how to control the current fast economic growth. The same question applies also to the growth of population but it was at least partly answered in the study of the effects of Malthusian positive checks (Nielsen, 2016c). The primary if not exclusive way of controlling the growth of human population is to improve the living conditions in developing countries.

The method of reciprocal values is so simple that it can be used by anyone and it is, therefore, expected that it will be of interest to many scientists who look for a simple method of analysis of empirical evidence, a method that does not involve any complicated mathematical formulae, any intricate mathematical algorisms or the use of powerful computers but a simple display of data and a remarkably simple fitting procedure. We have demonstrated that even a simple mathematical method can have a dramatic influence on scientific research.

It is essential to understand that by claiming that there was no stagnation in the economic growth or in the growth of population we are not claiming that there was no stagnation in the standard of living. We are only claiming that the two processes were decoupled. We might, if we insist, describe the past general living conditions as primitive or even stagnant, but there is no evidence that they were shaping the trajectories describing the growth of population or the economic growth.

It is also essential to understand that by claiming that there was no takeoff in the economic growth or in the growth of population we are not claiming that there was no takeoff in the technological development, or generally in the intellectual progress and in the dramatic changes in human experience and in living conditions. We are only claiming that these possible takeoffs had no impact on changing the economic growth trajectories or the trajectories describing the growth of population. There were no takeoffs in any of these two processes. Industrial Revolution can be linked with many changes in human living experience but all these changes had no impact on changing the economic or demographic growth trajectories.

There is no reason why scientific evidence presented here and in other related publications (Nielsen, 2016a, 2016b, 2016c, 2016d, 2016e, 2016f, 2016g, 2016h, 2016i, 2016j, 2016k, 2016l, 2016m, 2016n,

2016o, 2016p, 2017a, 2017b, 2017c, 2017d) should not be accepted by the scientific community. The only alternative option is to reject data but this would be no longer science.

Even Galor and his associates accept the same data and use them in their research. Their unfortunate mistake was only in choosing to support their investigations by the grossly distorted and self-misleading representations of data (Ashraf, 2009; Galor 2005a, 2005b, 2007, 2008a, 2008b, 2010, 2011, 2012a, 2012b; Galor & Moav, 2002; Snowdon & Galor, 2008). Consequently, the only way to reject scientific evidence and to accept the doctrines of stagnation and takeoffs, and all other associated erroneous explanations of the dynamics of the economic growth and of the growth of population, is to accept data but distort them in such a way as to make them to conform with preconceived ideas, but then again it is not science.

Evidence in data is overwhelming and leaves no room for accepting incorrect interpretations. In order to have progress in the demographic and economic research, incorrect interpretations of growth have to be abandoned and a new paradigm has to be developed. There is no other, scientifically justified, way. A serious mistake in scientific investigations is not in stumbling and in making mistakes but in refusing to learn from them and to correct them.

**3**

# Mathematical Analysis of the Historical Economic Growth with a Search for Takeoffs from Stagnation to Growth

## *By* Ron W. NIELSEN †

**Abstract.** Data describing historical economic growth are analysed. Included in the analysis is the world and regional economic growth. The analysis demonstrates that historical economic growth had natural tendency to follow hyperbolic distributions. Parameters describing hyperbolic distributions have been determined. A search for takeoffs from stagnation to growth produced negative results. This analysis throws a new light on the interpretation of the mechanism of the historical economic growth and suggests new lines of research.
**Keywords.** Historical economic growth; regimes of growth; stagnation; takeoffs; Malthusian trap; hyperbolic growth.
**JEL.** A10, B15, B16, B22, C29.

## 1. Introduction

T he latest publication of excellent data by the world-renown economist (Maddison, 2001, 2010) offers an unprecedented opportunity to study the mechanism of the historical economic growth. Earlier study (Nielsen, 2014), based on these data, indicated that historical economic growth can be described using hyperbolic distributions in much the same way as the growth of human population (von Foerster, Mora & Amiot, 1960). Unlike exponential growth, which is more familiar and which can be easier to understand, hyperbolic distributions are strongly deceptive because they appear to be made of two distinctly different components, slow and fast, joined perhaps by a certain transition component. This illusion is so strong that even the most experienced researchers can be easily deceived particularly if their research is based on a limited body of data, as it was in the past. Fortunately, Maddison's data solve this problem, and fortunately also their analysis is trivially simple because, as pointed out earlier (Nielsen, 2014), hyperbolic distributions can be easily identified and analysed using the reciprocal values of data.

Hyperbolic distribution describing *growth* is represented by a *reciprocal* of a linear function:

$$S(t) = \frac{1}{a - kt}, \tag{1}$$

where $S(t)$ is the size of the growing entity, in our case the Gross Domestic Product (GDP), while *a* and *k* are *positive* constants.

The reciprocal of such hyperbolic growth, $1/S(t)$, is represented by a *decreasing linear* function:

$$\frac{1}{S(t)} = a - kt. \tag{2}$$

† AKA Jan Nurzynski, Griffith University, Environmental Futures Research Institute, Gold Coast Campus, Qld, 4222, Australia.
☎. +61407201175
✉. r.nielsen@griffith.edu.au; ronwnielsen@gmail.com

Hyperbolic *distributions* should not be confused with hyperbolic *functions* ($\sinh(t)$, $\cosh(t)$ etc). Furthermore, *reciprocal* functions should not be confused with *inverse* functions. Thus, for instance, for the expression given by the eqn (1) the objective of finding the inverse function would be to calculate time *t* for a given size $S(t)$. The roles of the dependent and independent variables would be reversed. For the reciprocal function, the objective is to convert eqn (1) into eqn (2). The roles of dependent and independent variables are not changed.

Reciprocal values help in an easy and generally unique identification of hyperbolic growth because in this representation hyperbolic growth is given by a decreasing straight line. Apart from serving as an alternative way to analyse data, reciprocal values allow also for the investigation of even small deviations from hyperbolic distributions because deviations from a straight line can be easily noticed.

Reciprocal values allow also for an easy identification of different components of growth. This property can be used in comparing empirical information with theoretical interpretations (Galor, 2005, 2011), which are based on the assumption of the existence of different components of growth.

When comparing mathematically-calculated distributions with the reciprocal values of data, we have to remember that the sensitivity of the reciprocal values to small deviations increases with the decreasing size *S* of the growing entity.

Suppose we have two values of *S* at a given time: $S_1$ and $S_2$, representing, for instance, the empirical and calculated values. It is clear that

$$\Delta\left(\frac{1}{S}\right) = -\frac{\Delta S}{S_1 S_2}, \tag{3}$$

where $\Delta(1/S)$ is the difference between two inverse values and $\Delta S$ is the difference between *S* values.

For a given $|\Delta S|$, $|\Delta(1/S)|$ increases rapidly with the decreasing $S_1$ and $S_2$ values. The separation of small values of data from calculated distributions are magnified. Similar magnifications, though less pronounced, are also shown in the semilogarithmic displays of data. We shall use both displays to examine the quality of fits to the data.

It should be noted that the *decreasing* reciprocal values describe *growth*, while a deviation to *larger* reciprocal values describes decline. Consequently, a diversion to a faster trajectory will be indicated by a downward bending of a trajectory of the reciprocal values, away from an earlier observed trajectory, while the diversion to a slower trajectory will be indicated by an upward bending.

The data describing the historical economic growth (Maddison, 2001, 2010) do not allow for a detailed analysis below AD 1500 because there are two large gaps in the data: between AD 1 and 1000 and between AD 1000 and 1500. The best sets of data are from AD 1500. However, the compilation prepared by Magnuson appears to be the best and the most reliable source of data describing the historical economic growth.

Throughout the analysis presented here, the values of the Gross Domestic Product (GDP) will be expressed in billions of the 1990 International Geary-Khamis dollars. All diagrams are presented in the Appendix

Theories play an important role in scientific research because they crystallise interpretations of studied phenomena. However, theories have to be always tested by data. In science it is important to look for data confirming theoretical explanations but it is even more important to discover contradicting evidence, because data confirming a theory confirm only what we already know but contradicting evidence may lead to new discoveries.

Currently, the most complete theory describing the mechanism of the historical economic growth appears to be the Unified Growth Theory (Galor, 2005, 2008, 2011, 2012). One of the fundamental postulates of this theory is the postulate of the existence of three regimes of growth governed by three distinctly different mechanisms: (1) the Malthusian regime of stagnation, (2) the post-Malthusian regime, and (3) the sustained-growth regime.

According to Galor (2005, 2008, 2011, 2012), Malthusian regime of stagnation was between 100,000 BC and AD 1750 for developed regions and between 100,000 BC and AD 1900 for less-developed regions. The claimed starting time appears to be based entirely on conjecture because Maddison's data are terminated at AD 1 and even they contain significant gaps below AD 1500. The post-Malthusian regime was allegedly between AD 1750 and 1850 for developed regionsand and from 1900 for less-developed regions. The sustained-growth regime was supposed to have commenced around 1850 for developed regions.

Unified Growth Theory (Galor, 2005, 2008, 2011, 2012) can be tested in many ways but the easiest way to test it is to look for the dramatic takeoffs from stagnation to growth. These takeoffs are described as a "remarkable" or "stunning" escape from the Malthusian trap (Galor, 2005, pp. 177, 220). It is a signature, which cannot be missed.

This change in the pattern of growth is described as "the sudden take-off from stagnation to growth" (Galor, 2005, pp. 177, 220, 277) or as a "sudden spurt" (Galor, 2005, 177, 220). According to Galor, for developed regions, the end of the Malthusian regime of stagnation coincides with the Industrial Revolution. "The take-off of developed regions from the Malthusian Regime was associated with the Industrial Revolution" (Galor, 2005, p. 185). Indeed, the Industrial Revolution is considered to have been "the prime engine of economic growth" (Galor, 2005, p. 212).

This signature is characterised by three features: (1) it should be a prominent change in the pattern of growth, (2) it should be a transition from stagnation to growth and (3) it should occur at the time predicted by the theory. For developed regions, the postulated takeoffs should occur around AD 1750, or around the time of the Industrial Revolution, 1760-1840 (Floud & McCloskey, 1994). For less-developed regions, they should occur around 1900. The added advantage of using this simple test is that there are no significant gaps in the data around the time of the postulated takeoffs and consequently the stagnation and the expected prominent transitions from stagnation to growth should be easily identifiable.

A transition from growth to growth is not a signature of the postulated takeoff from stagnation to growth. Thus, a transition is from hyperbolic growth to another hyperbolic growth or to some other steadily-increasing trajectory is not a signature of the sudden takeoff from stagnation to growth. Likewise, a transition at a distinctly different time is not a confirmation of the theoretical expectations.

## 2. World economic growth

Results of mathematical analysis of the world economic growth are presented in Figures 1-3. Reciprocal values of historical data can be fitted using a straight line (representing hyperbolic growth) between AD 1000 and 1955. From around 1955, the world economic growth started to be diverted to a slower trajectory as indicated by the *upward* bending of the reciprocal values. This section is magnified in Figure 2. Global economic growth is now approximately exponential (Nielsen, 2014, 2015a).

Hyperbolic fit to the world GDP data (Maddison, 2010) is shown in Figure 3. The fit is remarkably good. The point at AD 1 is 77% away from the fitted curve. We would need more data between AD 1 and 1000 to decide whether such a difference is of any significance but it could reflect a pattern similar to the pattern observed for the growth of human population (Nielsen, 2016). Hyperbolic economic growth of the historical GDP has been uniquely identified by the straight-line fitting the reciprocal values of data.

Parameters describing hyperbolic trajectory fitting the data between AD 1000 and 1955 are: $a = 1.684 \times 10^{-2}$ and $k = 8.539 \times 10^{-6}$. Its singularity is at $t = 1972$. However, from around 1955, the world economic growth started to be diverted to a slower trajectory bypassing the singularity by 17 years (see Table 1).

The search for a takeoff in the world economic growth produced negative results. The data reveal a different pattern of growth than claimed by the Unified Growth Theory (Galor, 2005, 2008, 2011, 2012). The theory claims a long period of stagnation followed by a sudden takeoff. The data show a stable hyperbolic growth followed by a diversion to a slower trajectory.

The data also demonstrate that the Industrial Revolution had no impact on changing the economic growth trajectory. These results might not be surprising because the world economic growth is represented by the economic growth in developed and less-developed regions. However, even then, it would be hard to expect that the data would follow such a remarkably stable and specific trajectory. We would expect some distortions reflecting takeoffs around the time of the Industrial Revolution for developed regions and takeoffs around 1900 for less-developed regions. We see no signs of such distortions and no signs of the presence of such takeoffs.

The straight-line representing the reciprocal values of the GDP data shown in Figure 1 follows the data closely until 1955. There was no boosting in the economic growth, no unusual acceleration at *any time* between AD 1000 and 1955. The world economic growth was increasing monotonically before and after the Industrial Revolution as shown by either a steadily increasing hyperbolic distribution in Figure 3 or by the steadily-decreasing straight line (representing hyperbolic distribution) shown in Figure 1. Which point on a straight line should be selected to mark a boundary between different patterns of growth? How can we claim different patterns of growth on a straight line if the straight line shows clearly only one pattern? There was no takeoff in the world economic growth at any time, let alone around the time of the Industrial Revolution or around 1900.

Economic growth may have been slow over a long time but it was not stagnant. The growth was hyperbolic, and the characteristic feature of hyperbolic growth is a slow growth over a long time and a fast growth over a short time. Hyperbolic growth increases monotonically and it is *impossible* to locate a place marking a transition from a slow to fast growth because *such a transitions does not exist*.

Hyperbolic growth of the world economy is in harmony with the hyperbolic growth of the world population (Nielsen, 2016; von Foerster, Mora & Amiot, 1960). In both cases, the growth was indeed slow over a long time and fast over a short time. In both cases the growth creates an illusion of stagnation followed by a sudden takeoff. However, in both cases the growth was hyperbolic. There was no stagnation and no sudden takeoff. Furthermore, in both cases the growth started to be diverted, relatively recently, to slower trajectories.

## 3. Western Europe

The growth of the GDP in Western Europe is shown in Figures 4-6. Western Europe is represented by the total of 30 countries: Austria, Belgium, Denmark, Finland, France, Germany, Italy, the Netherlands, Norway, Sweden, Switzerland, the United Kingdom, Greece, Portugal, Spain and by 14 small, but unspecified countries. Ireland is missing in this list because it was included only from 1921.

The best hyperbolic fit to the data is between AD 1500 and 1900. Parameters for this distribution are $a = 9.859 \times 10^{-2}$ and $k = 5.112 \times 10^{-5}$. The point of singularity is at $t = 1929$. Between 1900 and 1910, economic growth started to be diverted to a slower, but still fast-increasing, trajectory bypassing the singularity by 29 years (see Table 1).

The most complete set of data for Western Europe is for Denmark, France, the Netherlands and Sweden. They are analysed separately and results are presented in Figures 7 and 8. According to Maddison (2010), these four countries accounted for 34% of the total GDP of the 30 countries of Western Europe in 2008.

Parameters describing the historical hyperbolic growth of the GDP in these four countries are: $a = 3.821 \times 10^{-1}$ and $k = 1.986 \times 10^{-4}$. The point of singularity is at $t = 1923$. From around 1875 economic growth in Denmark, France, the Netherlands and Sweden was diverted to a slower trajectory, bypassing the singularity by 48 years.

The quality of the hyperbolic fit to the data is virtually the same as for the total of the 30 countries but now the fitted curve passes also through the AD 1 point. However, it still does not reproduce the point at AD 1000. This point is only 41% below the fitted hyperbolic distribution.

The historical growth of the GDP in Western Europe was definitely hyperbolic from AD 1500 to 1900 but there is also a good indication that it might have been hyperbolic from AD 1 (see Figures 7 and 8). Even if we make allowance for this uncertainty, the search for a sudden takeoff around the expected time, i.e. around the time of the Industrial Revolution, produced negative results for the 30 countries of Western Europe and for the four (Denmark, France, the Netherlands and Sweden) characterised by the most complete sets of data.

The claim of a stunning or remarkable takeoff is contradicted by data. There was no takeoff of any kind and at any time, stunning or less stunning, remarkable or less remarkable, sudden or gradual – none at all. The Industrial Revolution, the alleged "prime engine of economic growth" (Galor, 2005a, p. 212), made no impression on changing the economic growth trajectory in regions where this engine should have been working most efficiently. Industrial Revolution brought many other important changes but, surprisingly perhaps, did not change the economic growth trajectory in the countries closest to this monumental development.

## 4. Eastern Europe

Systematic data for Eastern Europe are available only for seven countries: Albania, Bulgaria, Czechoslovakia, Hungry, Poland, Rumania and Yugoslavia. For other countries there are no data until 1990. The analysis of the historical data for Eastern Europe is summarised in Figures 9-11.

The best hyperbolic fit to the data is between AD 1000 and 1890. Hyperbolic parameters are: $a = 7.749 \times 10^{-1}$ and $k = 4.048 \times 10^{-4}$. The point of singularity is at $t = 1915$. From around 1890, economic growth in Eastern Europe was diverted to a slower trajectory, bypassing the singularity by 25 years.

There was no stagnation and no takeoff at any time. Industrial Revolution had no impact on changing the economic growth trajectory in the countries of Eastern Europe.

## 5. Former USSR

The analysis of the data for the countries of the former USSR is presented in Figures 12-14. The hyperbolic fit to the data is between AD 1 and 1870. Parameters fitting the data are: $a = 6.547 \times 10^{-1}$ and $k = 3.452 \times 10^{-4}$. The point of singularity is at $t = 1897$. From around 1870, or maybe even a little earlier (shortly after the Industrial Revolution) economic growth in the Former USSR was diverted to a slower trajectory, bypassing the singularity by at least 27 years.

There was no stagnation and no takeoff *at any time*. Industrial Revolution had no impact on changing the economic growth trajectory in the countries of former USSR.

## 6. Asia

Analysis of the historical economic growth in Asia (including Japan) is summarised in Figures 15-17. The best hyperbolic fit is between AD 1000 and 1950. Parameters fitting the data are: $a = 2.303 \times 10^{-2}$ and $k = 1.129 \times 10^{-5}$. The point of singularity is at $t = 2040$.

Asia is made primarily of less-developed countries (BBC, 2014, Pereira, 2011) and consequently, according to the Unified Growth Theory (Galor, 2005, 2008, 2011, 2012), economic growth in this region should have been characterised by stagnation until around 1900, the year marking the alleged stunning escape from the Malthusian trap, the escape, which was supposed to have been manifested by the postulated dramatic takeoff. (Until AD 1900, Japan's conrtibution to the total economy in Asia was on average only 5%.) The data and their analysis show that there was no stagnation, at least from AD 1000 and no expected takeoff. The data reveal a steadily increasing hyperbolic growth until around 1950. From around that year economic growth *was* diverted to a faster trajectory. This boosting can be seen clearly in Figures 16 and 17 and it occurred close to the time of the postulated takeoff from stagnation to growth. However, it was *not* a transition from stagnation to growth but from hyperbolic growth to a slightly faster trajectory of a different kind. It is, therefore, not the takeoff postulated in the Unified Growth Theory. Furthermore, it was only a temporary boosting, which is now returning to the original hyperbolic trajectory and, as indicated by the reciprocal values of the data, this new growth is likely to be slower than the original trajectory. Thus, it is a boosting of a completely different kind. It would be interesting to explain it but we cannot be helped by Unified Growth Theory because it discusses mechanisms, which are repeatedly contradicted by data. This transition is not even recognised in this theory

Reciprocal values of data presented in Figure 16 show that the economic growth became temporarily *slower* at the time overlapping the time of the Industrial Revolution, 1760-1840 (Floud & McCloskey, 1994), because while the point in 1820 is still located on the straight line, representing hyperbolic growth, the point in 1870 is above this line. The deceleration in the economic growth occurred sometime between 1820 and 1870.

This brief deceleration was followed by a transient growth between 1870 and 1940, which appears to have been also hyperbolic but a little faster than the earlier hyperbolic growth. This transition occurred earlier than the postulated takeoff around 1900 and it was not a transition from stagnation to growth but a transition from hyperbolic growth to hyperbolic growth. Furthermore, it was also a minor transition, which could be hardly noticed in the direct display of data shown in Figure 17. In summary, therefore, the examination of data for the economic growth in Asia demonstrates that the postulated takeoff (Galor, 2005, 2008, 2011, 2012) never happened. There was no stagnation and no sudden dramatic escape to a new and rapid growth.

## 7. Africa

Results of the analysis of the economic growth in the 57 African countries are presented in Figures 18-20. Reciprocal values of the GDP data, presented in Figures 18 and 19, show clearly that the economic growth was following *two* hyperbolic distributions. At first it was a slow hyperbolic growth between AD 1 and 1820 characterised by parameters $a = 1.244 \times 10^{-1}$ and $k = 5.030 \times 10^{-5}$ and by the singularity at $t = 2473$. Then, around 1820, this slow hyperbolic growth was replaced by a significantly faster hyperbolic growth characterised by parameters $a = 4.192 \times 10^{-1}$ and $k = 2.126 \times 10^{-4}$ and by the singularity at $t = 1972$. Defined by the parameter *k*, this new growth was 4.2 times faster than the earlier hyperbolic growth. From around 1950, this fast hyperbolic growth was diverted to a slower, non-hyperbolic trajectory, bypassing singularity by 22 years.

Africa is also made of less-developed countries (BBC, 2014; Pereira, 2011) so according to the Unified Growth Theory (Galor, 2005, 2008, 2011, 2012) it should have experienced stagnation in the economic growth until around 1900 followed by a clear takeoff around that year. These expectations are contradicted by the economic growth data because (1) economic growth was not stagnant but hyperbolic until 1950, (2) there was no takeoff from stagnation to growth around 1900 or around any other time and (3) shortly after the expected time of the takeoff, economic growth in Africa started to be diverted to a slower trajectory.

Acceleration in the economic growth in Africa occurred around 1820, but it was not a transition from stagnation to growth but *from growth to growth*. Even more specifically, it was a transition from the hyperbolic growth to another hyperbolic growth. It was also acceleration at a wrong time, not around 1900 but around the time of the Industrial Revolution. This acceleration can be explained by noticing that it appears to coincide with the intensified colonisation of Africa (Duignan & Gunn, 1973; McKay, et al. 2012; Pakenham, 1992). The fast-increasing GDP after 1820 was not reflecting the rapidly improving living conditions of African population brought about by the beneficial changes caused by the Industrial Revolution but the rapidly increasing wealth of new settlers and their countries of origin at the expense of the deploring living conditions of native populations.

The search for the takeoff from stagnation to growth, claimed by the Unified Growth Theory (Galor, 2005, 2008, 2011, 2012), produced negative results. The data show also that there was no stagnation in the economic growth over the entire range of time, from AD 1 to the present time.

## 8. Latin America

Results of the analysis of the economic growth in Latin America are presented in Figures 21 - 23. Data for Latin America are difficult to analyse because there was a significant decline in the economic growth between AD 1500 and 1600 but they also appear to follow two distinctly different hyperbolic trajectories. However, the identification of the first trajectory is not as clear as for Africa. The identification of the second hyperbolic trajectory is more convincing. Our tentative conclusion is that the economic growth in Latin America was following a slow hyperbolic distribution between AD 1 and 1500 and a fast hyperbolic distribution between AD 1600 and around 1870.

The tentatively assigned slow hyperbolic growth between AD 1 and 1500 is characterised by parameters $a = 4.421 \times 10^{-1}$ and $k = 2.093 \times 10^{-4}$. Its singularity is at $t = 2113$. The better determined fast hyperbolic growth between AD 1600 and 1870 is characterised by parameters $a = 1.570 \times 10^{0}$ and $k = 8.224 \times 10^{-4}$. Its singularity is at $t = 1910$. Defined by the parameter *k*, this growth was 3.9 times faster than the earlier hyperbolic growth. From around 1870, this fast hyperbolic growth started to be diverted to a slower trajectory bypassing the singularity by 40 years. The transition from the earlier apparent hyperbolic growth to a new and rapid hyperbolic growth, which occurred between around AD 1500 and 1600 appears to coincide with commencement of the Spanish conquest (Teeple, 2002).

Latin America is also made of less-developed countries (BBC, 2014; Pereira, 2011) so again, according to the Unified Growth Theory (Galor, 2005, 2008, 2011, 2012), the economic growth in this regions should have been stagnant until around 1900 and fast-increasing from around that year. This pattern of growth is not confirmed by data. The data show a diametrically different pattern: (1) there is no convincing evidence of the existence of stagnation over the entire range of time between AD 1 and 1870 but there is a sufficiently convincing indication of the hyperbolic growth particularly between AD 1600 and 1870, (2) there was no takeoff from stagnation to growth at any time, and (3) around the time of the postulated takeoff in 1900 there was a diversion to a slower trajectory in 1870.

Even if the identification of the hyperbolic growth between AD 1 and 1500 is questioned, the overall pattern of growth in Latin America is similar to the pattern in Africa: a slow hyperbolic growth is followed by a fast hyperbolic growth. However, in any case, there is no convincing evidence that the growth was ever stagnant. On the contrary, there is sufficiently convincing evidence that the growth was never stagnant. It was clearly not stagnant between AD 1600 and 1870.

There was also no takeoff, dramatic or modest, from stagnation to growth around the expected time of 1900, *first* because the growth before that year was not stagnant but hyperbolic and *second* because around the time of the expected remarkable takeoff the economic growth started to be diverted to a slower trajectory. The search for the postulated takeoff produced negative results.

### 9. Summary and conclusions

Results of mathematical analysis of the historical economic growth are presented in Table 1. The listed parameters *a* and *k* are for the fitted hyperbolic distributions. The last column shows the results of the search for the takeoffs from stagnation to growth claimed by the Unified Growth Theory (Galor, 2005, 2008, 2011, 2012).

**Table 1.** *Summary of the mathematical analysis or the historical economic growth*

| Region/Countries | $a$ | $k$ | Hyperbolic Range | Singularity | Proximity | Takeoff |
|---|---|---|---|---|---|---|
| World | $1.684 \times 10^{-2}$ | $8.539 \times 10^{-6}$ | 1000 – 1955 | 1972 | 17 | X |
| Western Europe | $9.859 \times 10^{-2}$ | $5.112 \times 10^{-5}$ | 1500 – 1900 | 1929 | 29 | X |
| Western Europe (4) | $3.821 \times 10^{-1}$ | $1.986 \times 10^{-4}$ | 1 – 1875 | 1923 | 48 | X |
| Eastern Europe | $7.749 \times 10^{-1}$ | $4.048 \times 10^{-4}$ | 1000 – 1890 | 1915 | 25 | X |
| Former USSR | $6.547 \times 10^{-1}$ | $3.452 \times 10^{-4}$ | 1 – 1870 | 1897 | 27 | X |
| Asia | $2.303 \times 10^{-2}$ | $1.129 \times 10^{-5}$ | 1000 – 1950 | 2040 | 90 | X |
| Africa | $1.244 \times 10^{-1}$ | $5.030 \times 10^{-5}$ | 1 – 1820 | 2473 | | |
|  | $4.192 \times 10^{-1}$ | $2.126 \times 10^{-4}$ | 1820 – 1950 | 1972 | 22 | X |
| Latin America | $4.421 \times 10^{-1}$ | $2.093 \times 10^{-4}$ | 1 – 1500 | 2113 | | |
|  | $1.570 \times 10^{0}$ | $8.224 \times 10^{-4}$ | 1600 – 1870 | 1910 | 40 | X |

**Notes:** *a* and *k* – Hyperbolic growth parameters [see eqn (1)]. *Hyperbolic Range* - The empirically-confirmed range of time when the economic growth can be described using hyperbolic distributions. *Singularity* - The time of the escape to infinity for a given hyperbolic distribution. *Proximity* - Proximity (in years) of the singularity at the time when the economic growth departed from the hyperbolic growth to a new trajectory. *Western Europe (4)* - Four countries of Western Europe: Denmark, France, the Netherlands and Sweden. *X* - No takeoff. The takeoff from stagnation to growth claimed by the Unified Growth Theory (Galor, 2005, 2008, 2011, 2012) never happened.

This analysis demonstrates that the natural tendency for the historical economic growth was to increase hyperbolically. In general, there is a remarkably good agreement between the data and the calculated hyperbolic distributions.

Unlike the more familiar exponential distributions, which are easier to understand because they show more readily a gradually increasing growth, hyperbolic distributions appear to be made of two or maybe even three components: a slow component, a fast component and perhaps even a transition component located between the apparent slow and fast components. This illusion is so strong that even the most experienced researchers can be deceived particularly if they have no access to good sets of data, which was in the past. Now, however, excellent data are available (Maddison, 2001, 2010) and we can use them not only to check the earlier interpretations of economic growth but also to expand the scope of the economic research.

The postulate of the existence of the epoch of Malthusian stagnation is suggested by a slow economic growth over a long time but this slow growth is just a part of the hyperbolic growth, which is convincingly identified using reciprocal values. Hyperbolic distributions create also the illusion of a sudden takeoff but this feature is also a part of the hyperbolic growth. Hyperbolic growth *is* slow over a long time and fast over a short time but the slow and fast growth are the integral features of the same monotonically increasing distribution, which is easier to understand by using the reciprocal values of the growing entity (Nielsen, 2014). In such displays, the illusion of distinctly different components disappears because hyperbolic growth is then represented by a decreasing straight line, which is easy to understand. It then becomes obvious that hyperbolic distribution cannot be divided into distinctly different sections governed by different mechanism because it makes no sense to divide a straight line into arbitrarily chosen sections and claim different mechanism to such arbitrarily-selected sections. It is also then clear that it is *impossible* to pinpoint the transition from a slow to a fast growth. Which point on a straight line should we select to identify such a transition? The transition does not happen at any specific time but gradually over the whole range of time.

Our search for the postulated takeoffs from stagnation to growth (Galor, 2005, 2008, 2011, 2012) produced negative results: *there were no takeoffs*. Galor's elaborate discussion revolving around his postulated three regimes of growth and the postulated takeoffs from stagnation to growth are irrelevant because there were no takeoffs in the growth of the GDP and in the growth of income per capita (GDP/cap) (Nielsen, 2015b). In science, just one contradicting evidence in data is sufficient to show that a theory advocating the contradicted postulate or postulates has to be either rejected or revised to bring it in the agreement with empirical evidence. In the case of the Unified Growth Theory (Galor, 2005, 2008, 2011, 2012), the postulated takeoffs from stagnation to growth are contradicted repeatedly by the economic growth in Western Europe, Eastern Europe, former USSR, Asia, Africa, and Latin America as well as by the world economic growth.

The data and their analysis suggest new lines of research of economic growth. They suggest that our attention should not be directed towards explaining the mechanism of stagnation and of the sudden takeoffs from stagnation to growth because these features are contradicted by data. What needs to be explained is why the historical economic growth was hyperbolic and why relatively recently it was diverted to a slower trajectory. Maddison published excellent data describing not only economic growth but also the growth of human population and these data can be used effectively in trying to explain the historical economic growth.

## Appendix

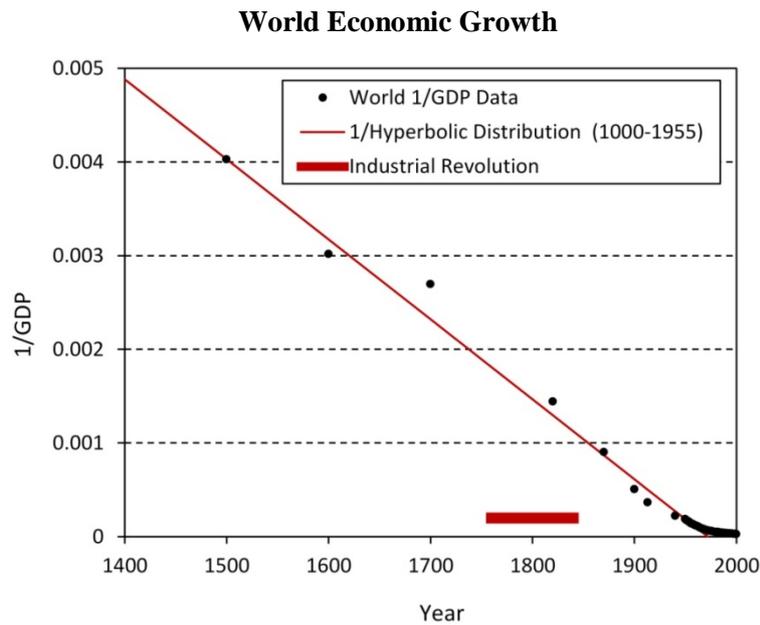

**Figure 1**. *Reciprocal values of the GDP data (Maddison, 2010) are fitted using straight line between AD 1000 and 1955 representing hyperbolic growth. There was no stagnation and no takeoff from stagnation to growth, claimed by the Unified Growth Theory (Galor, 2005, 2008, 2011, 2012). Industrial Revolution had no impact on changing the economic growth trajectory. From around 1955, the economic growth started to be diverted to a slower trajectory.*

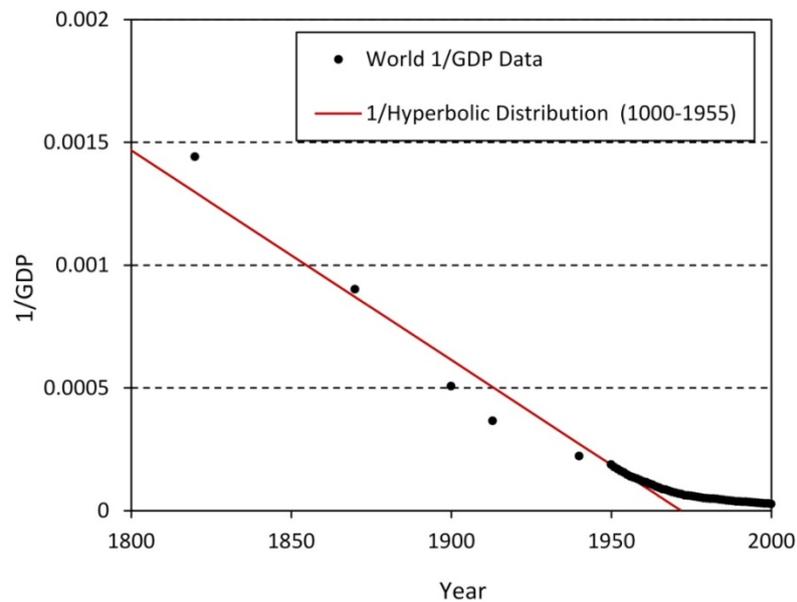

**Figure 2.** *Reciprocal values of the GDP data (Maddison, 2010) showing the diversion of the economic growth to a slower trajectory from around 1955, as indicated by the upward bending. The current global economic growth is approximately exponential (Nielsen, 2014, 2015a).*

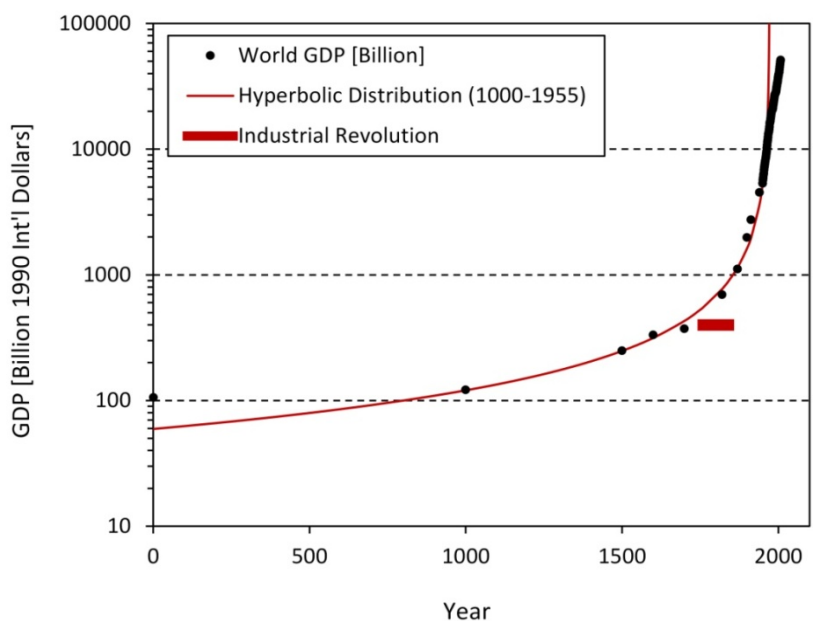

**Figure 3.** *World GDP data (Maddison, 2010) fitted using hyperbolic distribution. The point at AD 1 is 77% higher than the calculated distribution. There was no stagnation and no takeoff from stagnation to growth. Both features were incorrectly claimed by the Unified Growth Theory (Galor, 2005, 2008, 2011, 2012). Industrial Revolution had no impact on changing the economic growth trajectory. From around 1955, the world economic growth started to be diverted to a slower but still fast-increasing trajectory, which is now approximately exponential (Nielsen, 2014, 2015a).*

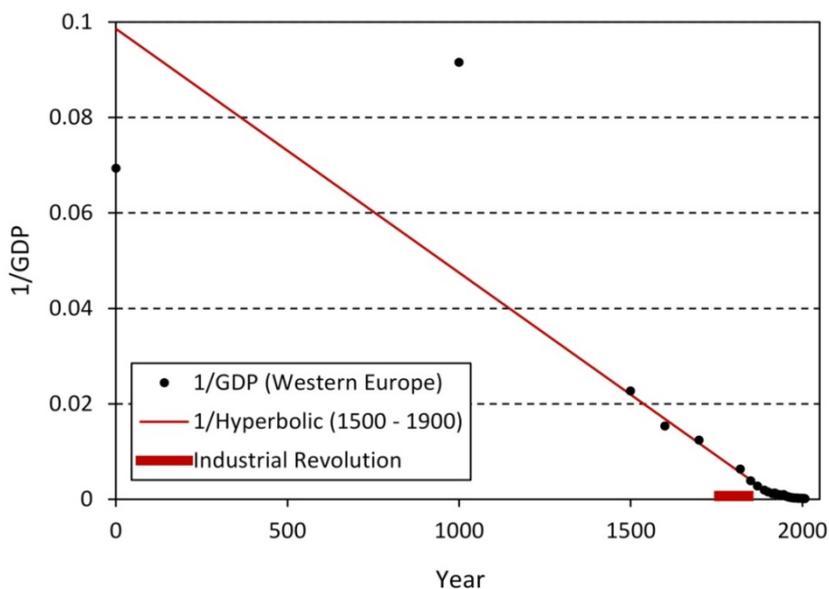

**Figure 4.** *Reciprocal values of the GDP data (Maddison, 2010) for Western Europe are compared with the hyperbolic distribution represented by the decreasing straight line. The growth was hyperbolic from at least AD 1500 to 1900. There was no takeoff from stagnation to growth. Industrial Revolution had no impact on changing the economic growth trajectory in Western Europe, the centre of this revolution. On the contrary, from around 1900, shortly after the Industrial Revolution, the economic growth in Western Europe started to be diverted to a slower trajectory as indicated by the upward bending of the trajectory representing the reciprocal values of data.*

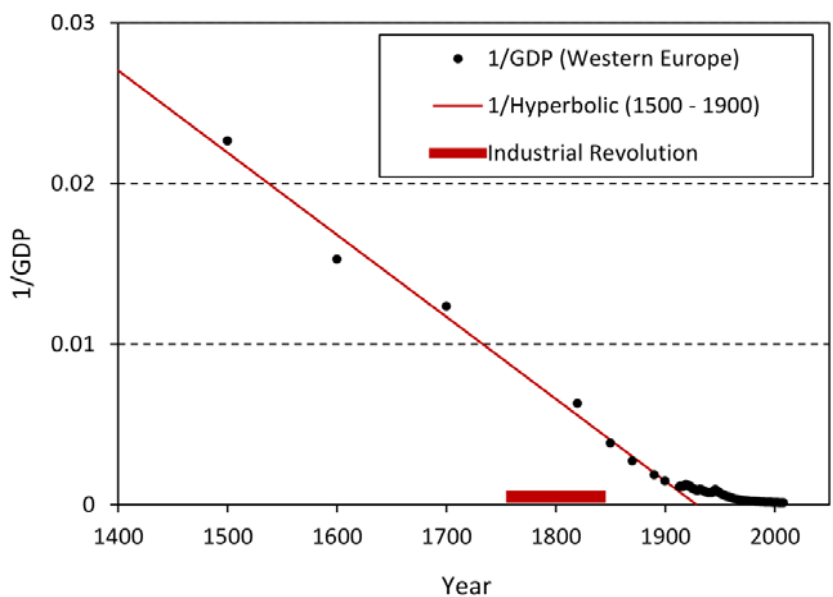

**Figure 5.** *Reciprocal values of the GDP data (Maddison, 2010) for Western Europe between AD 1500 and 2008 showing a diversion to a slower trajectory from around 1900. There was no takeoff from stagnation to growth, claimed incorrectly by the Unified Growth Theory (Galor, 2005, 2008, 2011, 2012). Industrial Revolution had absolutely no impact on changing the economic growth trajectory in Western Europe, the centre of this revolution.*

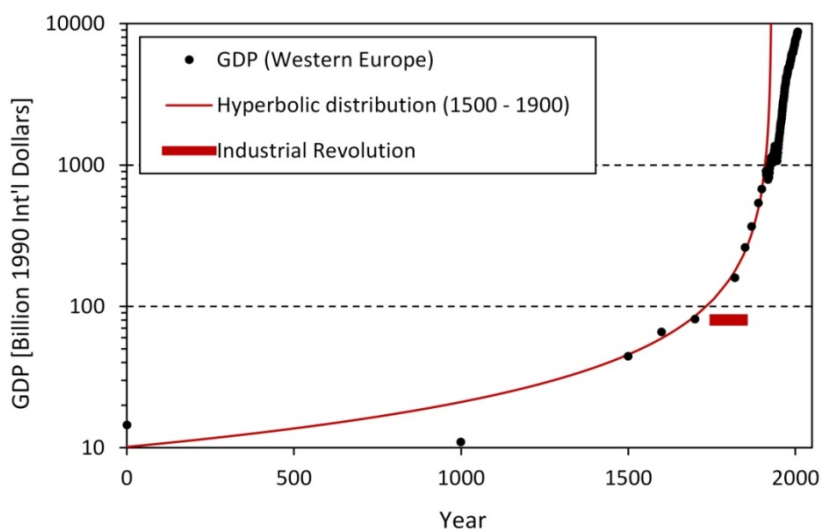

**Figure 6.** *Economic growth in Western Europe. The GDP data (Maddison, 2010) are compared with hyperbolic distribution. The growth was hyperbolic from at least AD 1500 to around 1900. The point at AD 1 is 42% higher than for the calculated distribution and 48% lower at AD 1000. There was no takeoff from stagnation to growth, claimed incorrectly by the Unified Growth Theory (Galor, 2005, 2008, 2011, 2012). Industrial Revolution had no impact on changing the economic growth trajectory in Western Europe, the centre of this revolution. From around 1900, economic growth in Western Europe started to be diverted to a slower trajectory.*

**Denmark, France, Netherlands and Sweden**

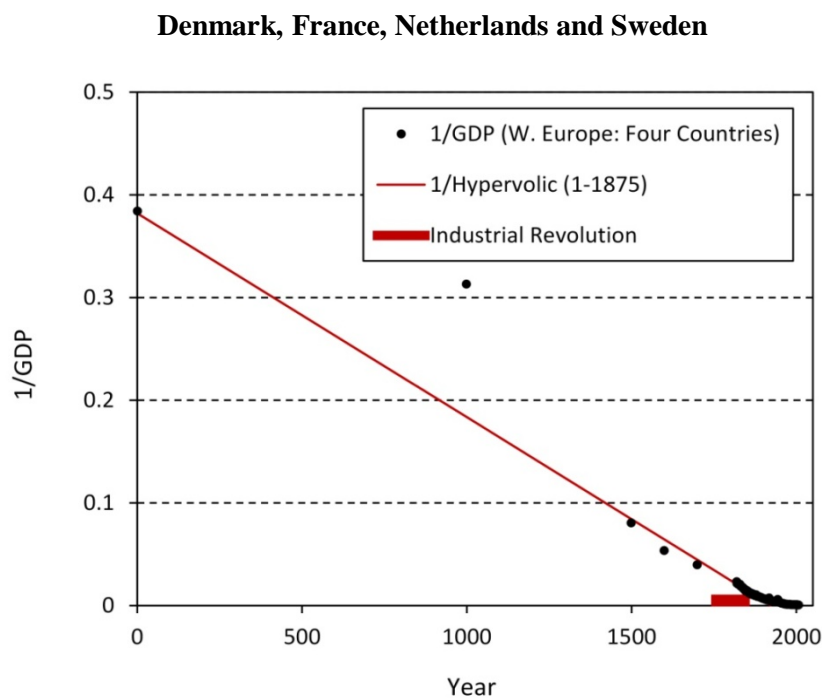

**Figure 7.** *Reciprocal values of the GDP data (Maddison, 2010) describing economic growth in four countries of Western Europe (Denmark, France, the Netherlands and Sweden) compared with the straight line representing hyperbolic growth fitting the data between AD 1 and 1875. From around 1875, or shortly after the Industrial Revolution, economic growth in these four countries started to be diverted to a slower trajectory. Industrial Revolution did not boost economic growth. There was no takeoff from stagnation to growth.*

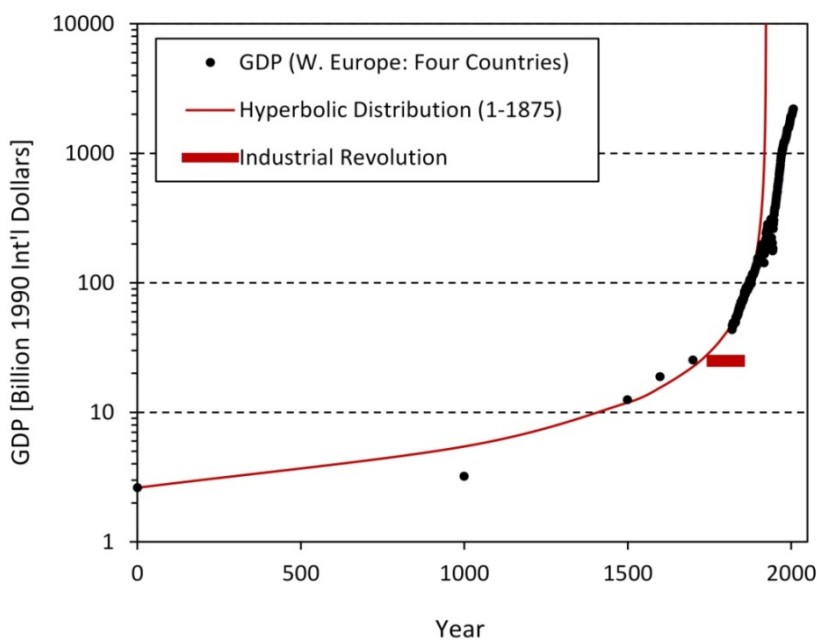

**Figure 8.** *Economic growth in Denmark, France, the Netherlands and Sweden. The data (Maddison, 2010) are compared with hyperbolic distribution. The point at AD 1000 is 41% lower than for the calculated distribution. From around 1875, the economic growth started to be diverted to a slower trajectory. There was no takeoff from stagnation to growth.*

**Eastern Europe**

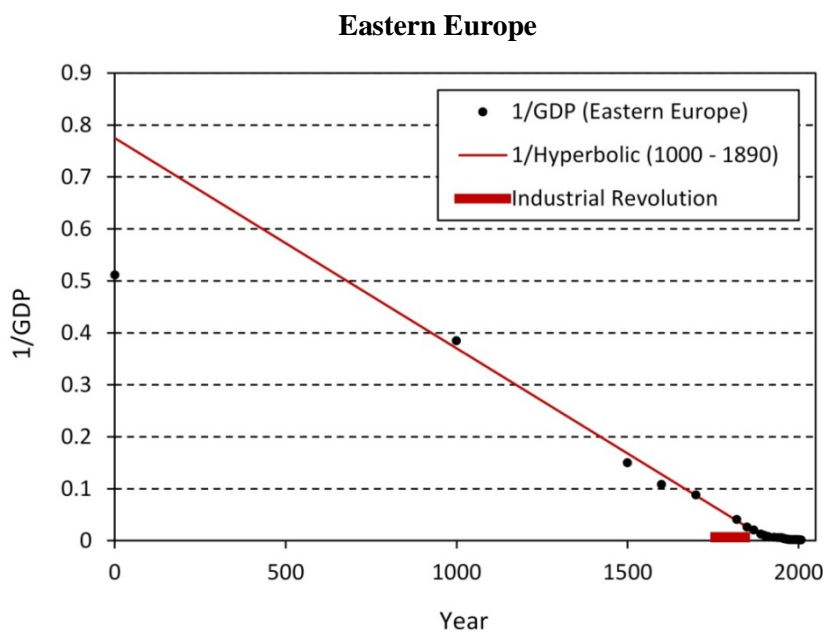

**Figure 9.** *Reciprocal values of the GDP data (Maddison, 2010) for Eastern Europe are compared with the hyperbolic distribution represented by the decreasing straight line. Economic growth was hyperbolic from at least AD 1000. The takeoff from stagnation to growth never happened because there was no stagnation. Industrial Revolution did not boost the economic growth in Eastern Europe.*

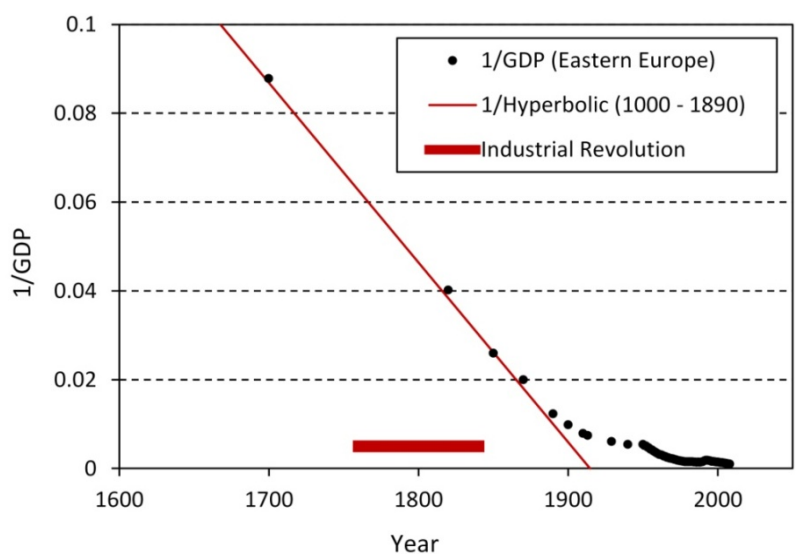

**Figure 10.** *Reciprocal values of the GDP data (Maddison, 2010) for Eastern Europe showing that from around 1890, shortly after the Industrial Revolution, the economic growth started to be diverted to a slower trajectory. There was no takeoff from stagnation to growth because there was no stagnation. Industrial Revolution did not boost the economic growth in Eastern Europe. Hyperbolic growth around that time remained undisturbed.*

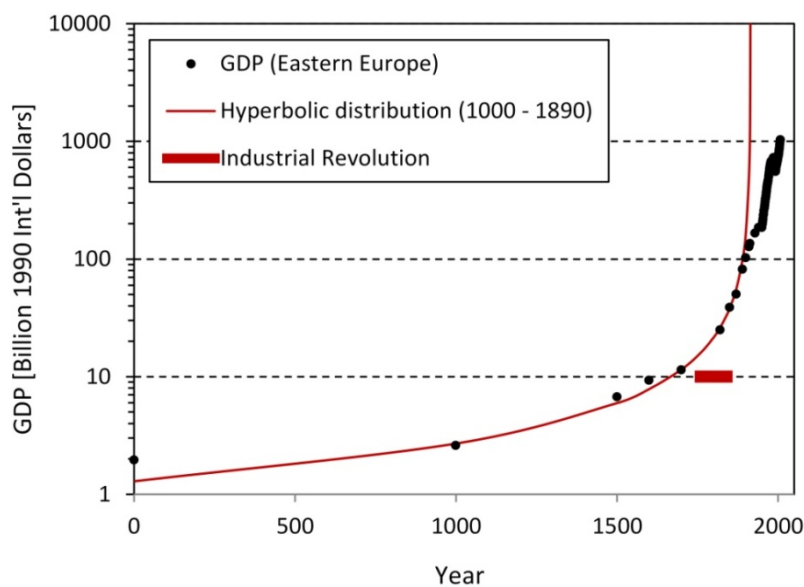

**Figure 11**. *Economic growth in Eastern Europe. GDP data (Maddison, 2010) are compared with the best hyperbolic fit. The point at AD 1 is 51% higher than for the calculated distribution. From around 1890, shortly after the Industrial Revolution, economic growth started to be diverted to a slower trajectory. Industrial Revolution did not boost the economic growth in Eastern Europe. Contrary to the Unified Growth Theory (Galor, 2005, 2008, 2011, 2012), there was no stagnation and no takeoff from stagnation to growth.*

**Former USSR**

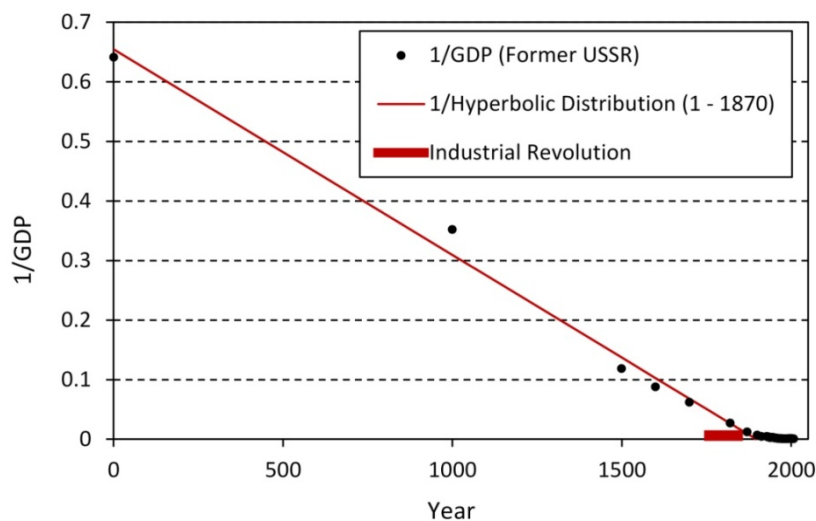

**Figure 12.** *Reciprocal values of the GDP data (Maddison, 2010) for the former USSR compared with the hyperbolic distribution represented by the decreasing straight line. Data indicate that the economic growth was hyperbolic from AD 1 to 1870. Industrial Revolution did not boost the economic growth. There was no stagnation and no takeoff from stagnation to growth. Shortly after the Industrial Revolution, the economic growth in Eastern Europe started to be diverted to a slower trajectory. Unified Growth Theory (Galor, 2005, 2008, 2011, 2012) is contradicted by the economic growth data.*

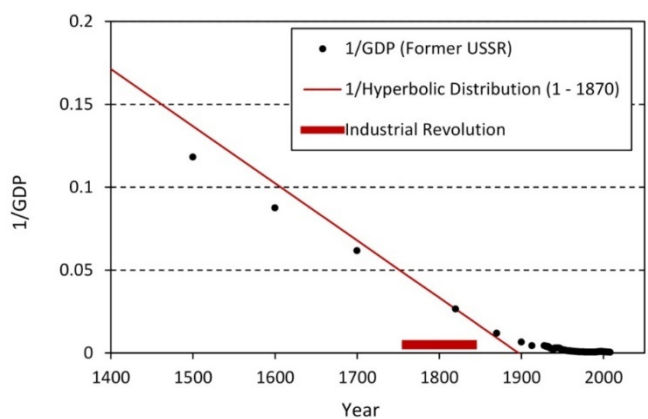

**Figure 13.** *Reciprocal values of the GDP data (Maddison, 2010) for the former USSR showing that from around 1870, shortly after the Industrial Revolution, economic growth started to be diverted to a slower trajectory.*

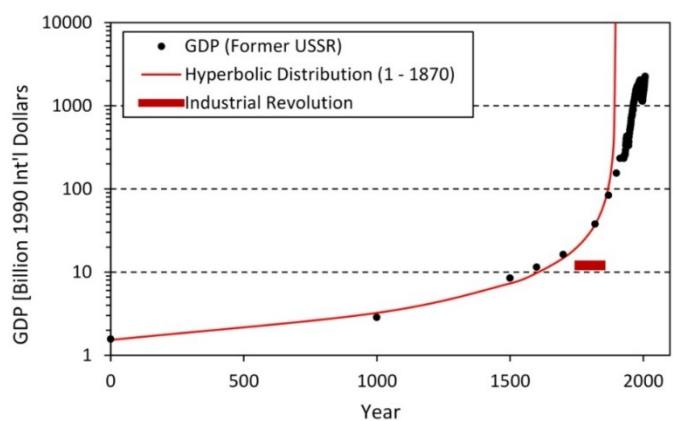

**Figure 14.** *Economic growth in the former USSR. GDP data (Maddison, 2010) are compared with the hyperbolic fit. The growth was hyperbolic from AD 1 to 1870. From around 1870, shortly after the Industrial Revolution, economic growth started to be diverted to a slower trajectory. Epoch of stagnation did not exist in the economic growth. Industrial Revolution did not boost the economic growth. There was no takeoff from stagnation to growth because there was no stagnation but a steadily-increasing growth. Unified Growth Theory (Galor, 2005, 2008, 2011, 2012) is contradicted by the economic growth data.*

### Asia (including Japan)

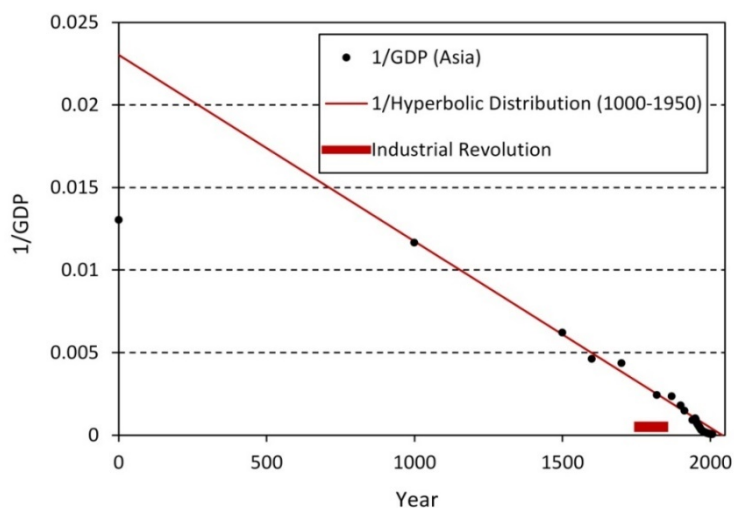

**Figure 15.** *Reciprocal values of the GDP data (Maddison, 2010) for Asia (including Japan) compared with the hyperbolic distribution represented by the decreasing straight line. Economic growth was hyperbolic from at least AD 1000. There was no expected transition from stagnation to growth.*

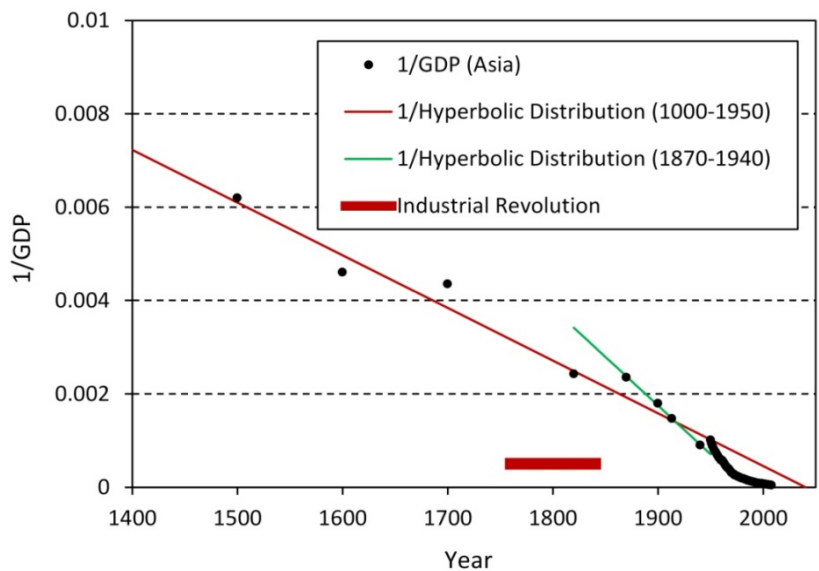

**Figure 16.** *Reciprocal values of the GDP data ([Maddison, 2010](#)) for Asia (including Japan). The data show a minor deceleration of growth towards the end of the time of the Industrial Revolution followed by a slightly faster hyperbolic growth between 1870 and 1940. The expected takeoff from stagnation to growth around 1900 ([Galor, 2005](#), [2008](#), [2011](#), [2012](#)) did not happen. The data show a small boosting around 1950 but it was not a transition from stagnation to growth. The search for the postulated takeoff ([Galor, 2005](#), [2008](#), [2011](#), [2012](#)) produced negative results.*

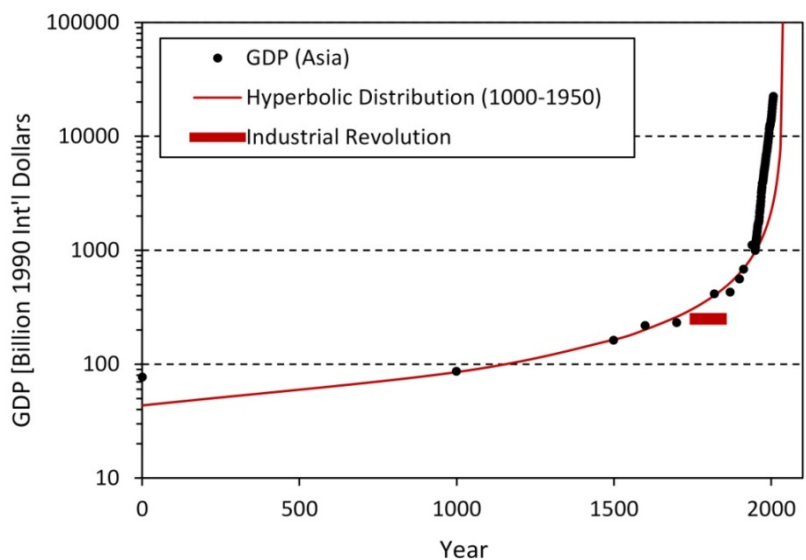

**Figure 17.** *Economic growth in Asia (including Japan). The data ([Maddison, 2010](#)) are compared with the hyperbolic distribution. The point at AD 1 is 76% higher than the calculated value. The data show a minor boosting around 1950 but it was not a transition from stagnation to growth but from the hyperbolic growth to a slightly faster trajectory, which is now coming closer to the earlier hyperbolic trajectory. The boosting was not only small but also it did not last long. The search for the postulated takeoff from stagnation to growth ([Galor, 2005](#), [2008](#), [2011](#), [2012](#)) produced negative results.*

**Africa**

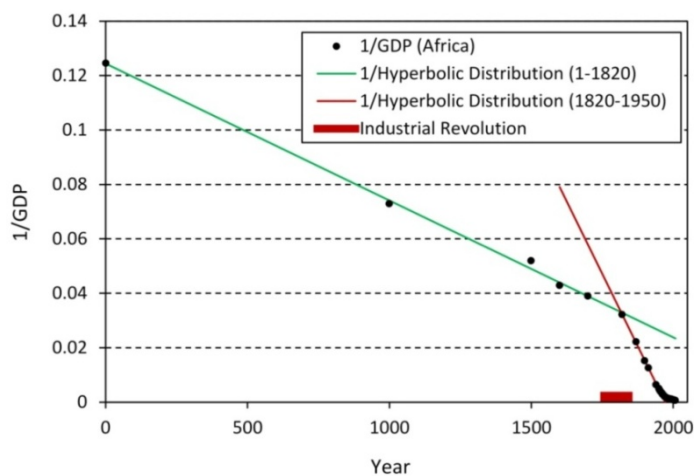

**Figure 18.** *Reciprocal values of the GDP data (Maddison, 2010) for Africa compared with hyperbolic distributions represented by the decreasing straight lines. There was no stagnation in the economic growth. Economic growth was increasing hyperbolically between AD 1 and around 1820 and again from 1820 to around 1950. The expected takeoff from stagnation to growth (Galor, 2005, 2008, 2011, 2012) never happened. The acceleration around 1820 was not a transition from stagnation to growth but from growth to growth. It also occurred earlier than expected (in 1820 rather than around 1900). Furthermore, close to the postulated takeoff in 1900, economic growth started to be diverted to a slower trajectory. The search for the takeoff from stagnation to growth around 1900 produced negative results. Unified Growth Theory (Galor, 2005, 2008, 2011, 2012) is contradicted by data.*

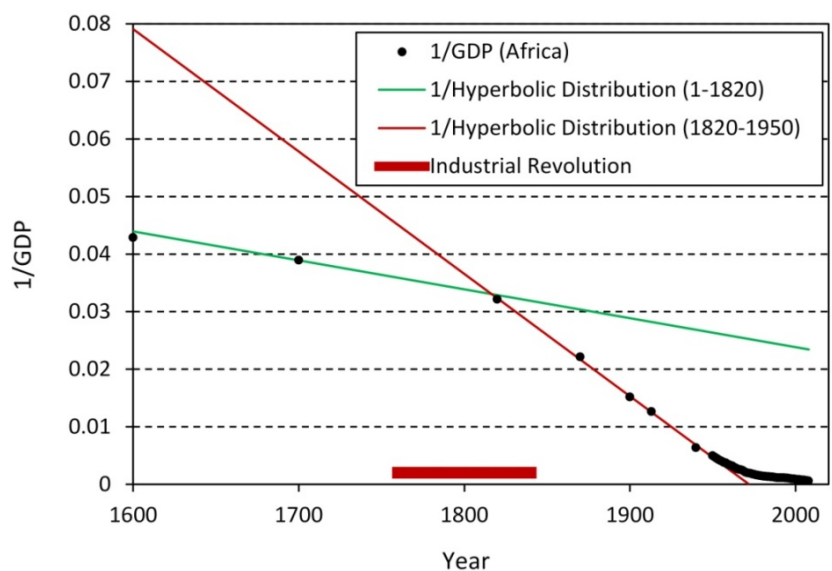

**Figure 19.** *Reciprocal values of the GDP data (Maddison, 2010) for Africa showing that from around 1950 economic growth started to be diverted to a slower trajectory. There was no takeoff around 1900, not even from growth to growth.*

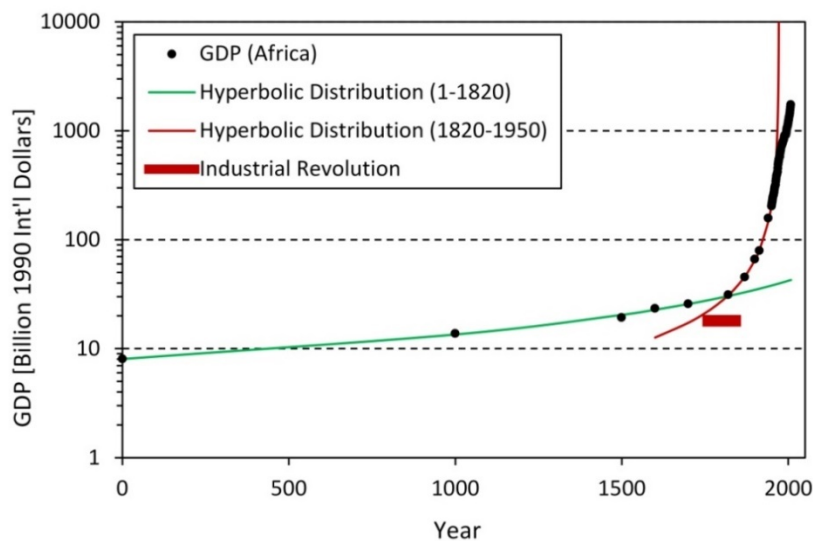

**Figure 20.** *Economic growth in Africa. Data (Maddison, 2010) are compared with hyperbolic distributions. The claimed takeoff from stagnation to growth (Galor, 2005, 2008, 2011, 2012) never happened because there was no stagnation. Furthermore, the transition from hyperbolic growth to hyperbolic growth occurred earlier (around 1820) than the postulated takeoff from stagnation to growth (around 1900). From around 1950, close to the claimed but non-existing takeoff from stagnation to growth, economic growth started to be diverted to a slower trajectory. Unified Growth Theory (Galor, 2005, 2008, 2011, 2012) is contradicted by data.*

**Latin America**

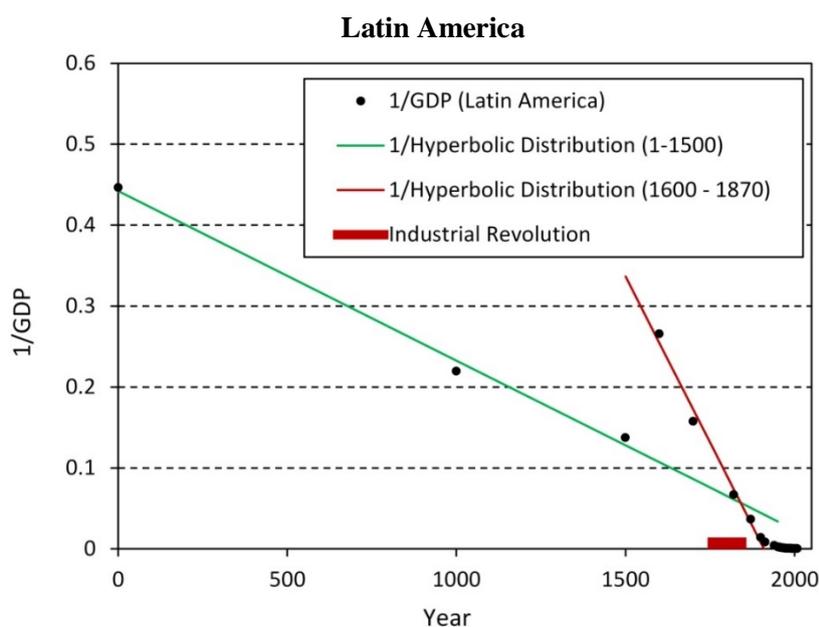

**Figure 21.** *Reciprocal values of the GDP data (Maddison, 2010) for Latin America are compared with hyperbolic distributions represented by the decreasing straight lines. The pattern of growth in Latin America is similar to the pattern of growth in Africa. The expected takeoff from stagnation to growth around 1900 (Galor, 2005, 2008, 2011, 2012) did not happen, because there was no stagnation and because, from around 1870, economic growth started to be diverted to a slower trajectory.*

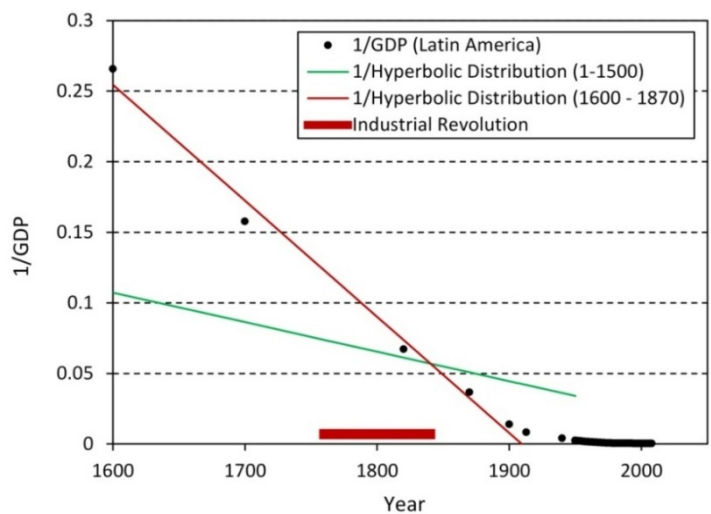

**Figure 22.** *Reciprocal values of the GDP data (Maddison, 2010) for Latin America showing that from around 1870, i.e. close to the time of the expected takeoff (around 1900) from stagnation to growth (Galor, 2005, 2008, 2011, 2012) economic growth started to be diverted to a slower trajectory. The data show also that the takeoff from stagnation to growth could not have happened because there was no stagnation.*

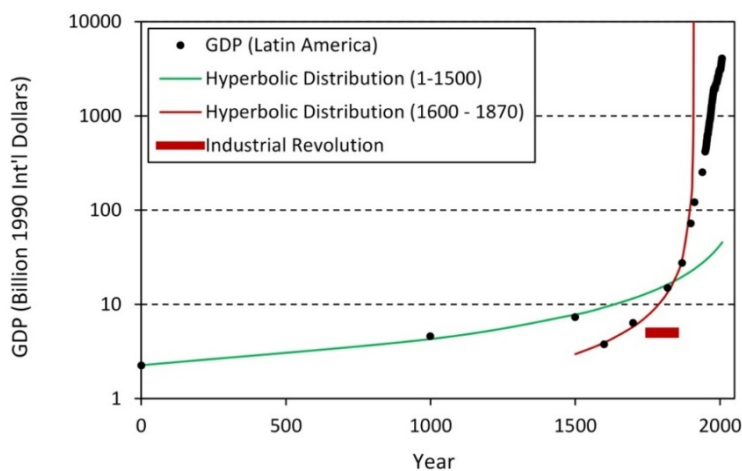

**Figure 23.** *Economic growth in Latin America. Economic growth data (Maddison, 2010) are compared with hyperbolic distributions. Unified Growth Theory (Galor, 2005, 2008, 2011, 2012) is contradicted by data. Economic growth was not stagnant before the postulated takeoff from stagnation to growth (around 1900) but hyperbolic. The growth was also stable and hyperbolic around the time of the Industrial Revolution in the Western world. The transition from stagnation to growth could not have happened because there was no stagnation. Furthermore, from around 1870, i.e. from around the time of the postulated takeoff, economic growth started to be diverted to a slower trajectory. The search for the takeoff from stagnation to growth produced negative results.*

**4**

# Growth of the World Population in the Past 12,000 Years and Its Link to the Economic Growth

## *By* Ron W. NIELSEN †

**Abstract.** Data describing the growth of the world population in the past 12,000 years are analysed. It is shown that population, if unchecked, does not increase exponentially, as expected by Malthus, but hyperbolically. This analysis reveals three episodes of hyperbolic growth: 10,000 - 500 BC, AD 500 - 1200 and AD 1400 -1950, representing the total of about 89% of the past 12,000 years. The remaining 11% were taken by three demographic transitions: 500 BC - AD 500, AD 1200 - 1400 and AD 1950 - present. The first two transitions were between sustained hyperbolic trajectories. The current transition is from a sustained hyperbolic growth to a yet unknown trajectory. The often claimed but never proven existence of Malthusian stagnation in the growth of human population is convincingly contradicted by the mathematical analysis of data. There was also never any form of a dramatic transition from stagnation to growth, described often as a takeoff, because there was no stagnation in the growth of the world population. The perceived takeoff is just the natural continuation of hyperbolic growth. Correct understanding of the historical growth of human population is essential in the correct interpretation of the historical growth of income per capita.
**Keywords.** Growth of human population, Economic growth, Growth of income per capita, Stagnation, Takeoffs, Hyperbolic growth, Demographic transitions.
**JEL.** A12, B22, B25, F01, N00, Y80.

## 1. Introduction

T he study of historical economic growth involves not only the study of the Gross Domestic Product (GDP) but also the study of the growth of population, because as pointed out by Galor (2005, 2011), it is important to understand the relationship between the these two process and particularly to understand income per capita (GDP/cap) distributions. It is for this reason that the latest and the most extensive compilations of historical GDP data, published by the world-renown economist, include also data describing the historical growth of human population (Maddison, 2001, 2010). Maddison lists not only the GDP data but also the calculated GDP/cap values.

About 50 years ago, von Foerster, Mora and Amiot (1960) demonstrated that human population was increasing hyperbolically during the AD era. We now have far better and more extensive sets of data compiled not only by Maddison (2001, 2010) but also by Manning (2008) and by the US Census Bureau (2016). The last two compilations are based on virtually the same primary sources but they are complimentary.

Maddison's compilation is useful in studying the growth of the population not only global but also regional and national. However, his data are terminated in AD 1. Furthermore, they also contain significant gaps below AD 1500. The data compiled by Manning and by the US Census Bureau are significantly richer but they are limited only to the description of the world population. However, they extend down to 10,000 BC.

Preliminary examination of Maddison's data indicates that economic growth and the growth of human population followed similar trajectories. Consequently, by using the rich set of data extending down to 10,000 BC we might gain better insight not only into the historical growth of human population but also of the economic growth.

† AKA Jan Nurzynski, Griffith University, Environmental Futures Research Institute, Gold Coast Campus, Qld, 4222, Australia.
☏ . +61407201175
✉ . r.nielsen@griffith.edu.au; ronwnielsen@gmail.com

## 2. The data

Procedures adopted in estimating historical populations are described by Durand (1977). The data for the AD era are of exceptionally good quality. Between AD 400 and 1850, independent estimates are within ±10% of their corresponding averaged values. The estimates after 1850 are within ±1.5%. The largest deviations of around ±30% are for the AD 1 data. The two estimates for AD 200 differ by ±15% from their average value. The BC data are less accurate and less consistent but when closely analysed they are also found to follow a well-defined trajectory.

## 3. Analysis of population data

In order to understand hyperbolic distributions, it is useful to compare them with the more familiar exponential distributions. The differential equation describing exponential growth is given by the following simple equation:

$$\frac{1}{S(t)}\frac{dS(t)}{dt} = k, \tag{1}$$

where $S(t)$ is the size of a growing entity, in our case the size of population, and $k$ is an arbitrary constant.

The left-hand side of this equation represents the growth rate. For $k > 0$ the eqn (1) describes growth, while for $k < 0$ it describes decay.

The solution of the eqn (1) is

$$S(t) = ae^{kt}, \tag{2}$$

where $a$ is a constant related to the constant of integration.

The eqn (2) gives

$$\ln S(t) = \ln a + kt. \tag{3}$$

The logarithm of the size of the growing entity increases linearly with time. Exponential growth can be easily identified by plotting data using semilogarithmic scales of reference because in such presentation exponential growth is represented by an increasing straight line.

Data for the growth of the population during the BC and AD eras (Manning, 2008; US Census Bureau, 2015) are shown in Figure 1. They are compared with the best fit obtained by using exponential function. The world population was not increasing exponentially.

Let us now examine hyperbolic growth. This type of growth is described by the following differential equation:

$$\frac{1}{S(t)}\frac{dS(t)}{dt} = kS(t), \tag{4}$$

where $k > 0$.

It is a slight modification of the eqn (1). Here, the growth rate is not constant but directly proportional to the size of the growing entity. The solution of this equation, which can be found by substitution $S(t) = Z^{-1}(t)$, is given by the following simple formula:

$$S(t) = \frac{1}{a - kt}. \tag{5}$$

It is just a reciprocal of a linearly-decreasing function. Consequently,

$$\frac{1}{S(t)} = a - kt \tag{6}$$

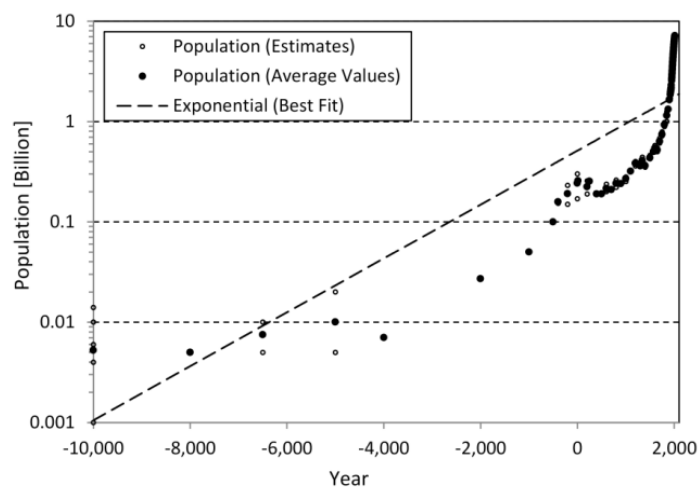

**Figure 1.** *Data describing the growth of the world population (Manning, 2008; US Census Bureau, 2016) are compared with the best fit obtained by using exponential function. The world population was not increasing exponentially. The BC time scale is identified by the negative numbers.*

Reciprocal values of the size of a growing entity follow a decreasing straight line. This representation simplifies the analysis of hyperbolic distributions. We can use this dependence to identify uniquely hyperbolic growth, in much the same way as the linearly increasing logarithm of a growing entity can be used to identify exponential growth.

It is useful to understand the difference between the exponential growth and the hyperbolic growth. For the exponential growth, the growth rate is constant. It does not matter how large is the size of the growing entity, the growth rate never changes. For this reason, exponential growth can be characterised and identified by using the growth rate or equivalently by using the doubling time. This approach is inapplicable to the hyperbolic growth or to any other type of growth. It is incorrect to use the doubling time to characterise any other type of growth. In particular, it is incorrect to use the so-called "rule of 70" for any other type of growth because in all other cases the growth rate and the doubling time are not constant. In order to characterise any other types of growth by the growth rate or by the doubling time we cannot just present a single value for any of these two quantities at a certain time but we have to show how their growth rate or the doubling time depends on time or on the size of the growing entity. For instance, if we look at the eqn (4) we can see that, for the hyperbolic growth, the growth rate is *directly* proportional to the size of the growing entity. This is a useful characteristic feature of hyperbolic growth. Another characteristic feature of hyperbolic growth is that the growth rate *per size* of the growing entity is constant.

As discussed elsewhere (Nielsen, 2014), hyperbolic distributions are confusing because they appear to be made of two distinctly-different components, slow and fast, leading to countless misconceptions and misinterpretations of the distributions describing the growth of human population or the economic growth. However, the analysis of these distributions and their interpretation becomes trivially simple if reciprocal values of data are used, as shown in Figure 2, because according to the eqns 5 and 6, if the reciprocal values of data follow a decreasing straight line, then the growth is hyperbolic. We can then fit the reciprocal values to find the mathematical expression for the hyperbolic growth given by the eqn (5).

© RON W. NIELSEN, 2017, Explaining the Mechanism of Growth in the Past Two Million Years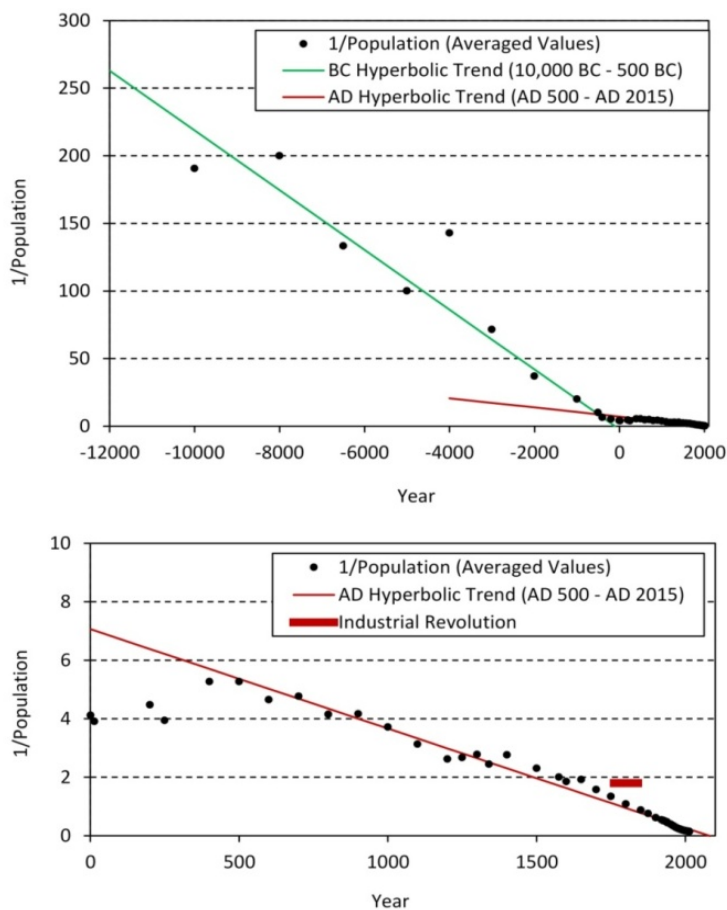

**Figure 2**. *Reciprocal values of the world population data (Manning, 2008; US Census Bureau, 2015) reveal two distinctly different hyperbolic trajectories (represented by the decreasing straight lines). They also show a dramatic demographic transition between around 500 BC and AD 500. Furthermore, they show that there was no takeoff around the time of the Industrial Revolution. In fact, there was no transition from stagnation to growth at any time. The size of the population is in billions.*

   Furthermore, if reciprocal values of data follow a decreasing straight line, the growth is not stagnant. However, the concept of stagnation is not supported even if the reciprocal values of data do not decrease linearly. Any monotonically-decreasing trajectory will show that the postulate of stagnation followed by a takeoff at a certain time is not supported by data. To prove the existence of the epoch of stagnation it is necessary to prove the presence of random fluctuations often described as Malthusian oscillations. Such random fluctuations should be clearly seen not only in the direct display of data but also in the display of their reciprocal values. It they are absent then there is no support in data for claiming the existence of the epoch of stagnation. However, if the reciprocal values of data follow a decreasing straight line, then they show, or at least strongly suggest, that the growth was hyperbolic. Positive identification of any type of growth depends on the range of available data.

   It should be also remembered that for the reciprocal values effects are reversed. A diversion to a *slower* trajectory will be indicated by an *upward* bending away from the earlier trajectory, while diversion to a *faster* trajectory will be indicated by a *downward* bending.

   Descriptions of economic growth involve frequent discussions of the so-called takeoffs (Galor, 2005, 2011) representing the assumed sudden and prominent change in the growth trajectory, a transition from the alleged stagnation to growth. For the economic growth or for the growth of human population represented by their reciprocal values, such sudden takeoffs should be indicated by a clear and strong downward bending of the growth trajectory.

   If the straight line representing the reciprocal values of data remains unchanged, then obviously, there is no change in the mechanism of growth. It makes no sense to divide a straight line into two or three arbitrarily selected sections and claim different regimes of growth controlled by different mechanisms for these arbitrarily-selected sections.

The analysis of data presented in Figure 2 reveals two distinctly different hyperbolic trajectories for the BC and AD eras. They are represented by two distinctly different straight lines fitting the reciprocal values of population data. In this representation, the growth during the AD era is dwarfed by the growth during the BC era but this part can be better examined by looking at the lower section of Figure 2.

The corresponding hyperbolic distributions are shown in Figure 3. Figures 1 and 2 make it clear that the growth of human population was not exponential, as claimed by Malthus (1798). Contrary to his expectations, data and their analysis show that *population, if unchecked, increases hyperbolically.* It shows that the growth of human population was increasing hyperbolically not only during the AD era, as observed by von Foerster, Mora and Amiot (1960), but also during the BC era. This analysis shows also that Industrial Revolution, 1760-1840 (Floud & McCloskey, 1994) did not boost the growth of human population, the result being in agreement with the analysis of the historical economic growth (Nielsen, 2016).

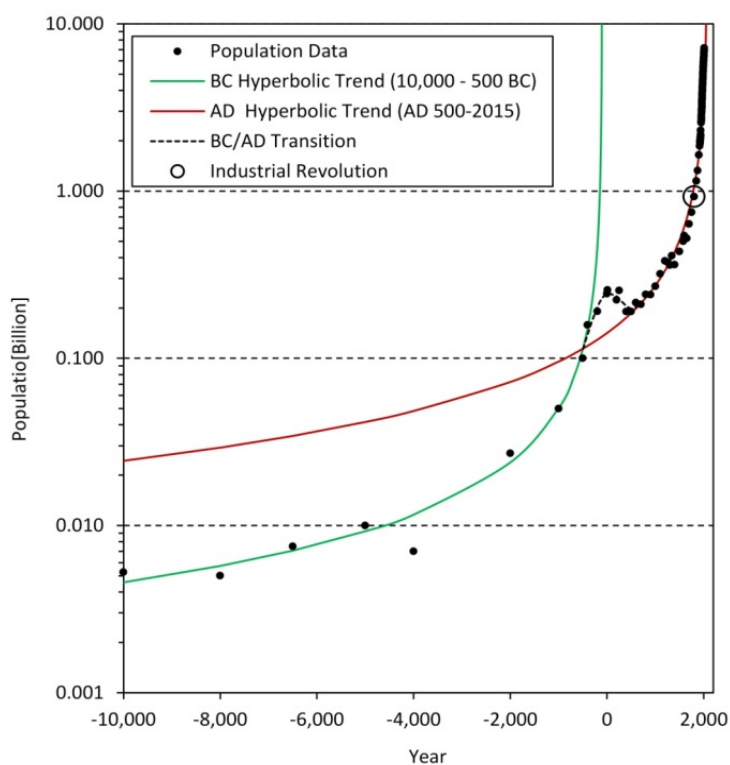

**Figure 3.** *Population, if unchecked, increases hyperbolically. This overall view shows that there was only one major demographic transition (between around 500 BC and AD 500) from a fast to a significantly slower hyperbolic trajectory. Industrial Revolution had no impact on the growth of human population. The perceived population explosion is just the natural continuation of hyperbolic growth.*

Results presented in Figures 2 and 3 show that from 10,000 BC to around 500 BC the growth of human population was hyperbolic. This hyperbolic growth was followed by a demographic transition between 500 BC and AD 500 from a fast BC hyperbolic trajectory to a significantly slower AD hyperbolic trajectory. It was not a transition from stagnation to growth because there was no stagnation in the growth of human population (Nielsen, 2013a).

Hyperbolic parameters fitting world population data are: $a = -2.282$ and $k = 2.210 \times 10^{-2}$ for the BC trajectory between 10,000 BC and 500 BC, and $a = 7.061$ and $k = 3.398 \times 10^{-3}$ for the AD trajectory between AD 500 and 2015. Characterised by the parameter $k$, the BC hyperbolic growth was 6.5 times faster than the AD growth.

Using the data (Manning, 2008; US Census Bureau, 2016), the fitted hyperbolic distributions (shown in Figure 3) and the eqn (4) we can now estimate the growth rates during the BC and AD eras. During the BC era, the growth rate was increasing hyperbolically (monotonically) with time or linearly (and again monotonically) with the size of the population from around $1.010 \times 10^{-4}$ (0.010%) per year in 10,000 BC to around $2.520 \times 10^{-3}$ (0.252%) per year in 500 BC. The growth was slow but not stagnant. During the AD era, the growth was again approximately hyperbolic between AD 500 and 1950, and the growth rate increased approximately monotonically from $6.337 \times 10^{-4}$ (0.063%, smaller than in 500 BC) to $7.805 \times 10^{-3}$ (0.781%) in 1950.

There was no stagnation but hyperbolic growth. The transition between 500 BC and AD 500 was not a transition from stagnation to growth but from growth to growth. It was not a dramatic takeoff but a transition to a *slower* hyperbolic trajectory. These features are important in relating the growth of population to the economic growth because contrary to the repeated claims presented in the Unified Growth Theory (Galor, 2005, 2011) there was no stagnation and no dramatic takeoff in the growth of the GDP (Nielsen, 2016) or in the growth of the GDP/cap (Nielsen, 2015a).

## 4. Detailed analysis of the AD data

Data for the AD era are of exceptionally good quality and they allow for a closer and minute examination of the pattern of growth. Even though the hyperbolic trajectory shown in Figures 2 and 3 fits the AD data well, the display of the reciprocal values presented in the lower part of Figure 2 shows that starting from around AD 1400, some data are systematically above the fitted straight line, suggesting a shift in the hyperbolic growth around that time.

Reciprocal values of data shown in Figure 4 reveal a clear delay in the growth of the population between around AD 1200 and 1400 followed by a new and slightly faster hyperbolic trajectory. Hyperbolic trajectory between AD 500 and 1200 is given by $a = 6.940$ and $k = 3.448 \times 10^{-3}$, and from AD 1400 by $a = 9.123$ and $k = 4.478 \times 10^{-3}$. For these new and improved fits to the data, the growth rate was $6.610 \times 10^{-4}$ (0.066%) in AD 500, $1.230 \times 10^{-3}$ (0.123%) in AD 1200, $1.568 \times 10^{-3}$ (0.157%) in AD 1400 and $1.142 \times 10^{-2}$ (1.142%) in 1950. The growth was hyperbolic (monotonic) between AD 500 and 1200 and again between AD 1400 and 1950. There was no stagnation and no dramatic takeoff from stagnation to growth at any time.

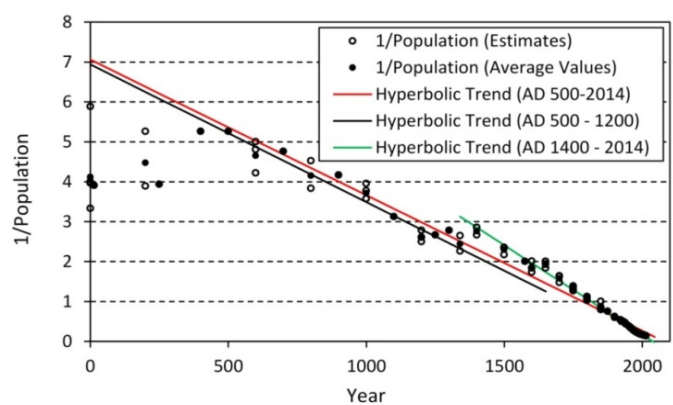

**Figure 4**. *Reciprocal values of data for the AD era show a clear but small disturbance in the growth of the population between AD 1200 and 1400. This disturbance caused a shift to a slightly faster hyperbolic trajectory. The size of the population in billions.*

Transition between AD 1200 and 1400 coincides with the unusual convergence of five strong lethal events. This minor, but noticeable delay in the growth of population appears to have been caused by a *combined impact of five* significant demographic catastrophes (Nielsen, 2013b): Mongolian Conquest (1260-1295) with the total estimated death toll of 40 million; Great European Famine (1315-1318), 7.5 million; the 15-year Famine in China (1333-1348), 9 million; Black Death (1343-1352), 75 million; and the Fall of Yuan Dynasty (1351-1369), 7.5 million. This is the only evidence in the data that demographic catastrophes might have had influence on the growth of the world population and if such is the case, not one but five of them were need to generate a small distortion.

There is no indication that exogenous conditions after AD 1400 were different than before AD 1200 so the slightly faster hyperbolic growth from around AD 1400 could be perhaps explained by the natural human response to crisis manifested in the intensified process of regeneration (Malthus, 1798; Nielsen, 2013c).

Closer view of the new growth trajectory, starting from around AD 1400, is displayed in Figure 5. The new hyperbolic growth was undisturbed until around 1950 when it experienced a *small but unsustained* acceleration, as indicated by a slight downward bending of the trajectory of the reciprocal values. This minor boosting lasted for only a short time and soon the growth of human population started to be diverted to a slower trajectory, as indicated by a change to an upward bending of the trajectory of reciprocal values.

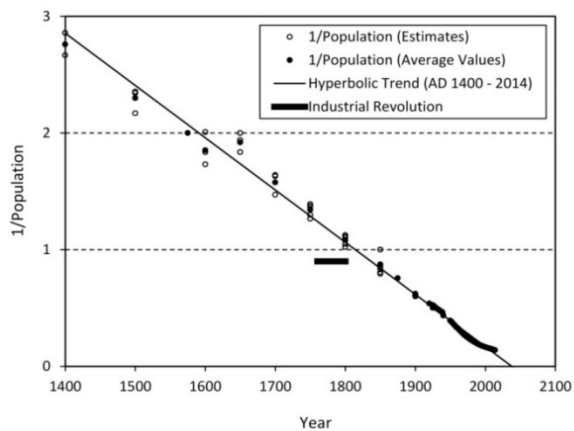

**Figure 5**. *Between AD 1400 and around 1950 the growth of human population was hyperbolic. Data show a minor boosting around 1950 followed quickly by a diversion to a slower trajectory. There was no takeoff from stagnation to growth at any time. Industrial Revolution had no impact on boosting the world population. The size of the population is in billions.*

Again, there was no dramatic takeoff and no transition from stagnation to growth as claimed repeatedly by Galor (2005, 2011). This hypothetical but non-existed takeoff, which was supposed to characterise not only the economic growth but also the growth of human population, is consistently contradicted by the analysis of the economic growth (Nielsen, 2015a, 2016) and by the presented here analysis of the growth of the world population.

It is remarkable that the growth of the world population was so hyperbolically stable over the past 12,000 years. The data show that during this long time there were only three transitions: 500 BC - AD 500, AD 1200 - AD 1400 and 1950 - present. Each of the two earlier transitions was a change-over between hyperbolic trajectories. The outcome of the current transition is unknown. The dynamics of growth in the past 12,000 years is summarised in Table 1.

**Table 1.** *Dynamics of growth of the world population in the past 12,000 years. Time intervals are approximate.*

| Hyperbolic Growth | Demographic Transitions |
|---|---|
| 10,000 BC – 500 BC $a = -2.282$, $k = 2.210 \times 10^{-2}$ | 500 BC – AD 500 Transition from a fast to much slower hyperbolic trajectory |
| AD 500 – 1200 $a = 6.940$, $k = 3.448 \times 10^{-3}$ | AD 1200 – 1400 Transition from a slow to a slightly faster hyperbolic trajectory |
| AD 1400 – 1950 $a = 9.123$, $k = 4.478 \times 10^{-3}$ | 1950 – present Transition from a hyperbolic trajectory to an unknown trend |
| Total time of hyperbolic growth: 10,750 years (~89% of the total combined time) | Total rime of transitions 1265 years (~11% of the total combined time) |

## 5. Implications for the economic growth

As mentioned earlier, preliminary analysis of Maddison's data (Maddison, 2001, 2010) shows close similarities between distributions describing economic growth and the growth of human population. Galor also commented that there was a "positive relationship between income per capita and population that existed throughout most of human history" (Galor, 2005, p. 177). The study of the economic growth goes hand in hand with the study of the growth of population.

Our analysis demonstrated that the growth of the world population was hyperbolic, and consequently monotonic. The frequently-claimed but never proven Malthusian stagnation in the growth of human population did not exist. There was also no transition from stagnation to growth, which could be described as a sudden takeoff. The fast-increasing growth of the world population in recent years is just the natural continuation of the hyperbolic growth.

Our analysis shows that with the exception of just two demographic transitions (500 BC - AD 500, and AD 1200 - 1400) the growth of human population was hyperbolic until around 1950, when it started to be diverted to a yet unknown trajectory. The first demographic transition (500 BC - AD 500) was from a faster to a slower hyperbolic growth. It was definitely not a takeoff from stagnation to growth. The second transition (AD 1200 – 1400) was from a slow to a slightly faster hyperbolic trajectory (only 30% faster, as indicated by the parameter $k$). It was also not a transition from stagnation to growth. The current transition, which commenced around 1950 was initially to a slightly faster trajectory, which was soon becoming progressively slower than the preceding hyperbolic trajectory. Here again, there was no transition from stagnation to growth. For about 89% of the past 12,000 years the growth of human population was hyperbolic and monotonic and *there was never a transition from stagnation to growth* because there was no stagnation. Our analysis shows that the growth of human population was remarkably stable and robust over the past 12,000 years.

Galor wonders "what is the origin of the sudden spurt in growth rates of output per capita and population?" (Galor, 2005, p. 177). This puzzle has now been solved: *there was no sudden spurt*.

Trying to explain this sudden spurt is like trying to explain why there is water in the middle of the desert when the image of water is created by a mirage. Such an exercise would be obviously ridiculous. Likewise, in the case of the growth of population or of the economic growth we can explain the *illusion* of the spurt but not existence of the spurt. The illusion of the spurt is explained by the hyperbolic properties but the sudden spurt has never happened. What we see as a sudden spurt is the natural continuation of the monotonically-increasing hyperbolic distribution and the simplest way to dispel the illusion of stagnation and of a sudden spurt is to use the reciprocal values of data (Nielsen, 2014) but we can also use other methods (Nielsen, 2015a). Data have to be rigorously analysed; otherwise it is easy to be distracted by illusions. Any perfunctory and hasty examination of data is likely to lead to incorrect conclusions and we can find many examples of such unprofessional use of data in the Unified Growth Theory (Galor, 2005, 2011).

We have demonstrated that there was no sudden spurt in the growth rate of the world population because the growth was hyperbolic, which means that the growth rate was also increasing hyperbolically with time or linearly with the size of the population, in both cases monotonically [see the eqn (4)]. Such an increase has no room for any form of spurts.

There were also no spurts during the past two demographic transitions. During the first transition (500 BC - AD 500), the growth rate *decreased* from 0.252% in 500 BC to 0.066% in AD 500. During the second transition (AD 1200 - 1400) the growth rate *increased only slightly* from 0.123% in AD 1200 to 0.157% in AD 1400.

So, our analysis eliminates at least one of Galor's spurts: the alleged spurt in the growth rate of human population. What remains to be explained is the alleged spurt in the growth rate of output per capita (GDP/cap) but the analysis of this ratio shows that the growth rate of the GDP/cap was also increasing monotonically (Nielsen, 2015a). There was no spurt at all. Furthermore, the analysis of the GDP data (Nielsen, 2016) shows that there were no spurts (takeoffs) in the growth of the GDP.

When data are properly analysed they show that what Galor saw as spurts in the growth rates represented just the natural features of *monotonically increasing* hyperbolic distributions describing the growth of population, the growth of the GDP, the growth of the GDP/cap and of their respective monotonically-increasing growth rates. All these distributions were slow over a long time and fast over a short time. These features are real but they represent nothing mysterious but the natural properties of monotonically-increasing hyperbolic distributions. They create strong illusions of stagnations followed by sudden spurts or takeoffs but when properly analysed they show that there was no stagnation and that the sudden spurts (takeoffs) never happened.

Galor wonders about the relationship between the income per capita (GDP/cap) and the population growth, but the answer to this apparent riddle is simple. When closely analysed, the growth of the population is found to be hyperbolic. The growth of the GDP is also hyperbolic (Nielsen, 2016) and hence, the growth of the GDP/cap is described by the ratio of hyperbolic distributions, which is just a linearly-modulated hyperbolic distribution (Nielsen, 2015a). The mystery is solved.

The only features, which need to be explained, are not the stagnation and sudden spurts (takeoffs) because they did not exist but why the growth of human population and the growth of the GDP were hyperbolic. This issue diverts our attention from the phantom problems, which do not need to be solved, and directs it to the problems, which need to be solved, because if we could explain why the growth of the population and the growth of the GDP were hyperbolic, we could also explain the mechanism of the historical income per capita.

Finally, we shall address a minor issue, which might help to understand at least one discrepancy between the fitted hyperbolic curve and the GDP data (Nielsen, 2016). In that analysis, we have found that one point, located at AD 1 was 77% higher than the fitted hyperbolic distribution. In Figure 3 we can see that something similar can be observed for the growth of human population. The size of the population in AD 1 was 71% higher than the size determined by the fitted hyperbolic distribution to the AD data, and the explanation of this discrepancy is simple: there was a maximum in the growth of the population around AD 1 caused by the transition from a fast-hyperbolic trajectory during the BC era to a significantly slower hyperbolic trajectory during the AD era. Close similarities between the growth of the GDP and the growth of the population displayed by Maddison's data (Maddison, 2001, 2010) suggest that the 77% difference between the GDP value and the fitted hyperbolic distribution at AD 1 (Nielsen, 2016) might reflect a similar maximum in the growth of the GDP as observed in the growth of the population.

## 6. Summary and conclusions

We have analysed the world population data (Manning, 2008; US Census Bureau, 2015) between 10,000 BC and AD 2015. We have found that the growth was hyperbolic during the BC and AD eras.

We have also found that there were just three, relatively, brief demographic transitions during that time: between 500 BC and AD 500, between AD 1200 and 1400 and currently from around 1950. These transitions were of a different kind than usually discussed in academic publications. Contrary to the frequently-repeated but never proven claims, there was never a transition from stagnation to growth because there was no stagnation.

The first transition was from a fast-hyperbolic trajectory to a significantly slower hyperbolic trajectory; the second from a slow hyperbolic trajectory to a slightly faster hyperbolic trajectory; and the current transition from the latest hyperbolic trajectory to a yet unknown trend. The total fraction of time characterising hyperbolic growth was about 89% of the past 12,000 years and the total time taken by transitions was only about 11%. Thus, the analysis shows that population, if unchecked, does not increase exponentially as believed by Malthus but hyperbolically. There was also no stagnation in the growth of the world population (Nielsen, 2013a), not only during the AD era but also during the BC era.

The correct understanding of the growth of human population is essential for the correct understanding of economic growth because, as pointed out by Galor (2005, 2011), and as can be easily checked using Maddison's data (Maddison, 2001, 2010), there is a close relationship between the growth of the population and the growth of the GDP. We have demonstrated that the growth of the world population was hyperbolic. Growth of the GDP was also hyperbolic (Nielsen, 2016). The growth of the world GDP/cap can be, therefore, described using hyperbolic distributions (Nielsen, 2015a). It is simply a ratio of hyperbolic distribution describing the growth of the GDP and the hyperbolic distribution describing the growth of human population. The ratio of two hyperbolic distributions is simply a linearly-modulated hyperbolic distribution (Nielsen, 2015a). There is nothing profoundly mysterious, perplexing or mind-boggling about such distributions as repeatedly claimed by Galor (2005, 2011).

It would be a waste of time, money and human resources to try to explain the mechanism of the historical economic growth, by using phantom features of stagnation and takeoffs. These erroneous and frequently-used postulates in the Unified Growth Theory (Galor, 2005, 2011) and in other publication are misleading and unhelpful. Explanations of the mechanism of the historical economic growth and of the growth of human population have to be based on accepting hyperbolic growth.